\RequirePackage{fix-cm}
\documentclass[twocolumn]{svjour3}          

\pdfoutput=1 

\usepackage{cite}

\usepackage{preprintcover}  
\PreprintCoverPaperTitle{Multi-channel search for squarks and gluinos in $\sqrt{s}=7$~\TeV\ $pp$ collisions with the ATLAS detector}  
\PreprintIdNumber{CERN-PH-EP-2012-330}  
\PreprintCoverAbstract{
A search for supersymmetric particles in final states with zero, one, and two leptons, with and without jets identified as originating from $b$-quarks, in 4.7~fb$^{-1}$ of $\sqrt{s}=7$~\TeV\ $pp$ collisions produced by the Large Hadron Collider and recorded by the ATLAS detector is presented.
The search uses a set of variables carrying information on the event kinematics transverse and parallel to the beam line that are sensitive to several topologies expected in supersymmetry.
Mutually exclusive final states are defined, allowing a combination of all channels to increase the search sensitivity.
No deviation from the Standard Model expectation is observed.
Upper limits at 95\,\% confidence level on fiducial cross-sections for the production of new particles are extracted.
Results are interpreted in the context of the constrained minimal supersymmetric extension to the Standard Model and in supersymmetry-inspired models with diverse, high-multiplicity final states.
}  
\PreprintJournalName{Eur. Phys. J. C}  

\smartqed  
\usepackage{graphicx}
\journalname{Eur. Phys. J. C}

\usepackage{atlasphysics}

\usepackage{amsmath}
\usepackage{amssymb}
\usepackage{lscape}
\usepackage{rotating}
\usepackage{slashed}
\usepackage{cite}
\usepackage[hyperindex,breaklinks]{hyperref}
\usepackage{xspace}

\newcommand{\ie}{\text{i.e.}\xspace}

\newcommand{\AKT}{anti-\ensuremath{k_{t}}\xspace}
\newcommand{\DeltaR}{\ensuremath{\Delta R_{\mathrm{cone}}}\xspace}

\begin{document}


\title{Multi-channel search for squarks and gluinos in $\sqrt{s}=7$~\TeV\ $pp$ collisions with the ATLAS detector at the LHC}

\titlerunning{Multi-channel search for squarks and gluinos with ATLAS}        

\author{The ATLAS Collaboration}
\authorrunning{The ATLAS Collaboration} 

\date{Received: date / Accepted: date}

\maketitle

\begin{abstract}
A search for supersymmetric particles in final states with zero, one, and two leptons, with and without jets identified as originating from $b$-quarks, in 4.7~fb$^{-1}$ of $\sqrt{s}=7$~\TeV\ $pp$ collisions produced by the Large Hadron Collider and recorded by the ATLAS detector is presented.
The search uses a set of variables carrying information on the event kinematics transverse and parallel to the beam line that are sensitive to several topologies expected in supersymmetry.
Mutually exclusive final states are defined, allowing a combination of all channels to increase the search sensitivity.
No deviation from the Standard Model expectation is observed.
Upper limits at 95\,\% confidence level on fiducial cross-sections for the production of new particles are extracted.
Results are interpreted in the context of the constrained minimal supersymmetric extension to the Standard Model and in supersymmetry-inspired models with diverse, high-multiplicity final states.
\end{abstract}

\section{Introduction}
\label{sec:intro}

One of the most promising extensions of the Standard Model, supersymmetry (SUSY)~\cite{Miyazawa:1966,Ramond:1971gb,Golfand:1971iw,Neveu:1971rx,Neveu:1971iv,Gervais:1971ji,Volkov:1973ix,Wess:1973kz,Wess:1974tw}, has been the target of a large number of searches at the LHC. Prompted by the large predicted production cross-section of coloured SUSY particles (sparticles), ATLAS and CMS have performed inclusive searches for strongly produced squarks and gluinos, the superpartners of quarks and gluons~\cite{0leptonATLAS,0leptonATLASMJ,0leptonATLASMBJ,multiLepMultiJetATLAS,CMSMultiJetMET,CMSOSLep,CMSSSLep,CMSSSLepWithB}.
Assuming R-parity conservation~\cite{Fayet:1976et,Fayet:1977yc,Farrar:1978xj,Fayet:1979sa,Dimopoulos:1981zb}, these sparticles are produced in pairs and decay into energetic jets, possibly leptons, and the lightest SUSY particle (LSP, typically the lightest neutralino $\tilde{\chi}_1^0$), which escapes detection and results in missing transverse momentum.
For these searches, the selections adopted to discriminate the signal processes from the background typically include requirements on the missing transverse momentum (\MET) and the scalar sum of transverse momenta of all selected physics objects ($\HT$) plus the scalar \MET.

This paper presents a search for strongly produced sparticles that makes use of a variety of final states including high transverse momentum jets and zero, one, or two leptons (electrons or muons).
The events are also separated according to the presence of a jet identified as originating from a $b$-quark ($b$-tagged jets).
Several mutually exclusive search channels are defined, facilitating a simultaneous search in all of the typical final states and increasing the search sensitivity.
The search employs a set of observables, called the ``razor variables''~\cite{Crogan}, which make use of both longitudinal and transverse event information.
These variables were first employed in SUSY searches by CMS~\cite{CMS35pb,CMS47fb}. 
Their use makes this search sensitive to different regions of kinematic phase space relative to the other \MET-based searches.
Thus, these search results complement those already performed by ATLAS.

This paper is organised as follows. 
The main features of the ATLAS detector are presented in Sect.~\ref{sec:atlas}.
Section~\ref{sec:variables} introduces the razor variables.
Section~\ref{sec:data-selection} describes the data sample, basic event selection, and the Monte Carlo simulation used to model the data.
Section~\ref{sec:objects} defines the basic physics objects and event-level variables that are used through the analysis.
The search technique is described in Sect.~\ref{sec-search}, and the background estimation is presented in Sect.~\ref{sec:bg}.
The performance of the search and interpretation of the results are presented in Sect.~\ref{sec:exclusion}.
Finally, Sect.~\ref{sec:summary} includes a summary of the analysis and of its findings. 

\section{ATLAS detector}
\label{sec:atlas}

The ATLAS detector comprises an inner tracking detector, a calorimeter, and a muon system~\cite{techpaper}.
The inner detector includes a silicon pixel detector, a silicon microstrip detector, and a transition radiation tracker.
It is immersed in a 2 T axial field and precisely measures the tracks of charged particles in the pseudorapidity region\footnote{
ATLAS uses a right-handed coordinate system with its origin at the nominal interaction point (IP) in the centre of the detector and the $z$-axis along the beam pipe. 
The $x$-axis points from the IP to the centre of the LHC ring, and the $y$-axis points upward. 
Cylindrical coordinates $(r,\phi)$ are used in the transverse plane, $\phi$ being the azimuthal angle around the beam pipe. 
The pseudorapidity is defined in terms of the polar angle $\theta$ as $\eta=-\ln\tan(\theta/2)$.} $|\eta|<2.5$.
The calorimeter covers the region $|\eta|<4.9$ and is divided into electromagnetic and hadronic compartments.
The electromagnetic calorimetry in the central ($|\eta|<3.2$) region is provided by liquid argon sampling calorimeters with lead absorbers.
In the barrel region ($|\eta|<1.4$), the hadronic calorimetry is provided by scintillator tiles with steel absorbers, and the more forward ($1.4<|\eta|<3.2$) region is covered by a liquid argon and copper sampling hadronic calorimeter.
The forward calorimetry ($|\eta|>3.2$) uses liquid argon and copper or tungsten absorbers.
The muon spectrometer covers $|\eta|<2.7$ and includes a system of air-core toroidal magnets.
A variety of technologies are used to provide precision muon tracking and identification for $|\eta|<2.7$ and rapid response for triggering for $|\eta|<2.4$.

ATLAS uses a three-tier trigger system to select events.
The first-level (L1) trigger is hardware-based and only uses coarse calorimeter information and muon system information.
The calorimeter information available at the lowest level includes basic objects with rough calibration and simple identification of electromagnetic objects (electrons and photons) as distinct from hadronic objects (jets). 
The second-level (L2) trigger and event-filter (EF) compose the software-based high-level trigger (HLT), in which full event reconstruction is run, similar to that used offline, in order to accurately identify and measure objects. The L2 only examines $\eta/\phi$ regions that triggered the L1. The EF fully reconstructs events that pass L2.

\section{Razor variable definitions}
\label{sec:variables}

Searches for sparticles in R-parity-conserving scenarios generally make the
assumption that the sparticles are pair-produced and decay subsequently to an
LSP that is invisible in the detector.  The heavy sparticles produced are 
either the same type of particle (pair-production) or are at the
same mass scale.  Thus, the production and visible energy in the decays are fairly
symmetric.  Most analyses make use of the transverse balance of typical
$pp$ collision events, or exploit the event symmetry in the
transverse plane.  The razor variables attempt to also include longitudinal
information about the event by making several assumptions motivated by the
kinematics of the models of interest.

In the rest frame of each heavy sparticle, called the $R$-frame, the
sparticle decays are symmetric. In an attempt to reconstruct the
primary produced sparticle pair, the razor calculation clusters all
final-state particles into a pair of objects with four-momenta called ``mega-jets.'' 
Each of these mega-jets is associated with one of the two SUSY decay 
chains and represents the visible energy-momentum of that 
produced sparticles.  All possible combinations of the four-vectors 
of the visibly reconstructed/selected objects (signal jets and leptons) are 
considered when constructing the two mega-jets.
The pair of mega-jets, $j_{1}$ and $j_{2}$, that minimises the
sum of the squared masses of the four-vectors is selected.  
Following the prescription in Ref.~\cite{Crogan} and for consistency 
with Ref.~\cite{CMS35pb}, all jets and the mega-jets are forced to be
massless by setting their energy equal to the magnitude of their three
momenta. Studies indicate that neither this choice nor the mega-jet
selection, based on minimizing the mega-jet mass squared, have a 
significant impact on the reach of the razor-based search.

In the $R$-frames, each heavy sparticle should be nearly at rest with some mass $m_\text{Heavy}$. The sparticle decay may then be approximated as a two-body decay to some visible object (a mega-jet) and the invisible, stable LSP. The final visible decay products (\ie\ the final-state quarks and gluons, or the observable jets and leptons) have masses far below the SUSY mass scale and can therefore be approximated as being massless. Then the energy of each mega-jet in the $R$-frame, $E_1$ and $E_2$, becomes:

\begin{equation}
E_1 = E_2 = \frac{ m_\text{Heavy}^2 - m_\text{LSP}^2 } { 2 \times m_\text{Heavy} },
\end{equation}

\noindent where $m_\text{LSP}$ is the mass of the LSP. This leads to a characteristic mass, $M_R$, in the $R$ frame of $M_{R}=2\times E_1 = 2\times E_2$, which for $m_\text{Heavy} \gg m_\text{LSP}$ is identical to $m_\text{Heavy}$. Therefore, in events where heavy particles are pair-produced, $M_R$, which is a measure of the scale of the heaviest particles produced, should form a bump. In \ttbar\ or $WW$ events, for example, the characteristic mass $M_R\approx m_\text{top}$ or $m_{W}$.  The product of $M_R$ and the Lorentz factor for the boost from the lab to $R$-frame, $M_R'=\gamma_{R} \times M_{R}$, is useful for characterisation of the sparticle mass scale, in part because of its close relation to $m_\text{Heavy}$, and in part because Standard Model backgrounds tend to have small values of $M_R'$. When expressed in terms of the mega-jet quantities in the lab frame, the expression is given by:

\begin{equation}
M_{R}' = \sqrt{ ( j_{1,E} + j_{2,E} )^2 - ( j_{1,z} + j_{2,z} )^2 } ,
\end{equation}

\noindent where $j_{i,E}$ and $j_{i,z}$ are the energy and
longitudinal momentum, respectively, of mega-jet $i$.  The transverse
information of the system is taken into account by constructing a
transverse mass for the mega-jets, assuming half of the \MET\ is
associated with each jet:

\begin{equation}
\begin{split}
M_\text{T}^R = \left[ \frac{1}{2} \times |\vec{\MET}| \times ( |\vec{j_{1,T}}| + |\vec{j_{2,T}}| ) \right. \\
        \left. - \frac{1}{2} \times \vec{\MET} \cdot ( \vec{j_{1,T}} + \vec{j_{2,T}} ) \right]^{1/2},
\end{split}
\end{equation}

\noindent where $\vec{\MET}$ is the two-dimensional vector of the \MET\ in the transverse plane. 
When an event contains ``fake'' \MET\ from a detector defect or mismeasurement, the system will tend to have back-to-back mega-jets. 
In such cases, the vector sum of the two mega-jet momenta will be small.
If, on the other hand, there is real \MET, the mega-jets may not be back-to-back and may even point in the same direction.
In these cases, the vector sum, and thus $M_T^R$, will have a large value.
$M_\text{T}^R$ is another measure of the scale of the event that only uses transverse quantities in contrast to longitudinal quantities in $M_{R}'$.

Finally a razor variable is defined to discriminate between signal and background: 

\begin{equation}
R = \frac{ M_\text{T}^R } { M_{R}' } .
\end{equation}

\noindent This variable takes low values for multijet-like events and tends to be uniformly distributed between 0 and 1 for sparticle decay-like events, providing good discrimination against backgrounds without genuine \MET. 
The impact of some important experimental uncertainties, like the jet energy scale uncertainty, are reduced in this ratio.
In an analysis based on the razor variables, a cut on $R$ can be used to eliminate these backgrounds before a SUSY search is made in the distribution of the variable $M_R'$.

\section{Data and Monte Carlo samples}
\label{sec:data-selection}

The data included in this analysis were collected between March and October 2011.
After basic trigger and data quality requirements, the full dataset corresponds to $4.7\pm0.2$~fb$^{-1}$~\cite{luminosityUnc,luminosityUnc2}.

Events in the zero-lepton channels are selected using a trigger that requires a jet with transverse momentum $\pt>100$~\GeV\ at L1.  In the event filter, $\HT>400$~\GeV\ is required, where $\HT$ is calculated through a scalar sum of the $\pt$ of all calorimeter objects with $\pt>30$~\GeV\ and $|\eta|<3.2$.  With the exception of a cross-check of the multijet background estimate, which uses prescaled single-jet triggers, this trigger requirement is fully efficient for the offline selection used in the analysis.

The one- and two-lepton channels make use of the lowest-$\pt$ single-lepton triggers available for the entire running period.  The muon triggers require a muon with $\pt>18$~\GeV, and the electron triggers require an electron with $\pt>22$~\GeV.  Offline, the leading lepton in the event is required to have $\pt>20$~\GeV\ ($\pt>25$~\GeV) if it is a muon (electron), in order to ensure that the triggers are fully efficient with respect to the offline event selection.  For the two-lepton analysis, where there are overlaps in the triggers, the electron trigger takes priority over the muon trigger.

Offline, an event is required to have at least one vertex with at least five tracks associated to it, each with $\pt^{\rm track}>400$~\MeV. This requirement reduces cosmic ray and beam-related backgrounds. The primary vertex is defined as the one with the largest $\sum (\pt^{\rm track})^2$ of the associated tracks. Events that suffer from sporadic calorimeter noise bursts or data integrity errors are also rejected.


Monte Carlo (MC) simulated events were used to develop the analysis and assist 
in estimations of background rates. All MC samples are processed through ATLAS's full detector 
simulation~\cite{SimJINST} based on {\sc Geant4}~\cite{geant}, which was run 
with four different configurations corresponding to detector conditions of four 
distinct operating periods of 2011. The fractions of MC simulation events in 
these four periods match the fractions of data in each period.  
During the data collection, the average number of proton--proton collisions per bunch 
crossing in addition to the one of interest (``event pile-up'' or simply ``pile-up'') 
increased from approximately two to twelve.
To mimic the effect of pile-up, additional inelastic proton--proton collisions are 
generated using {\sc Pythia}~\cite{pythia} and overlaid on top of every MC 
event.  Within each period, the profile of the average number of events per 
bunch crossing ($\langle\mu\rangle$) is re-weighted to match the data in that period. The 
same trigger selection is applied to the MC simulation events, which are then passed 
through the same analysis code as the data.  Reconstruction and trigger 
efficiency scale factors are applied to the MC simulation in order to take into 
account small discrepancies between the data and the MC simulation.

Table~\ref{table:backgroundSummary} lists the major backgrounds along
with the chosen estimation method (described in Sect.~\ref{sec:bg})
and the primary and alternative MC generators used in this analysis.  In all cases, {\sc MC@NLO} 
and {\sc Alpgen} are interfaced to {\sc Herwig} and {\sc Jimmy} for the 
parton shower, hadronisation, and underlying event modelling.  The multijet
background is normalised to the leading order generator cross-section predicted by {\sc Pythia}.
The \ttbar\ production cross-section of 166.8~pb is calculated at approximate NNLO in
QCD using Hathor~\cite{Hathor} with the MSTW2008 NNLO PDF
sets~\cite{MSTW}.   The calculation is
cross-checked with an NLO+NNLL calculation~\cite{CTopXSec} implemented 
in Top++~\cite{Top++}.  The single-top 
production cross-sections are calculated separately for
$s$-channel, $t$-channel, and $Wt$ production at
NNLO~\cite{Kidonakis1,Kidonakis2,Kidonakis3}.  

The $W$ and $Z$ (including Drell--Yan with $m_{\ell\ell}>40$~\GeV) production
cross-sections of 10.46~nb and 0.964~nb are calculated at NNLO using
FEWZ~\cite{FEWZ}.  For the production of vector bosons in association with
heavy flavour, in accordance with ATLAS measurements~\cite{WHFatlas},
the production cross-section for $W+\bbar$ and $W+\ccbar$ are scaled
by 1.63, and the cross-section for $W+c$ is scaled by $1.11$ compared
to the NLO cross-section~\cite{HFXsec}.  Additional
uncertainties on the production of $W$ and $Z$ bosons in association
with heavy flavour of 45\,\% for $W+\bbar$ and $W+\ccbar$, 32\,\% for
$W+c$, and 55\,\% for $Z+b\bbar$ are included. {\sc Alpgen} describes the
jet multiplicity and inclusive $M_R'$ distributions well, but it does
not correctly model the vector boson $\pt$ distribution.  Therefore,
the boson $\pt$ in the {\sc Alpgen} samples is re-weighted according
to the distribution produced {\sc Sherpa}. Half of the difference between the weight 
and unity is applied as a systematic uncertainty on the re-weighting
procedure.  Further systematic uncertainties
on the shapes of {\sc Alpgen} samples are derived by systematically
varying the generator parameters, including matching and factorisation
scales.  Diboson production cross-sections of 44.92~pb, 17.97~pb, and 9.23~pb
for $WW$, $WZ$, and $ZZ$ (including off-shell production with
$m_{\ell\ell}>12$~\GeV) are calculated at NLO using MCFM~\cite{MCFM}.
In order to avoid low-mass resonances, all dilepton events are required
to have the invariant mass $m_{\ell\ell}>20$~\GeV. These cross-sections provide the
starting normalisations for all background processes.

\begin{table*}[tbp]
\caption{
Background estimation methods, primary and alternative MC event generators, and normalisation uncertainties for each of the major backgrounds.
The backgrounds are constrained using various Control Regions (CRs) that are enriched in certain samples (see Sect.~\ref{sec-search}). 
The \ttbar\ background estimate includes small contributions from $\ttbar W$ and $\ttbar Z$, generated with MadGraph.
The diboson $WW$ background estimate also includes $W^\pm W^\pm jj$ generated with MadGraph.
The last column of the table indicates the uncertainty on the normalization in the simultaneous fit used to test signal hypotheses.
``None'' indicates that the normalization is fully constrained in the fit.  
The grouping indicates the samples that are collected together in the fit.  Within a group, the relative normalizations are fixed.
\label{table:backgroundSummary}}
\begin{center}
\scalebox{0.92}{
\begin{tabular}{lllllll}
Background             & 0-lepton    & 1-lepton      & 2-lepton      & Generator                   & Alternate                  & Normalisation Uncertainty \\
\hline\noalign{\smallskip}
Multijets              & MJ CRs      & Matrix method & Matrix method & {\sc Pythia}~\cite{pythia}  & {\sc Alpgen}~\cite{Alpgen} & None \\
$W\rightarrow\ell\nu$  & $W$ CRs     & $W$ CRs       & Matrix method & {\sc Alpgen}~\cite{Alpgen}  &                            & None (grouped with $Z$) \\
$Z\rightarrow\ell\ell$ & $Z$ CRs     & $Z$ CRs       & $Z$ CRs       & {\sc Alpgen}~\cite{Alpgen}  &                            & None (grouped with $W$) \\
Drell--Yan              & $Z$ CRs     & $Z$ CRs       & $Z$ CRs       & {\sc Alpgen}~\cite{Alpgen}  &                            & None (grouped with $W$) \\
$Z\rightarrow\nu\nu$   & $Z$ CRs     & Matrix method & Matrix method & {\sc Alpgen}~\cite{Alpgen}  &                            & None (grouped with $W/Z$) \\
\ttbar (had)           & \ttbar\ CRs & \ttbar\ CRs   & \ttbar\ CRs   & {\sc MC@NLO}~\cite{McAtNlo} &                            & None \\
\ttbar (leptonic)      & \ttbar\ CRs & \ttbar\ CRs   & \ttbar\ CRs   & {\sc Alpgen}~\cite{Alpgen}  & {\sc MC@NLO}~\cite{McAtNlo,McAtNlo1,McAtNlo2,McAtNlo3} & None \\
Single top             & \ttbar\ CRs & \ttbar\ CRs   & \ttbar\ CRs   & {\sc MC@NLO}~\cite{McAtNlo} &                            & None (grouped with \ttbar) \\
$WW$ diboson          & MC          & MC            & MC            & {\sc Herwig}~\cite{herwig}  & {\sc Alpgen}~\cite{Alpgen} & NLO $\pm$ 30\,\% \\
Other diboson         & $Z$ CRs     & $Z$ CRs       & $Z$ CRs       & {\sc Herwig}~\cite{herwig}  & {\sc Alpgen}~\cite{Alpgen} & None (grouped with $W/Z$) \\
\noalign{\smallskip}\hline
\end{tabular}
}
\end{center}
\end{table*}


Two SUSY-inspired simplified models are used for the interpretation of
the results from this search.  The first considers gluino
pair-production, with the gluino decaying to a \ttbar\ pair and the
LSP via an off-shell stop.  This model is generated using {\sc
Herwig++}~\cite{Herwigpp}, with the gluino and LSP masses being the only
free parameters.  The top quarks are required to be on-shell, limiting the
mass splitting between the gluino and the LSP to greater than $2\times
m_\text{top}$.

The second considers gluino pair-production, with the gluino decaying
to two quarks and a chargino via an off-shell squark.  The chargino then
decays to a $W$ boson and the LSP.  The free parameters of this model
are the masses of the gluino, chargino, and LSP.  For convenience, two
two-dimensional planes are generated: one with the chargino mass
exactly between the masses of the gluino and the LSP and one
with the mass of the LSP fixed to 60~\GeV.  Because initial-state
radiation can be important for the acceptance of these models when the
mass splitting between the gluino and LSP is small, this model is generated
using {\sc MadGraph}~\cite{MadGraph} with at most one additional jet in the
matrix element.  {\sc Pythia} is used for the parton shower and 
hadronisation.  Systematic uncertainties on matrix element matching
and initial-state radiation modelling are included, leading to 20\,\%
uncertainties for small mass splittings and small gluino masses, but no
uncertainty for mass splittings above 200~\GeV\ and masses above
400~\GeV.

Additionally, the results are interpreted in terms of SUSY signal models based on the constrained minimal supersymmetric model (CMSSM or MSUGRA)~\cite{Fayet:1976et,Fayet:1977yc,Farrar:1978xj,Fayet:1979sa,Dimopoulos:1981zb}.
The parameters of this model are the high-energy-scale universal scalar mass, $m_0$, the universal gaugino mass, $m_{1/2}$, the ratio of the vacuum expectation values of the two Higgs fields, $\tan(\beta)$, the tri-linear coupling strength, $A_0$, and the sign of the Higgsino mass parameter, $\mu$.
Samples are generated in a two-dimensional grid of the  $m_0$--$m_{1/2}$ parameters where $\tan(\beta)=10$ and $A_0=0$ are fixed and $\mu$ is set positive.
This MC data grid is generated using {\sc Herwig++}~\cite{Herwigpp}, with a more dense population of points at low mass. IsaSUSY~\cite{Isajet} is used to run the high-energy-scale parameters down to the weak-scale. 

Signal cross-sections are calculated to next-to-leading order in the
strong coupling constant, adding the resummation of soft gluon
emission at next-to-leading-logarithmic accuracy
(NLO+NLL)~\cite{Beenakker:1996ch,Kulesza:2008jb,Kulesza:2009kq,Beenakker:2009ha,Beenakker:2011fu}. The
nominal cross-section and the uncertainty are taken from an envelope
of cross-section predictions using different PDF sets and
factorisation and renormalisation scales, as described in
Ref.~\cite{Kramer:2012bx}.  For each of these signal models, the
luminosity systematic uncertainty of 3.9\,\%~\cite{luminosityUnc,luminosityUnc2} and
statistical uncertainty, typically of order 10\,\%, is included.

\section{Physics object identification and selection}
\label{sec:objects}

Events are categorised into six exclusive samples defined by the
presence of zero, one, or two leptons, with or without $b$-tagged jets. The
particle candidate selections that define these samples are referred
to as the ``baseline'' object selection. Since a particle may
simultaneously satisfy multiple particle hypotheses (e.g. electron
and jet), an overlap removal procedure (described below) assigns
a unique interpretation to each candidate. The selections are then
further refined to define ``signal'' candidates with higher signal-to-background ratio in each sample.

Baseline electrons are required to have $\ET>10$ \GeV, be within the
fiducial acceptance of the inner detector ($|\eta|<2.47$), and 
pass a version of the ``medium'' selection
criteria~\cite{electronID} updated for 2011 running conditions, which
requires hadronic calorimeter energy deposition and a calorimetric shower shape consistent
with an electron and a match to a good quality inner detector track. Signal electrons
are required to be isolated from other objects and satisfy ``tight''
selections. The tight selection applies
stricter track quality and matching than medium and ensures the number
of hits in the transition radation tracker is consistent with the
electron hypothesis. The isolation requirement is that the sum of the $\pt$ of
all charged particle tracks associated with the primary vertex within $\Delta R=0.2$,
where $\Delta R = \sqrt{(\Delta \eta)^2 + (\Delta \phi)^2}$, of the electron is less
than 10\,\% of the electron $\ET$. 
In the leptonic channels, if the leading lepton in a data event is an electron, it
is additionally required to match to an EF trigger electron.
MC simulation events are re-weighted to compensate for mis-modelling of the single-lepton
trigger efficiency. The energy of electrons in simulated events is also smeared
prior to object selection in order to reproduce the resolution in
$Z$ and $J/\psi$ data.  Finally, in order to account for differences in electron
reconstruction efficiency, $\eta$- and $\ET$-dependent 
scale factors, derived from $Z$, $W$, and $J/\psi$ events in the data, are applied 
to each simulated electron satisfying overlap removal and selection requirements.

Baseline muons are reconstructed as either a combined track in the
muon spectrometer and inner detector, or as an inner detector
track matching with a muon spectrometer segment~\cite{MuonReco}. Tracks are
required to have good quality, and the muon is required to have
$\pt>10$ \GeV\ and $|\eta|<2.4$. Signal muons are required to be
isolated by ensuring that the sum of the $\pt$ of all charged
particle tracks associated with the primary vertex within $\Delta R=0.2$ of the muon is less than
1.8~\GeV. Matching to EF trigger muons in data, MC event trigger re-weighting,
muon momentum smearing, and MC/data efficiency scaling are performed in
a similar way for muons as electrons (described
above)~\cite{MuonRes,MuonEff1,MuonEff2}.

Calorimeter jets are reconstructed from topological clusters of energy 
deposited in the calorimeter
calibrated at the electromagnetic (EM) scale~\cite{topocluster} using the
\AKT jet algorithm~\cite{Cacciari:2008gp,FastJet} with a four-momentum
recombination scheme and a distance parameter of $0.4$.  Jets reconstructed 
with an EM-scale $\pt>7$~\GeV\ are calibrated to
the hadronic scale (particle level) using $\pt$ and $\eta$-dependent
factors, derived from simulation and validated with test beam and collision
data~\cite{JetPaper}. In order to remove specific
non-collision backgrounds, events are rejected if they contain a
reconstructed jet that does not pass several quality and selection
criteria~\cite{JetPaper}.  Signal jets are selected if they lie within $|\eta|<2.5$ 
with a jet vertex fraction (JVF) of at least 75\,\%, 
where the JVF is the fraction of summed $\pt$ of the tracks associated
with the jet that is carried by tracks consistent with the primary vertex of the event, thus
associating the jet with the $pp$ collision of interest. 
Jets are tagged as heavy flavour using the combined neural network ``jet fitter''
algorithm~\cite{JetFitterBTagging} with the 60\,\% efficiency working
point. Scale factors for heavy flavour jets are used in MC simulation in order to
reproduce the expected $b$-jet identification performance in data.

In order to ensure that objects are not double counted, 
overlaps between objects are removed
using a hierachical procedure.  If any two baseline electrons lie
within a distance of $\Delta R=0.1$ of one another, the electron with
the lower calorimeter $\ET$ is discarded.  Next, jets passing basic
selections are required to be at least 0.2 units away from all surviving
baseline electrons in $\eta$--$\phi$.  Electrons are then required to be
at least 0.4 units away from surviving jets. Finally, in order to mitigate the effect of
jets which have deposited significant energy in the muon spectrometer on mass measurements 
and reduce the number of events with badly measured missing transverse momentum, muons with
$\pt>250$~\GeV\ within $\Delta R=0.2$ of a jet with $\pt>500$~\GeV\ are
removed. A negligible number of events in the data are removed by this
cut.

Following these overlap removal procedures, the missing transverse
momentum and razor variables are calculated. 
The determination of the missing transverse momentum uses all baseline
electrons with
$\ET>20$~\GeV, all baseline muons, all calibrated jets with
$\pt>20$~\GeV, and EM scale topological calorimeter clusters not
belonging to any object. Note that in the MC simulation, objects enter
this calculation after the energy or $\pt$ smearing described above. 

In counting leptons for event classification, baseline electrons and muons
are then required to be at least 0.4 units away from all good jets in
$\eta$--$\phi$. If an electron and muon are separated by $\DeltaR < 0.1$, 
neither is counted.

In order to remove
events with large missing transverse momentum due to cosmic rays,
events are vetoed if they contain a muon in which the transverse 
and longitudinal impact track parameters are greater than $0.2$ mm 
and $0.1$ mm with respect to the primary vertex, respectively.
Also vetoed are events with badly measured, non-isolated muons.
These muons with large momentum uncertainties are rare in both the 
signal and background events and can have significant impact on 
the \MET\ and razor variables.

During a portion of the run period, a hardware failure resulted in a 
region of the calorimeter not being read out.  For
data collected during this period, and for a corresponding fraction of
the MC samples, events are rejected if they fail the ``smart LAr hole
veto''~\cite{multiLepMultiJetATLAS}.  This ensures that if an event contains
one or more jets pointing to the dead region and those jets may
contribute substantially to the missing transverse momentum in the
event, the event is discarded.

Signal regions are defined after all overlap removal is complete.
Events with no baseline leptons and events with the highest-$\pt$
lepton below the leading lepton requirement (25~\GeV\ for electrons,
20~\GeV\ for muons) are accepted into the zero-lepton regions.  Events
with one leading lepton satisfying all requirements, including that on leading lepton $\pt$ (above),
and no other baseline leptons with $\pt>10$~\GeV\ are accepted into the
one-lepton regions.  Events with exactly one additional signal lepton
above 10~\GeV\ and no other baseline leptons are accepted into the
two-lepton regions.

\section{Search technique}
\label{sec-search}

\begin{figure*}[htp]
\centering
\includegraphics[width=0.47\textwidth]{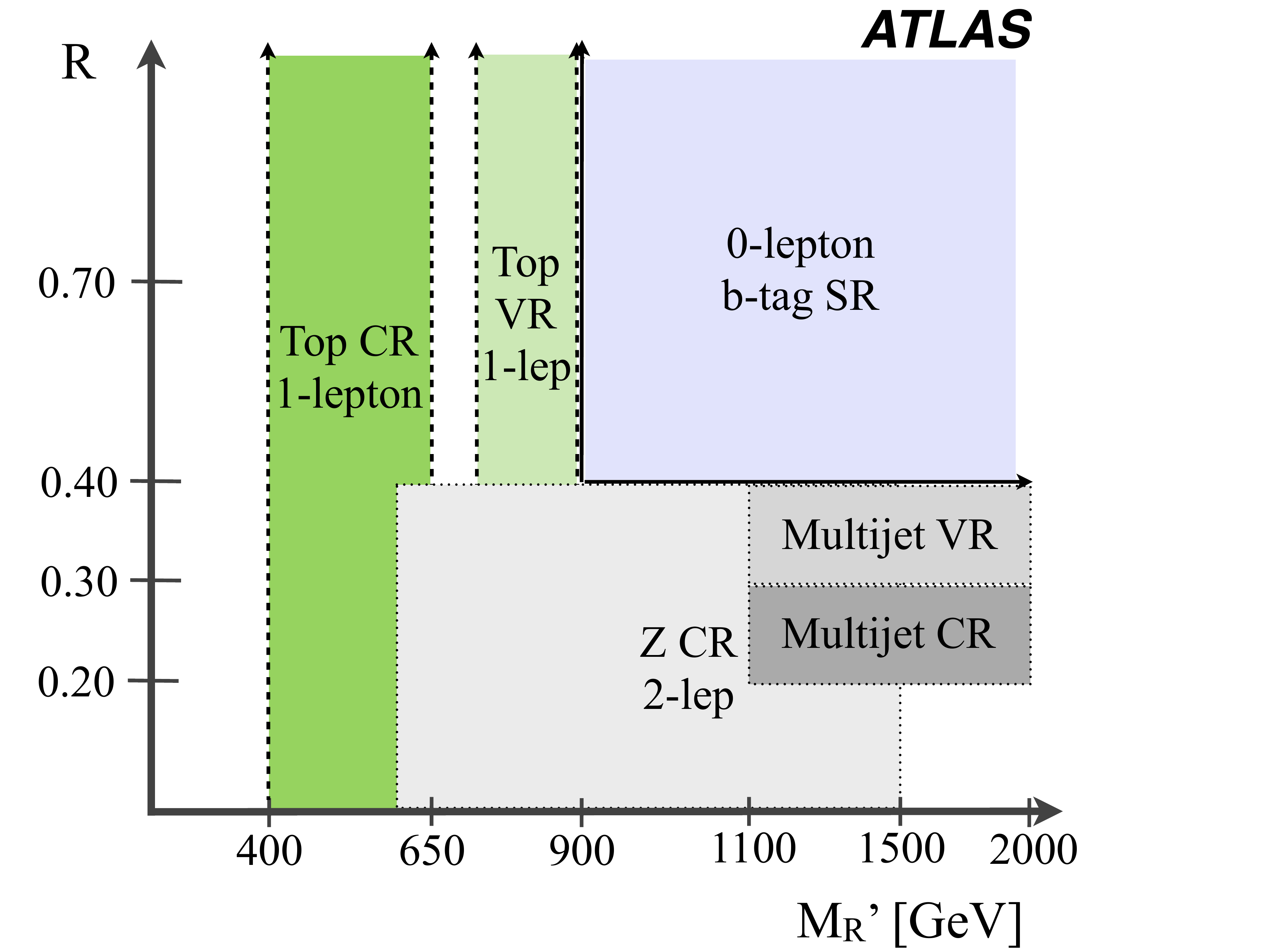}
\includegraphics[width=0.47\textwidth]{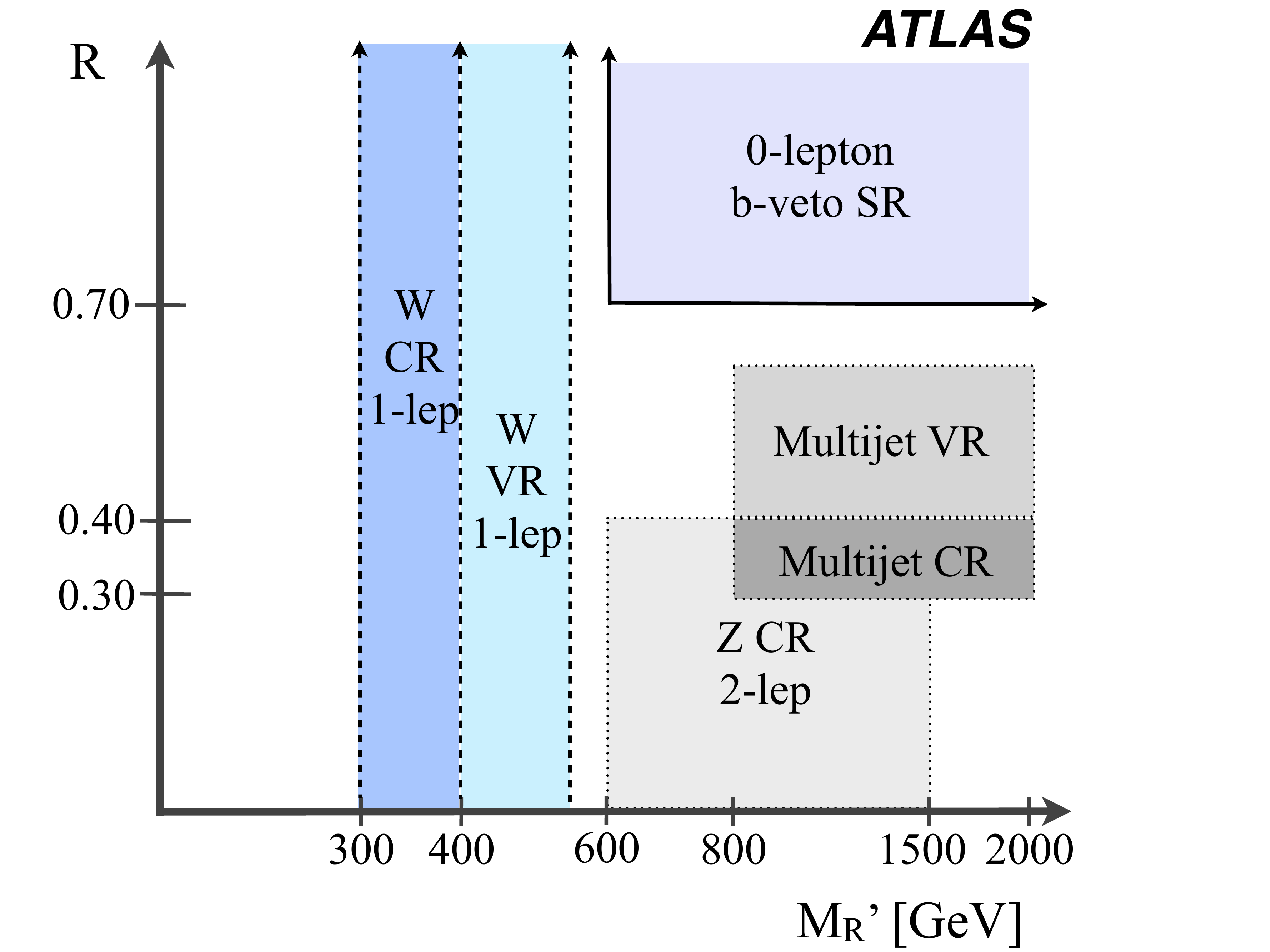}
\smallskip
\vspace{1 cm}
\includegraphics[width=0.47\textwidth]{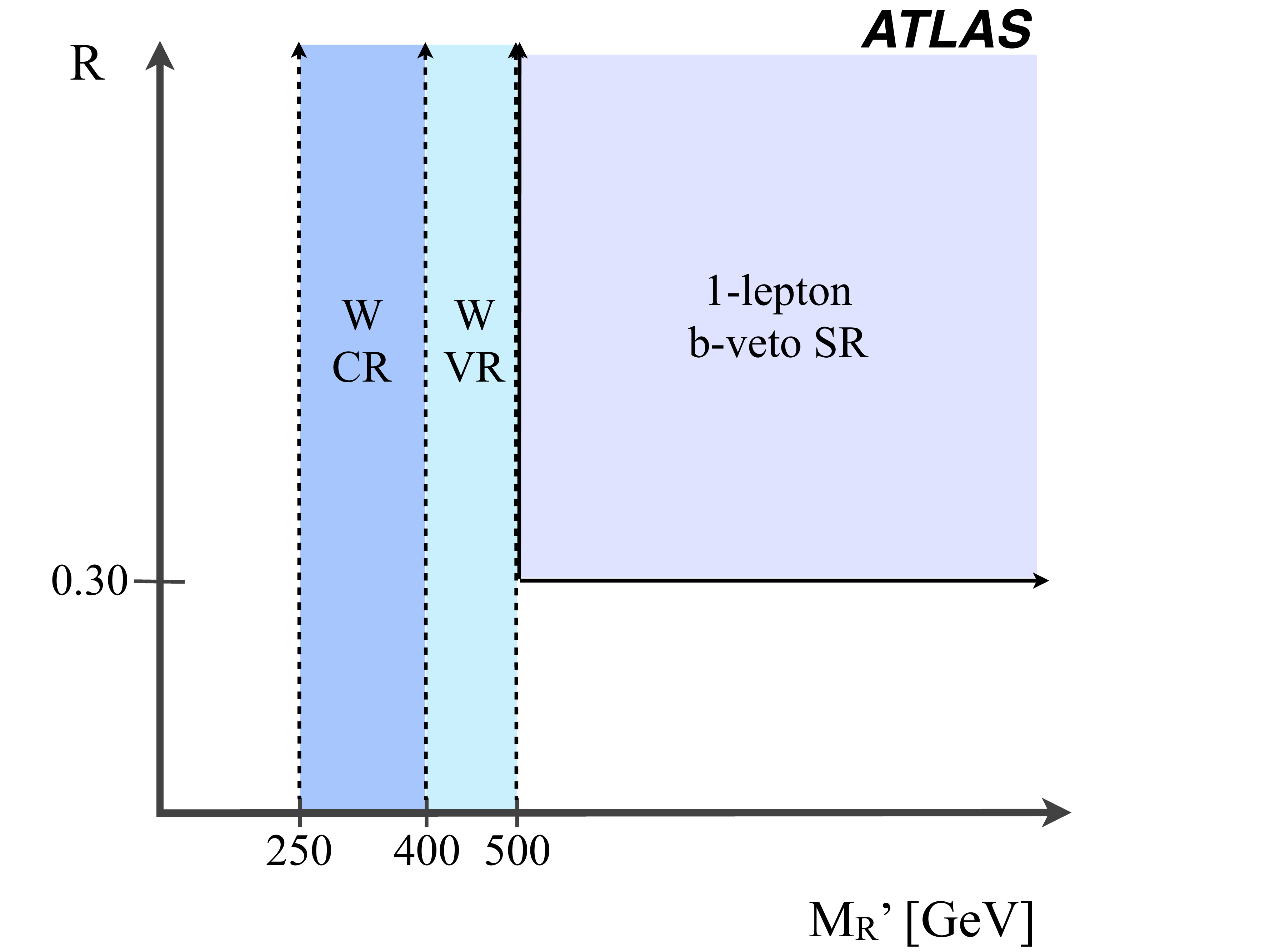}
\includegraphics[width=0.47\textwidth]{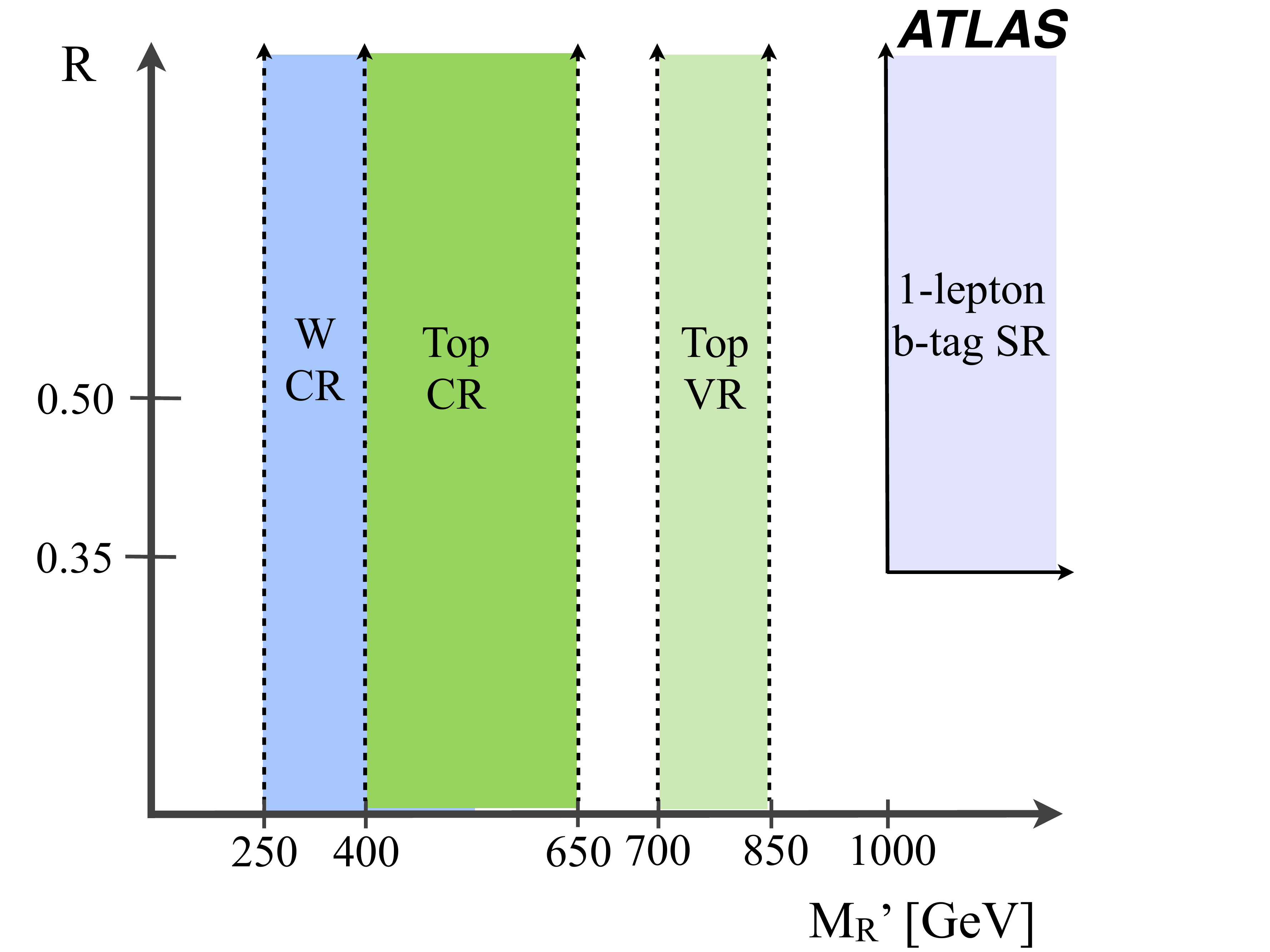}
\smallskip
\vspace{1 cm}
\includegraphics[width=0.3\textwidth]{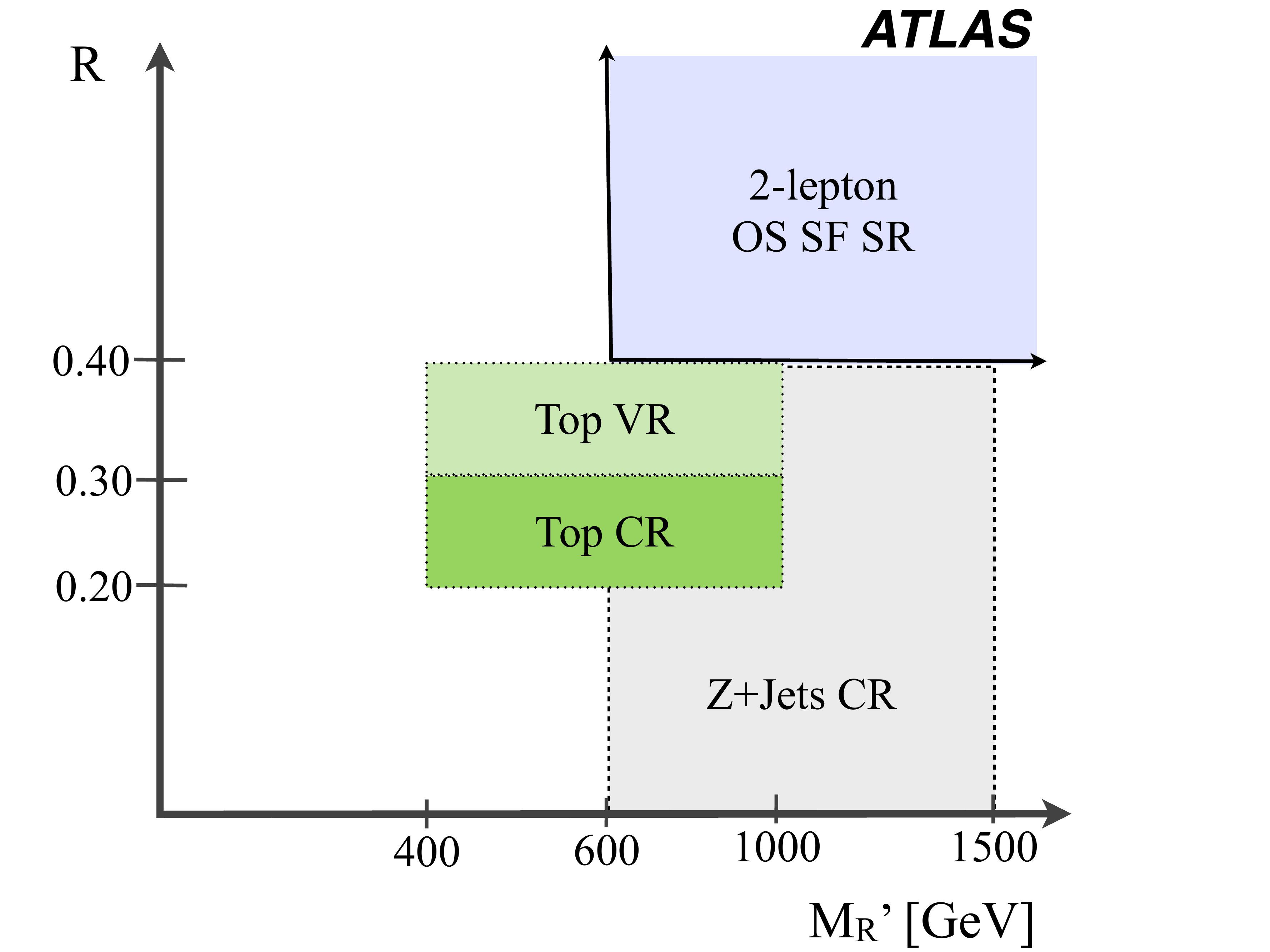}
\includegraphics[width=0.3\textwidth]{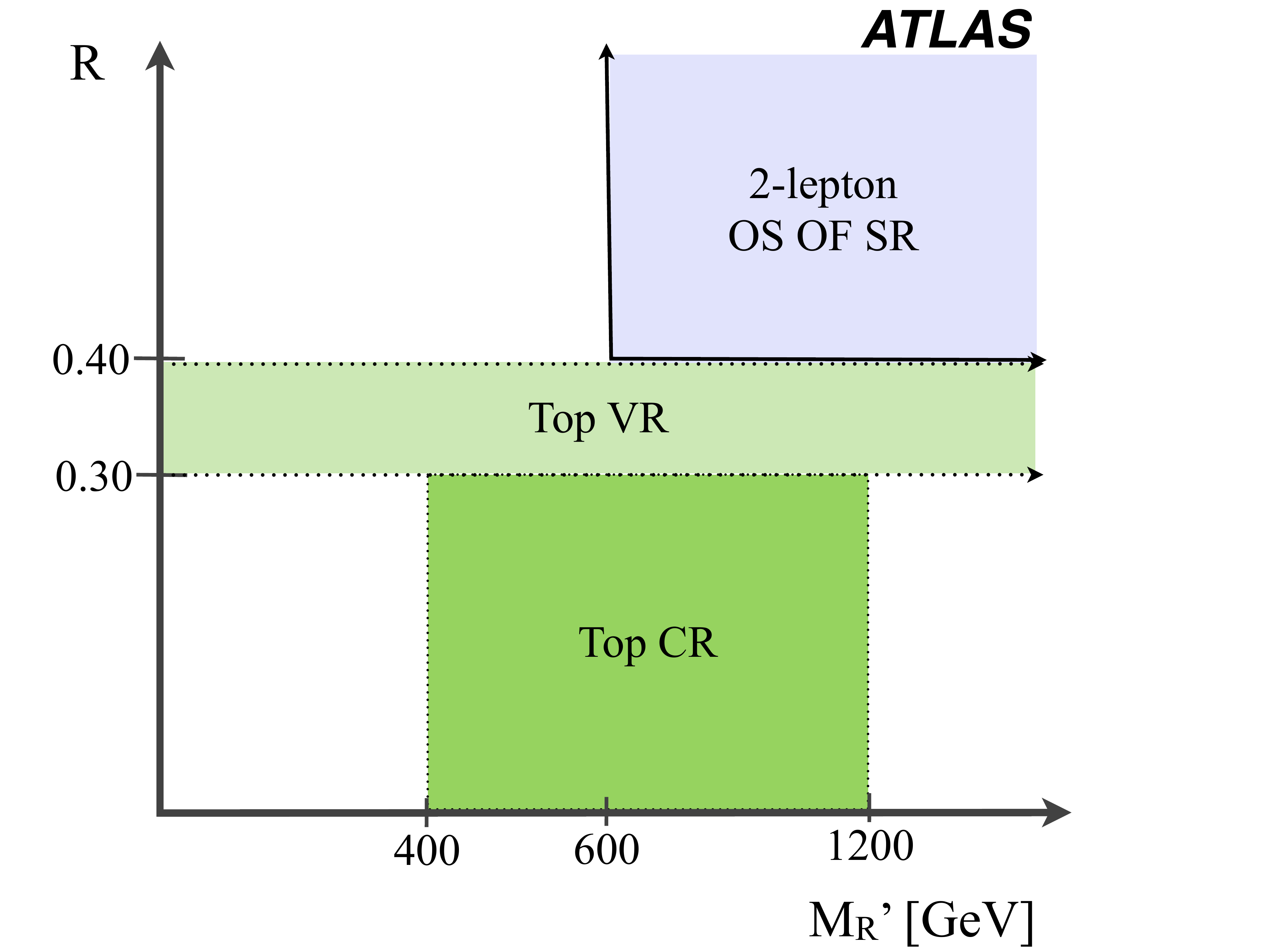}
\includegraphics[width=0.3\textwidth]{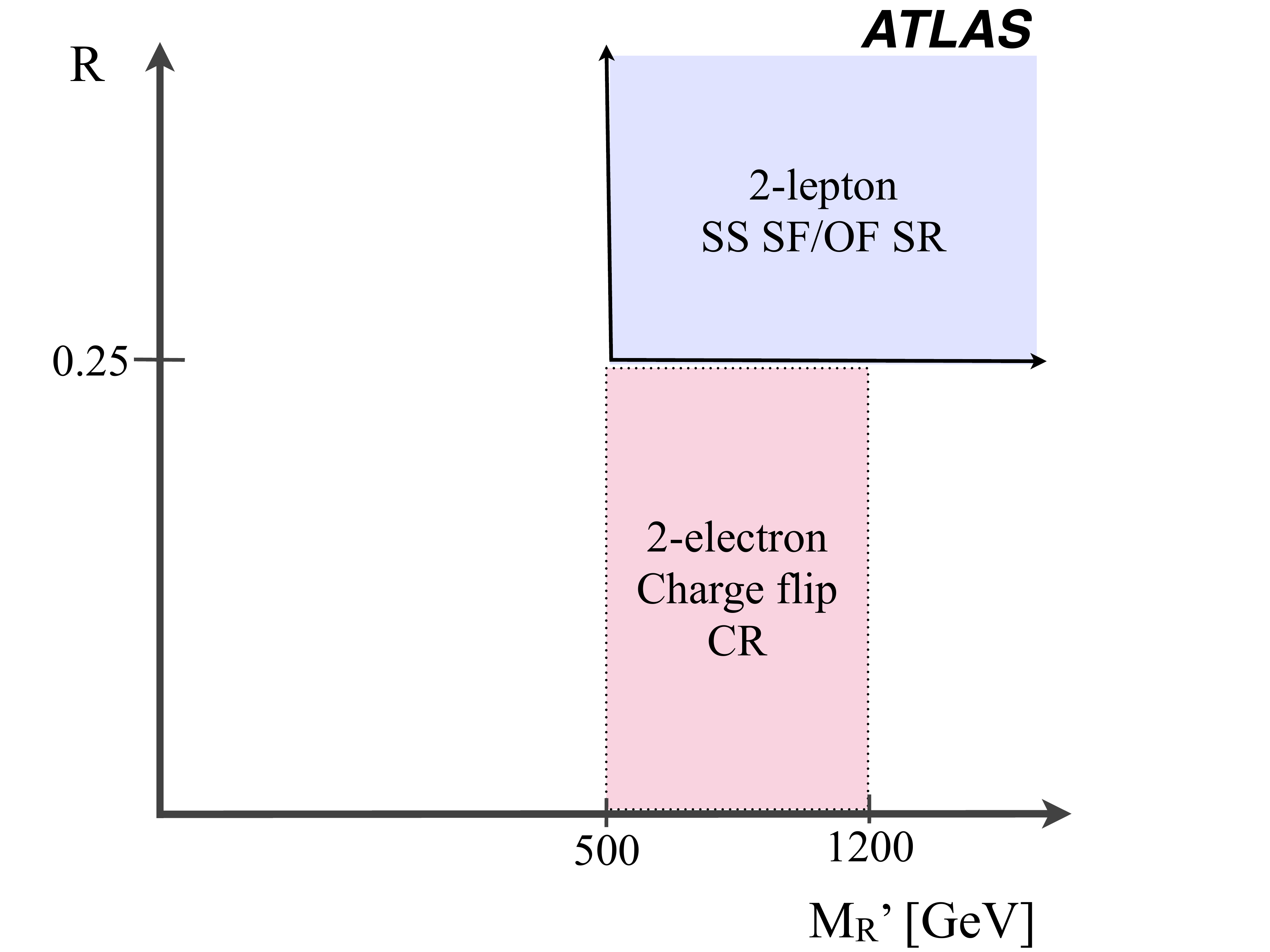}

\caption{
\label{fig:012LepSRCRVR}
A visual representation of the zero-lepton (top), one-lepton (middle) signal regions, and two-lepton (bottom) control (CR), validation (VR), and signal (SR) regions.}
\end{figure*}

After sorting events into the six samples described in the previous
section, each sample is further divided in the $R$--$M_R'$ plane 
into control regions (CR), which are choosen so that they are dominated by a 
specific background, and signal regions (SR). Additionally, validation regions (VR)
are constructed, which do not constrain the background but are used to
evaluate the agreement between data and MC simulation.
Table~\ref{table:CRSRs} lists these regions, which are also
visualised in the $R$--$M_R'$ plane in Fig.~\ref{fig:012LepSRCRVR}. 
These regions are binned in either $R$ or
$M_R'$ and then simultaneously fit to MC estimates for
background and signal rates with correlations from sample to sample
and region to region taken into account. The hadronic (had.) and one-lepton
signal regions are divided into events with and without 
$b$-tagged jets (``b-tag'' and ``b-veto,'' respectively).  The two-lepton 
events are divided into
regions with opposite-sign (OS) and same-sign (SS) leptons and regions with
opposite-flavour (OF) and same-flavour (SF) leptons. While some background
components are sufficiently constrained by the CRs to be left free in
the fit, others are constrained to estimates derived from other
techniques or MC simulation. Table~\ref{table:backgroundSummary} summarises the
backgrounds, the estimation technique, the source of the estimate,
and the normalization uncertainty used in the fit. 
Finally, systematic uncertainties on all backgrounds are included 
as nuisance parameters.  The result is a maximum likelihood fit that 
encapsulates all knowledge about the background and signal consistently 
across all channels. 

When evaluating a signal sample for exclusion, any signal 
contamination in the control regions is taken into account for each
signal point, as the control region fits are performed for each signal
hypothesis.  Separately, each signal region (one at a time), along 
with all control regions, is also fit under the background-only 
hypothesis. This fit is used to characterise agreement in each 
signal region with the background-only hypothesis and to extract 
fiducial cross-section limits and upper limits on the production of 
events from new physics ($N_\text{BSM}$).

\begin{table*}[htbp]
\caption{
Background control, signal, and validation regions.  All signal regions include the overflow in the highest bin.
``N/A'' means that there is no requirement.
The binning of the validation regions does not affect the results, so they are listed as ``N/A.''
\label{table:CRSRs}}
\begin{center}
\begin{tabular}{lllllll}
Name                 & Leptons    & $b$-jets & $N_{\rm Jets}$ & $R$ Range   & $M_R'$ Range         & Number of bins \\
\hline\noalign{\smallskip}
Control Regions \\
\hline
Had. $b$-veto Multijet      & 0 leptons  & $=0$       & $>5$          & $0.3<R<0.4$ & $800< M_R'<2000$~\GeV & 12 in $M_R'$ \\
Had. $b$-tag Multijet       & 0 leptons  & $>0$      & $>5$          & $0.2<R<0.3$ & $1000<M_R'<2000$~\GeV & 10 in $M_R'$  \\
$e$ $W$+jets                & 1 electron & $=0$       & $>5$          & $0<R<0.7$   & $300< M_R'<400$~\GeV  & 7 in $R$ \\
$\mu$ $W$+jets              & 1 muon     & $=0$       & $>5$          & $0<R<0.7$   & $300< M_R'<400$~\GeV  & 7 in $R$ \\
$e$  \ttbar\                & 1 electron & $>0$      & $>5$          & $0<R<0.7$   & $400< M_R'<650$~\GeV  & 7 in $R$ \\
$\mu$ \ttbar\             & 1 muon       & $>0$      & $>5$          & $0<R<0.7$   & $400< M_R'<650$~\GeV  & 7 in $R$ \\
$ee$ \ttbar\              & 2 OS electrons  & $>0$ & N/A           & $0.2<R<0.3$ & $400< M_R'<1000$~\GeV & 6 in $M_R'$  \\
$\mu\mu$ \ttbar\          & 2 OS muons      & $>0$ & N/A           & $0.2<R<0.3$ & $400< M_R'<1000$~\GeV & 6 in $M_R'$  \\
$e\mu$  \ttbar\           & 2 OS OF leptons & $>0$ & N/A           & $R<0.3$     & $400< M_R'<1200$~\GeV & 8 in $M_R'$  \\
$ee$ $Z$                  & 2 OS electrons  & N/A & N/A           & $R<0.4$     & $600< M_R'<1500$~\GeV & 9 in $M_R'$  \\
$\mu\mu$ $Z$              & 2 OS muons      & N/A & N/A           & $R<0.4$     & $600< M_R'<1500$~\GeV & 9 in $M_R'$  \\
$ee$ Charge flip          & 2 SS electrons  & N/A & N/A           & $R<0.25$    & $500< M_R'<1200$~\GeV & 7 in $M_R'$  \\
\hline\noalign{\smallskip}
Signal Regions\\
\hline
Had. $b$-veto     & 0 leptons  & $=0$        & $>5$          &$R>0.70$     &$600< M_R'<1200$~\GeV  & 3 in $M_R'$ \\
Had. $b$-tag      & 0 leptons  & $>0$        & $>5$          &$R>0.40$     &$900< M_R'<1500$~\GeV  & 3 in $M_R'$ \\

$e$ $b$-veto      & 1 electron  & $=0$        & $>5$          &$R>0.55$     &$500< M_R'<1000$~\GeV  & 3 in $M_R'$ \\
$e$ $b$-tag       & 1 electron  & $>0$        & $>5$          &$R>0.35$     &$1000< M_R'<1600$~\GeV  & 6 in $M_R'$ \\

$\mu$ $b$-veto    & 1 muon      & $=0$        & $>5$          &$R>0.55$     &$500< M_R'<1000$~\GeV  & 3 in $M_R'$ \\
$\mu$ $b$-tag     & 1 muon      & $>0$        & $>5$          &$R>0.35$     &$1000< M_R'<1400$~\GeV  & 4 in $M_R'$ \\

OS-$ee$           & 2 OS electrons & N/A  & N/A           &$R>0.40$     &$600< M_R'<1000$~\GeV &  4 in $M_R'$ \\ 
OS-$\mu\mu$       & 2 OS muons     & N/A  & N/A           &$R>0.40$     &$600< M_R'<1000$~\GeV &  4 in $M_R'$ \\

SS-$ee$           & 2 SS electrons & N/A  & N/A           &$R>0.25$     &$500< M_R'<900$~\GeV &  4 in $M_R'$ \\
SS-$\mu\mu$       & 2 SS muons     & N/A  & N/A           &$R>0.25$     &$500< M_R'<900$~\GeV &  4 in $M_R'$ \\

OS-$e\mu$         & 2 OS OF leptons & N/A  & N/A           &$R>0.40$     &$600< M_R'<1000$~\GeV &  4 in $M_R'$ \\ 
SS-$e\mu$         & 2 SS OF leptons & N/A  & N/A           &$R>0.25$     &$500< M_R'<900$~\GeV &   4 in $M_R'$ \\ 
\hline\noalign{\smallskip}
Validation Regions\\
\hline
Had. $b$-veto Multijet      & 0 leptons  & $=0$       & $>5$          & $0.4<R<0.6$ & $800< M_R'<2000$~\GeV & N/A \\
Had. $b$-tag Multijet       & 0 leptons  & $>0$       & $>5$          & $0.3<R<0.4$ & $1100<M_R'<2000$~\GeV & N/A \\
1-lep $b$-veto $W$+jets    & 1 lepton   & $=0$       & $>5$          & N/A   & $400< M_R'<550$~\GeV  & N/A \\
1-lep $b$-tag \ttbar\      & 1 lepton   & $>0$       & $>5$          & N/A   & $700< M_R'<850$~\GeV  & N/A \\
OS-$ee/\mu\mu$ \ttbar\       & 2 OS SF leptons & >0  & N/A           & $0.3<R<0.4$ & $400~\GeV < M_R'$  & N/A \\
OS-$e\mu$ \ttbar\            & 2 OS OF leptons & >0  & N/A           & $0.3<R<0.4$ & N/A & N/A \\
\noalign{\smallskip}\hline\hline
\end{tabular}
\end{center}
\end{table*}

The fit considers several independent background components:
\begin{itemize}

\item {\em \ttbar\ and single top.}  A total of five top control
  regions are defined in the one- and two-lepton channels.  The
  normalisation of this component is allowed to vary freely in the fit.

\item {\em Bosons, except diboson $WW$.}  The inclusion
  of the $WZ$ and $ZZ$ diboson samples is motivated by the dominance of
  leptonic $Z$ decays in the two-lepton signal regions, the
  dominance of $Z\rightarrow\nu\nu$ in the zero-lepton signal
  regions, and the dominance of $WZ\rightarrow\ell\nu\qqbar$ in the one-lepton
  signal regions.  In all of these cases, the experimental uncertainties
  affect the samples in the same way as they do $W$+jets or $Z$+jets, and
  therefore they are combined in order to treat them as fully correlated.
  The normalisation of this sample is allowed to vary 
  freely in the fit. Independent validation of the $Z$+jets
  background is carried out in two-lepton control regions.  The agreement 
  is good between data and MC simulation in both normalisation and shape.

\item {\em Diboson $WW$.}  This sample is constrained with a 30\,\%
  cross-section systematic uncertainty.  The constraint is necessary
  because of the relatively small contribution of the sample in most
  signal and control regions and because no $WW$-dominated control
  region can be constructed; if the background were allowed to vary 
  freely, then the fit may find a minimum with an unreasonably large 
  or small contribution from diboson $WW$ events and hide some 
  other effect with an artificial $WW$ normalisation.

\item {\em Charge flip.}  Charge mis-identification can occur due to 
  physical effects, like lepton Bremsstrahlung, and detector effects, 
  especially for high-$\pt$ leptons with almost straight tracks. These effects generate
  background in the same-sign dielectron and electron-muon
  channels. This background is negligable in the dimuon channel, where 
  the contribution from both physical and detector effects is far smaller.
  The electron charge-flip rate is measured as a function of $\eta$ in
  the data~\cite{twoLepATLAS}, allowing MC simulation to model the 
  lesser dependence on $\pt$.  These charge-flip rates are applied to
  opposite-sign MC simulation events, providing an estimate of the
  overall contribution from charge flip in these channels.  The
  electron $\pt$ is additionally shifted and smeared to mimic the effect of charge
  mis-identification.  This shift in the $\pt$ is propagated through
  to the razor variables. The uncertainty from the charge flip
  probabilities dominates the uncertainty of this background.

\item {\em Fake leptons.} The multijet background in the one-lepton
  signal regions, as well as the $W$+jets, semi-leptonic \ttbar, and
  multijet background in the two-lepton signal regions, comes
  predominantly from had\-rons faking electrons and muons.  This
  background is estimated using the ``matrix
  method''~\cite{multiLepMultiJetATLAS,twoLepATLAS}, using the number of
  baseline leptons not passing signal lepton requirements.  The
  efficiency for a real lepton passing the baseline lepton
  requirements to pass the signal lepton requirements is estimated
  using $Z$ MC simulation events.  The rejection rate for fake leptons
  is estimated in data, using samples enriched in fake leptons.  For
  electrons, the factors are derived and applied separately for
  inclusive samples of events and samples requiring a $b$-tagged jet.
  Because this background accounts for \emph{all} fake background, MC
  events in the one-lepton (two-lepton) channels are required to have
  at least one (two) prompt lepton(s) from a $\tau$ lepton, $W$ boson, $Z$ boson,
  or sparticle.  The uncertainty on this background estimate has a
  statistical component from the number of events in the control
  region and a systematic component from the uncertainty on the scale
  factors.

  Some fraction of the events with
  same-sign, baseline leptons in the data may be due to charge
  flip. Thus, the matrix method overestimates somewhat the fake lepton
  background in the dilepton channels. In order to correct for this 
  overlap, opposite-sign
  events in data containing baseline leptons that do not pass the
  signal lepton requirements are used. Each event is assigned a
  weight representing the likelihood of that event being subject to
  charge mis-identification. The weighted events are then presented
  as a negative component to the same-sign fake background
  distribution, such that the contribution to the same sign fake
  background from originally oppositely charged leptons is subtracted.

\item {\em Multijets in zero-lepton channel.} Two specific control regions
  constrain this background, and its normalisation is allowed to
  vary freely in the fit. Several different approaches are used to
  cross-check this estimate. Pre\-scaled single jet triggers are used to
  construct independent multijet-enriched control regions at low
  $M_R'$ that is free from the inefficiency of the
  $\HT$-based trigger. The observed number of events in this region are
  then projected into the signal region using transfer factors from MC
  simulation. Alternatively, in order to model the mis-measurement of
  jets in the calorimeter, jets in events collected with these
  single-jet triggers are smeared according to response functions
  estimated using data~\cite{0leptonATLAS}. Both of these methods
  result in an estimate consistent with that derived in the main fit.

\end{itemize}

The systematic effects included as nuisance parameters in the fit are:
the jet energy scale and resolution uncertainties; $b$-tagging
uncertainties; uncertainty on the MC simulation modelling of the JVF; the
additional cross-section uncertainty on the production of heavy flavour in
association with a vector boson; the uncertainties on trigger
efficiency and matching and reconstruction efficiency; a systematic
uncertainty on the re-weighting of the $W$-boson $\pt$; uncertainties
on the missing transverse momentum pile-up dependence and the
calibration of energy not associated with an object in the event; the
matrix method statistical and systematic uncertainties; the charge
flip systematic uncertainties; the diboson $WW$ shape systematic
uncertainty taken from comparing {\sc Herwig} to {\sc Alpgen}.  Where
the systematic uncertainties affect object definitions, corrections
are propagated to the missing transverse momentum and razor variable
calculations. The effects of other uncertainties on the final results are negligible.
These uncertainties affect the signal yield and shape in the signal 
regions, as well as the allowed variation in signal-region background 
estimates after the control region constraints.  In most signal regions,
the jet energy scale uncertainty is the dominant experimental uncertainty
(from 10\,\% to 25\,\%).

\section{Background fit}
\label{sec:bg}

\begin{figure*}[tbp]
\centering
\includegraphics[width=0.47\textwidth]{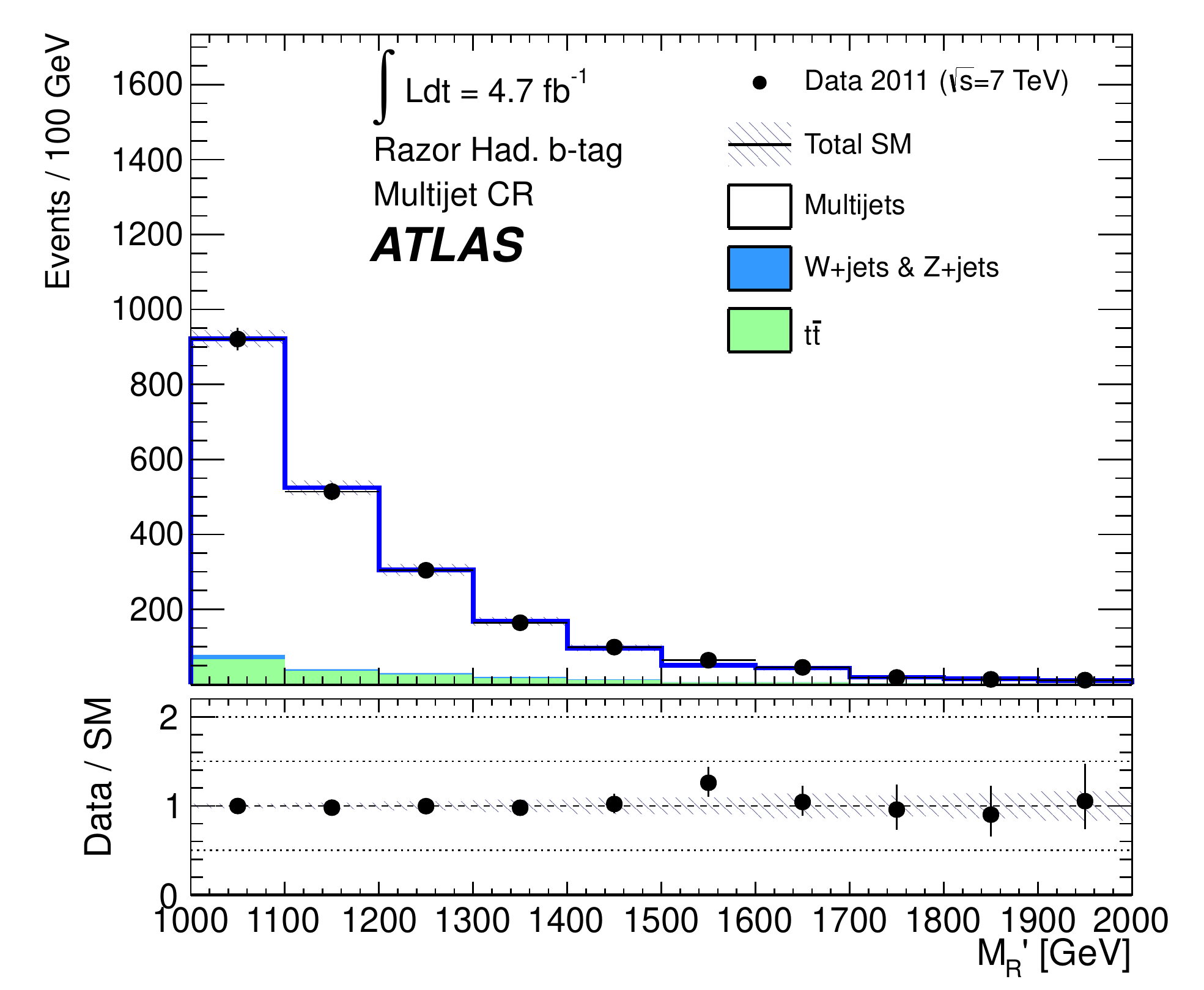}
\includegraphics[width=0.47\textwidth]{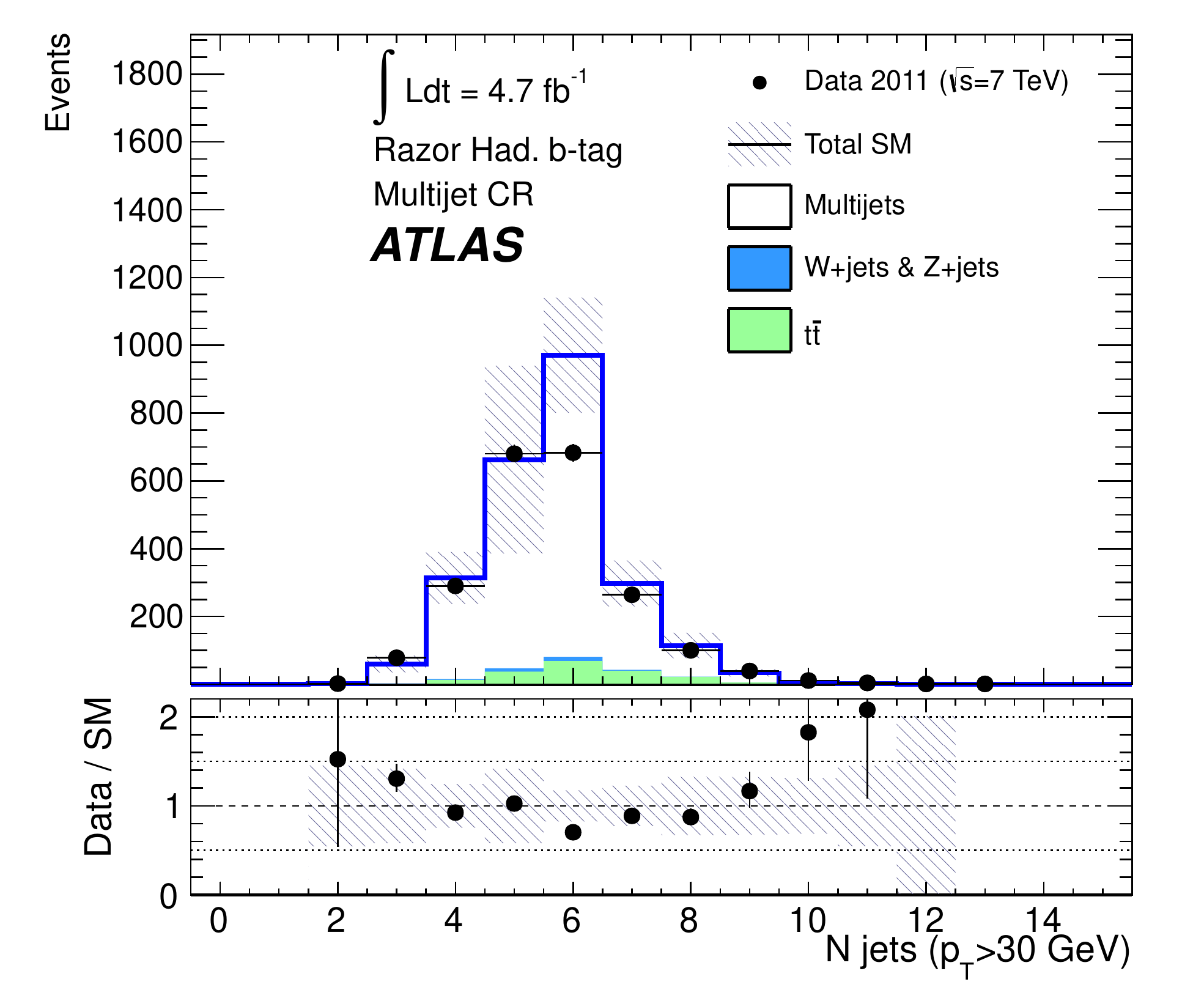}
\caption{
The distribution of $M_R'$ (left) and the number of jets with $\pt>30$~\GeV\ (right) in the multijet control region with a $b$-tagged jet requirement (dots with error bars), the expectation from the control region fit for various backgrounds (filled), and the systematic uncertainty (hatched).
\label{fig:mjcr_mr}}
\end{figure*}

Figure~\ref{fig:mjcr_mr} shows the distributions of $M_R'$ 
and jet multiplicity in the zero-lepton multijet control region 
with a $b$-tagged jet requirement, with results from the fit to the 
control regions overlayed. By design, the multijet background is dominant in
these regions. The small contribution from \ttbar\ and $W$+jets
backgrounds are constrained by other control regions in the
simultaneous fit. The hatched area indicates the total systematic
uncertainty after the constraints imposed by the fit.  

\begin{figure*}[tbp]
\centering
\includegraphics[width=0.47\textwidth]{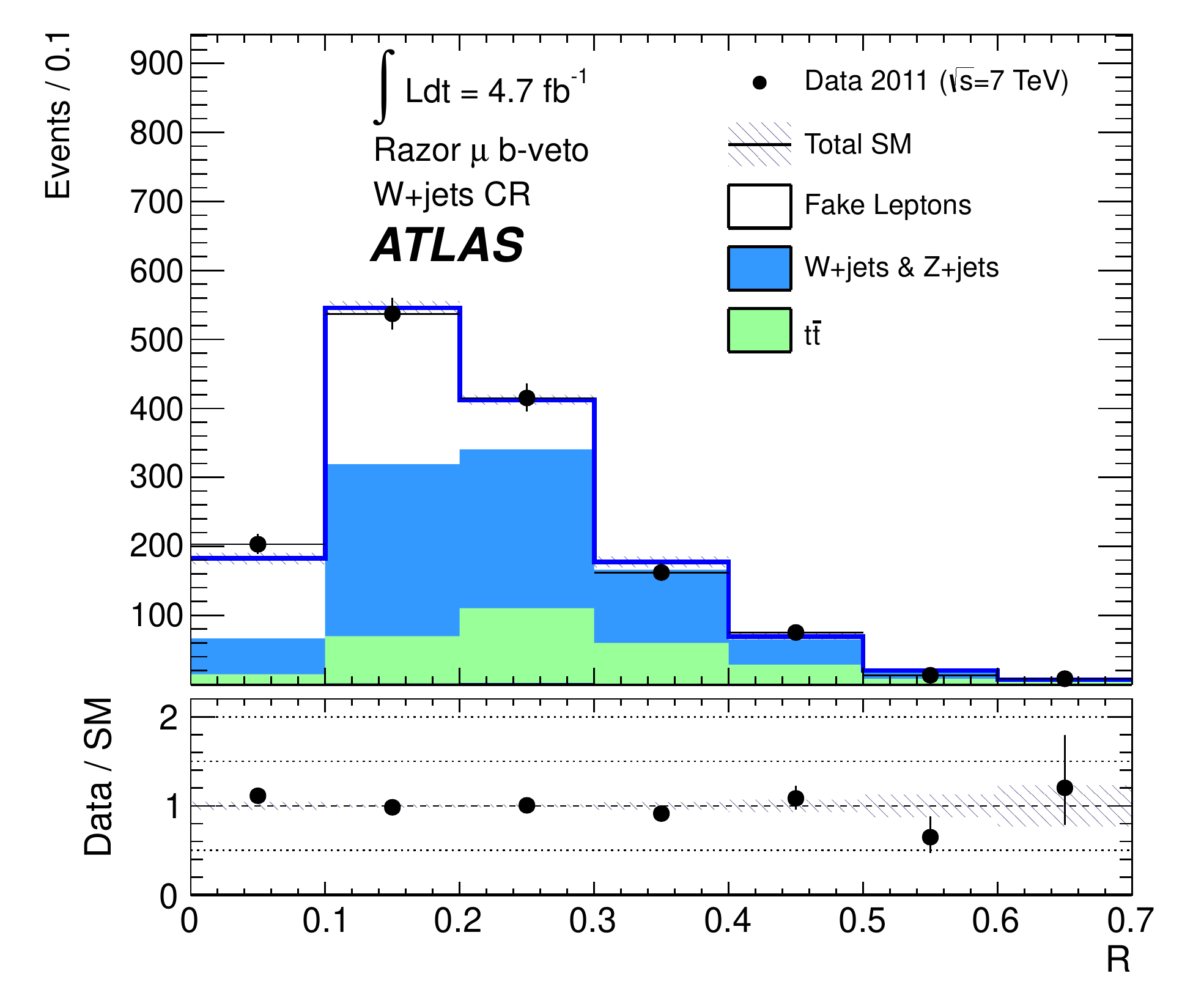}
\includegraphics[width=0.47\textwidth]{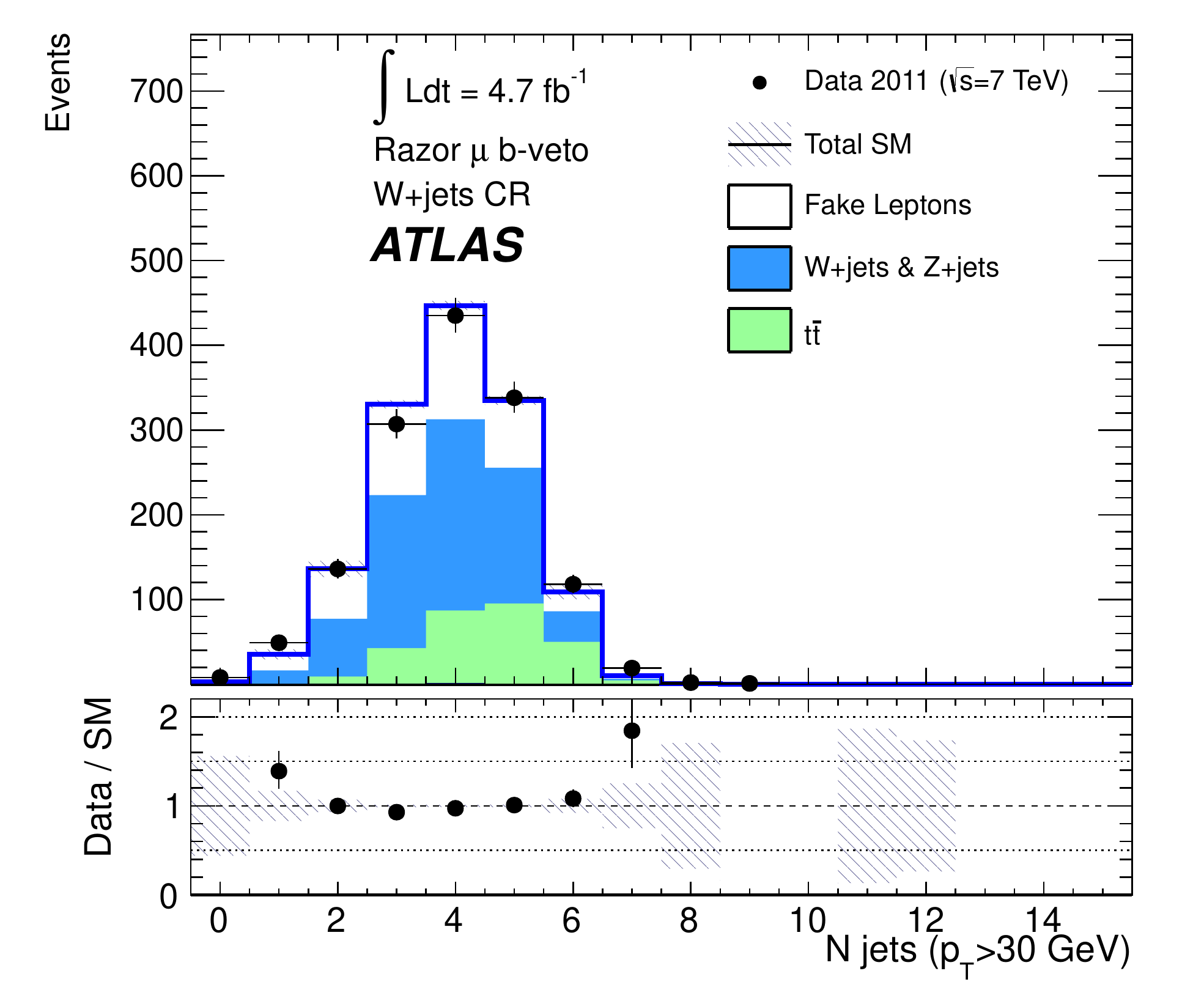}
\caption{
The distribution of $R$ (left) and the number of jets with $\pt>30$~\GeV\ (right) in the $W\rightarrow\mu\nu$+jets control region (dots with error bars), the expectation from the control region fit for various backgrounds (filled), and the systematic uncertainty (hatched).
\label{fig:wcr_r}}
\end{figure*}

The distributions of $R$ and jet multiplicty for the backgrounds after the control region fit in the 
$W\rightarrow\mu\nu$+jets control region are shown in Fig.~\ref{fig:wcr_r}.  The control region at low $R$ is
dominated by fake backgrounds, and at moderate-to-high $R$ they are dominated by $W$+jets. The use of an alternate control region with a cut on transverse mass, which significantly
reduces the fake contribution, results in a negligible change in the final search results.

\begin{figure*}[tbp]
\centering
\includegraphics[width=0.47\textwidth]{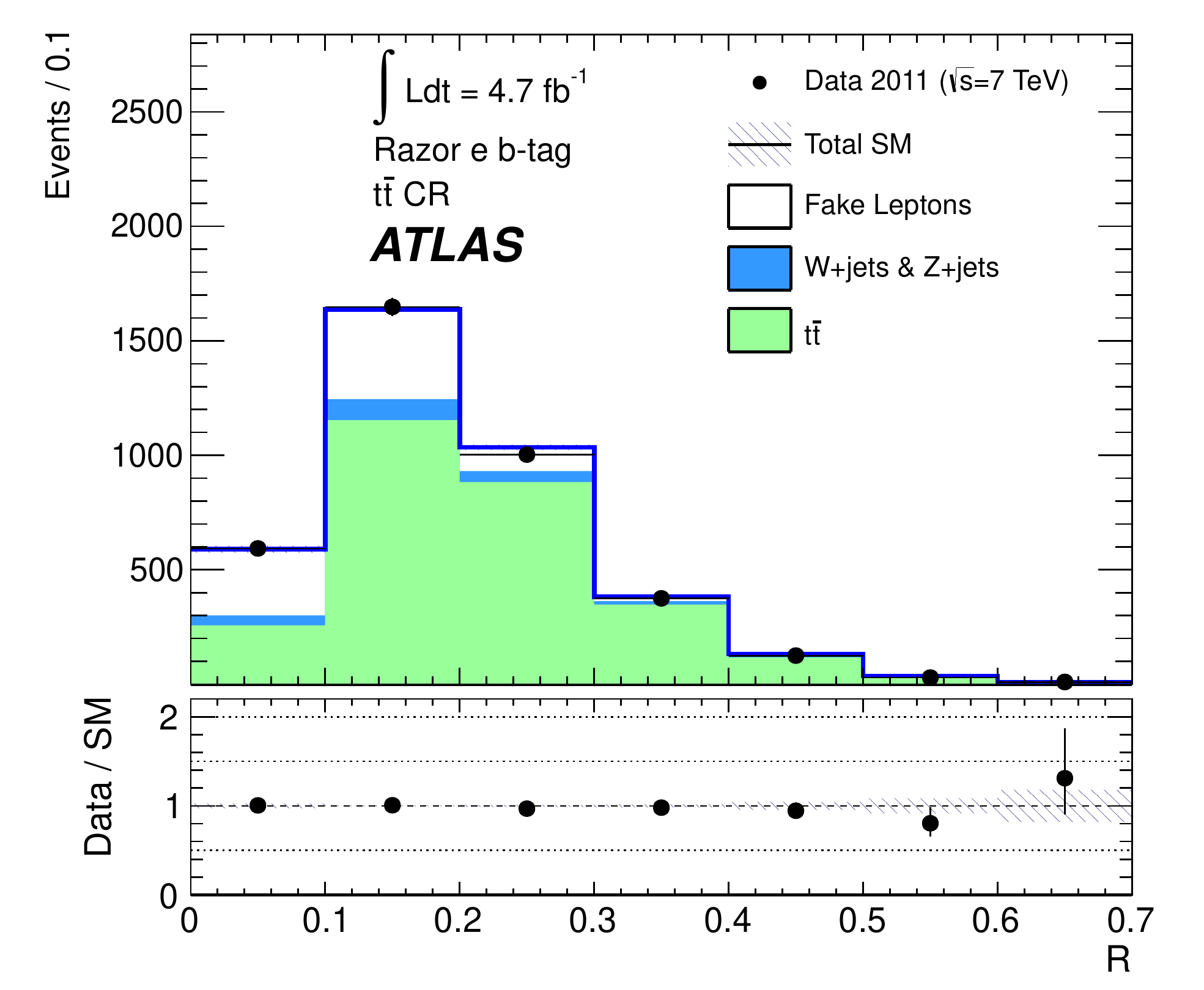}
\includegraphics[width=0.47\textwidth]{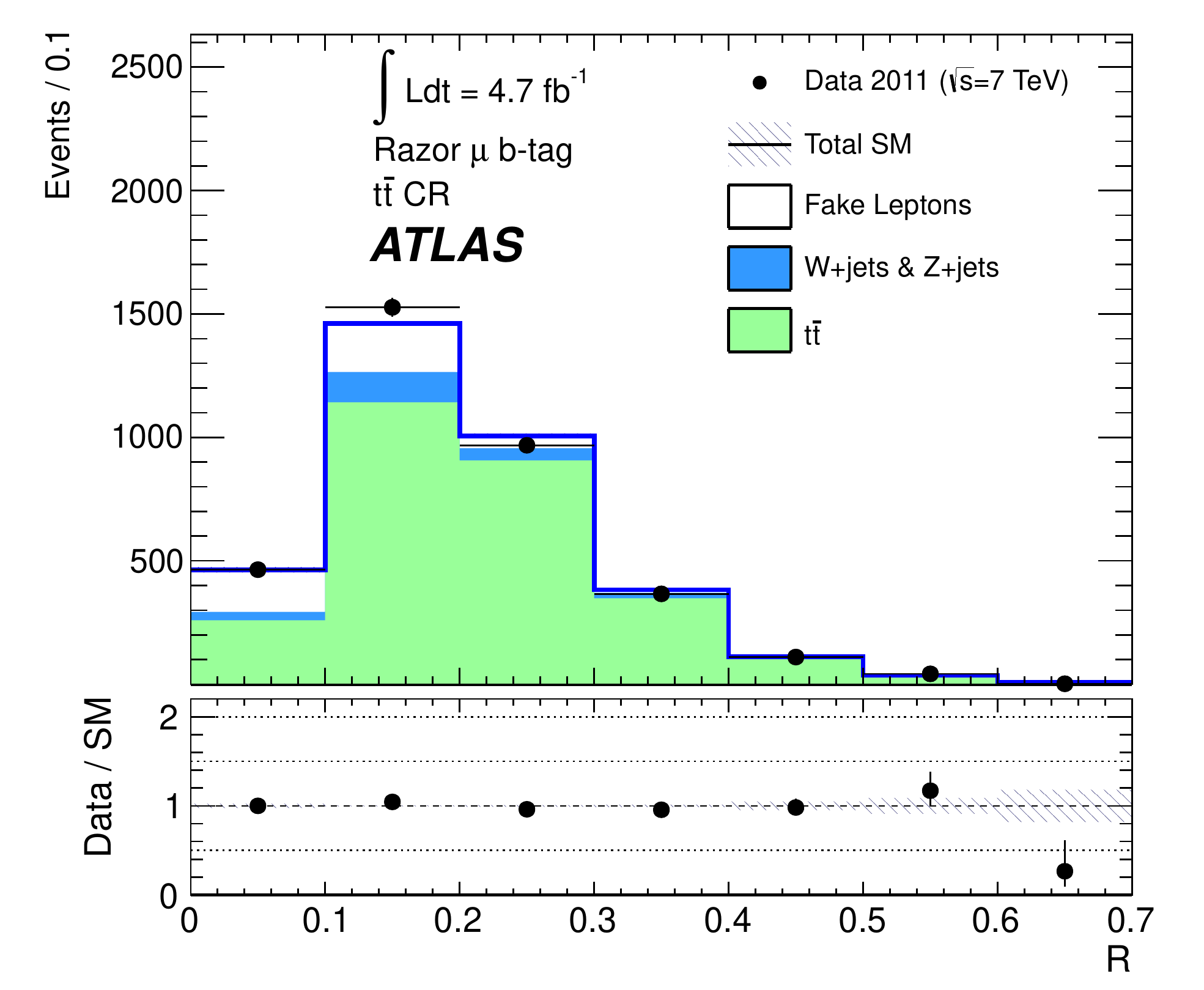}
\caption{
The distribution of $R$ in the one-lepton \ttbar\ control regions (dots with error bars), the expectation after the control region fit for various backgrounds (filled), and the systematic uncertainty (hatched).
\label{fig:tcr_r}}
\end{figure*}

\begin{figure*}[p]
\centering
\includegraphics[width=0.47\textwidth]{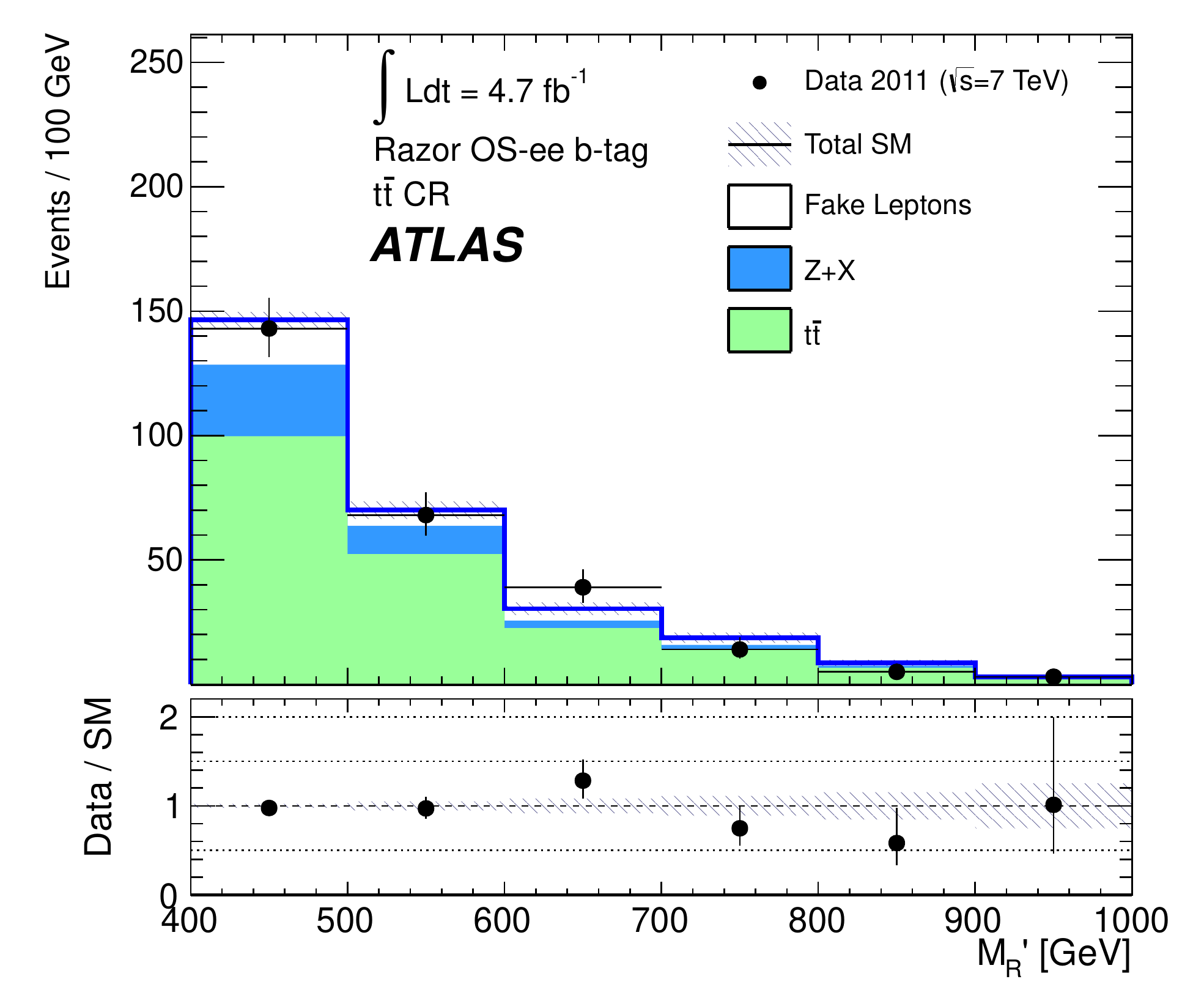}
\includegraphics[width=0.47\textwidth]{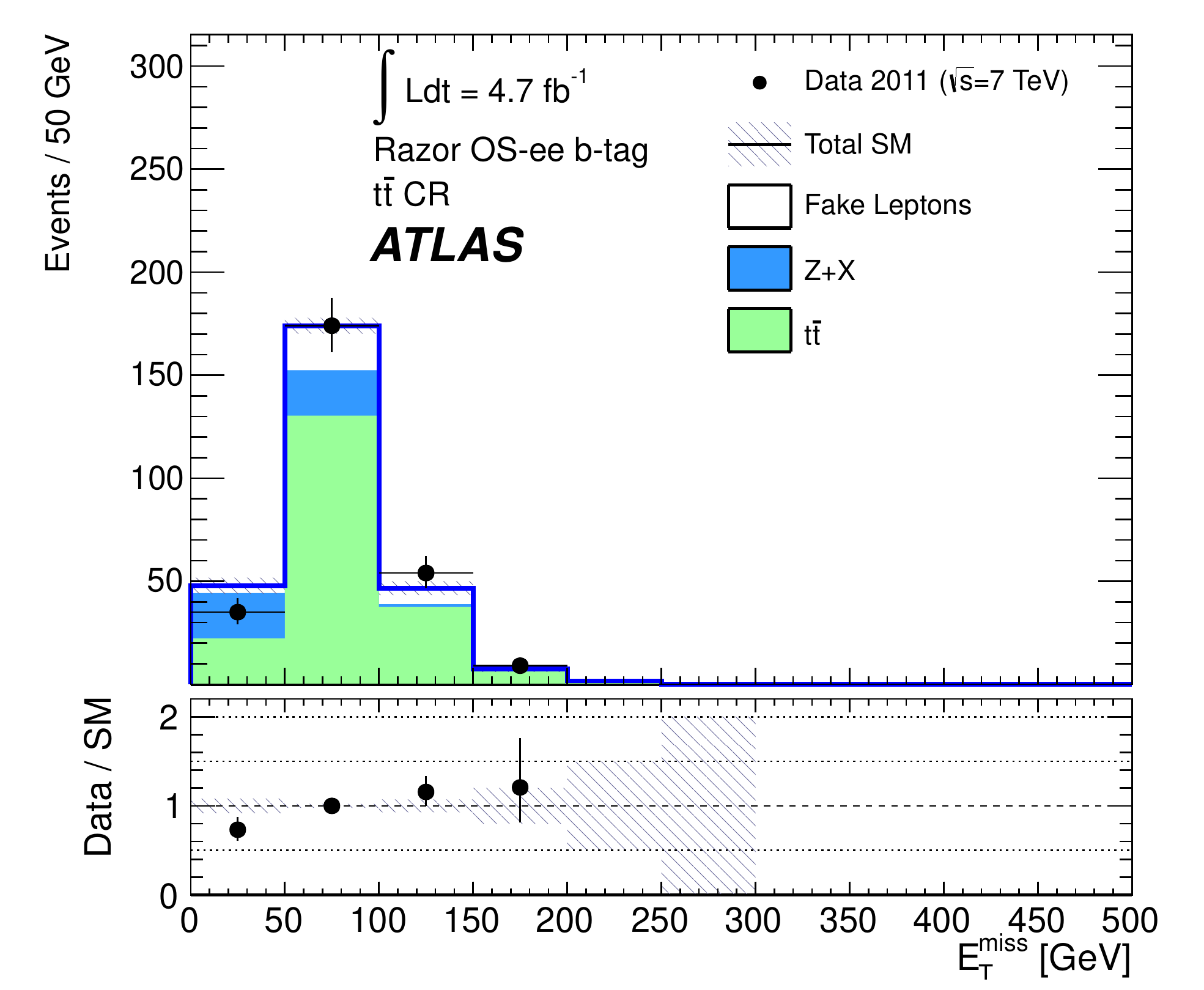}
\includegraphics[width=0.47\textwidth]{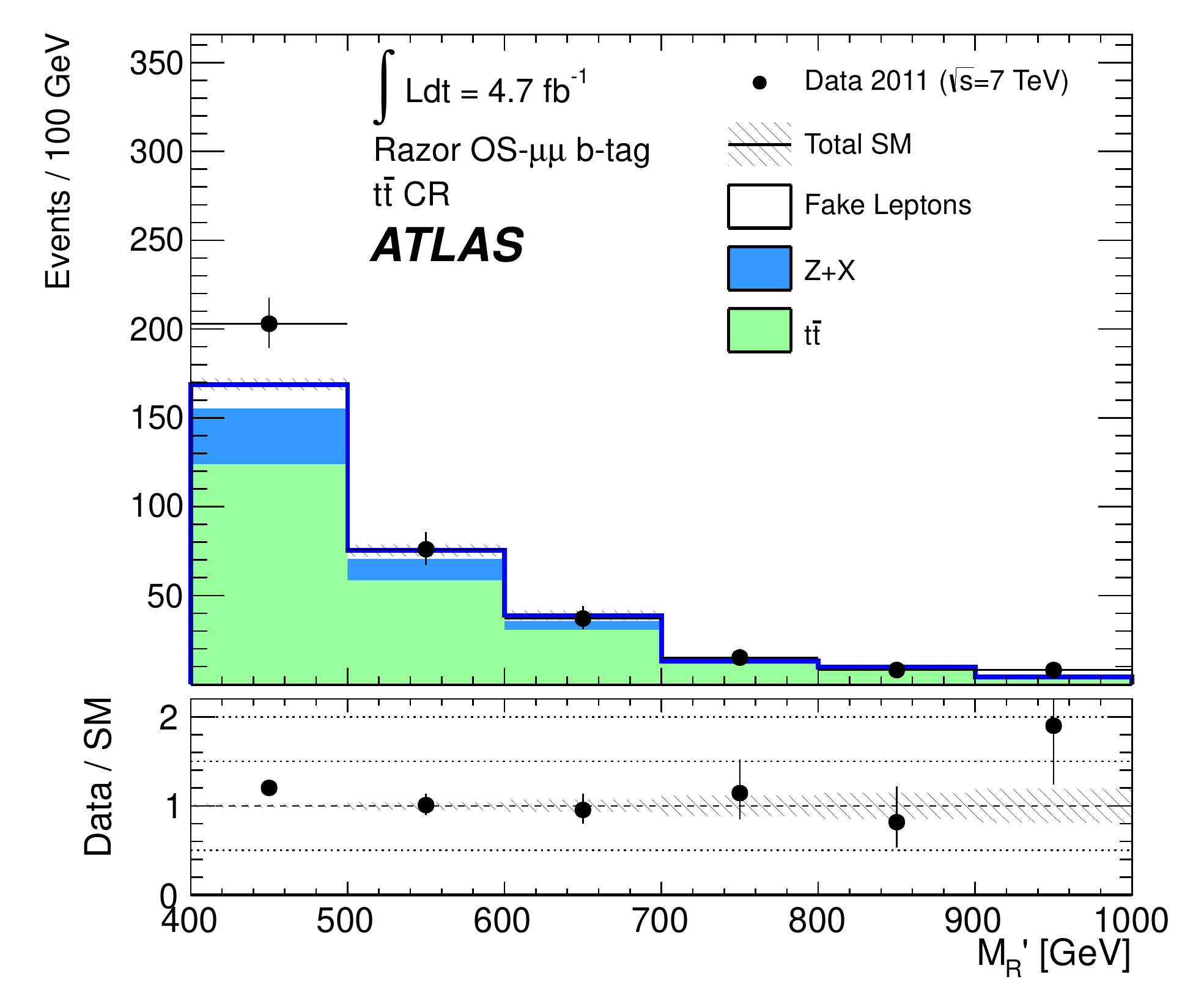}
\includegraphics[width=0.47\textwidth]{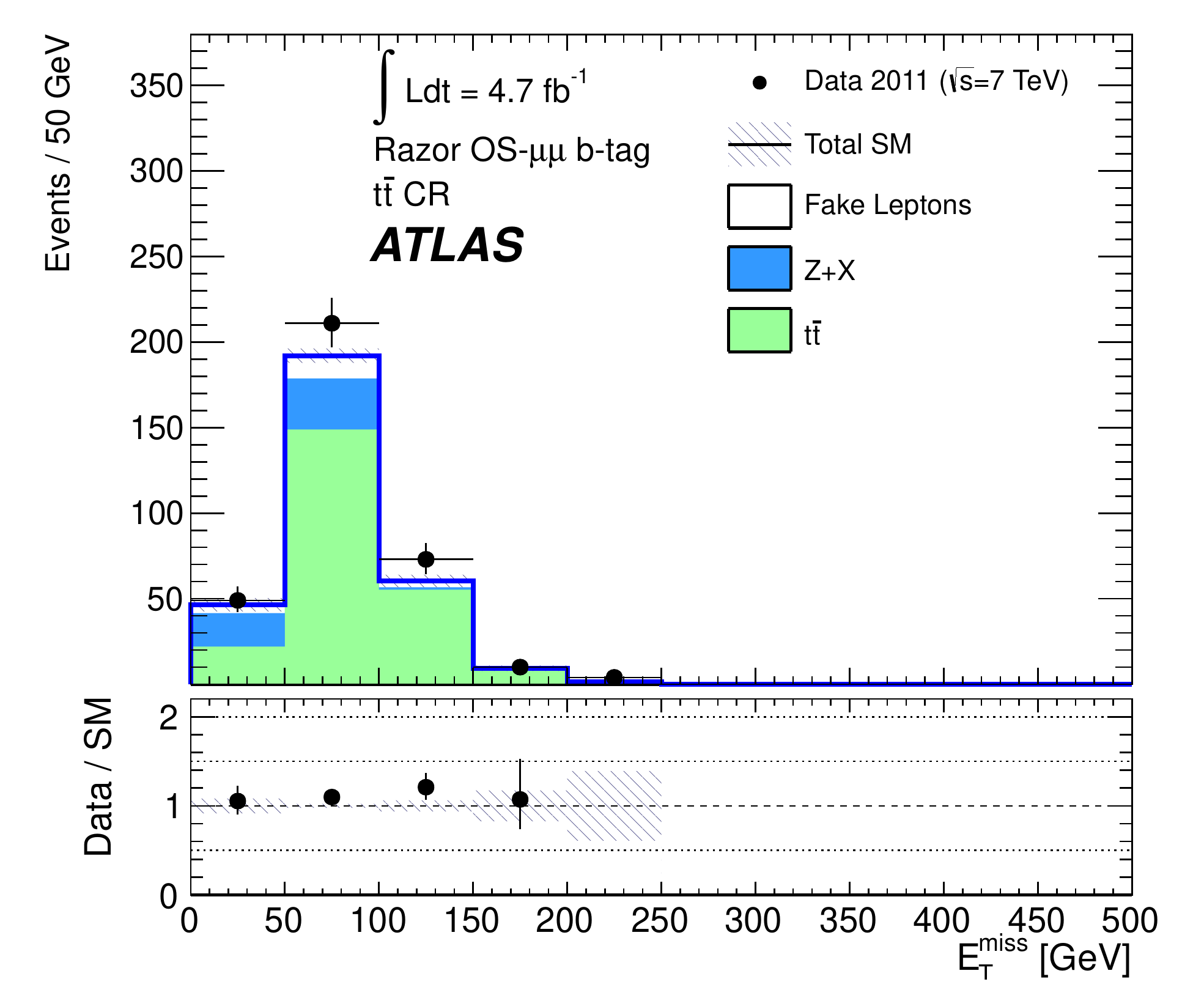}
\includegraphics[width=0.47\textwidth]{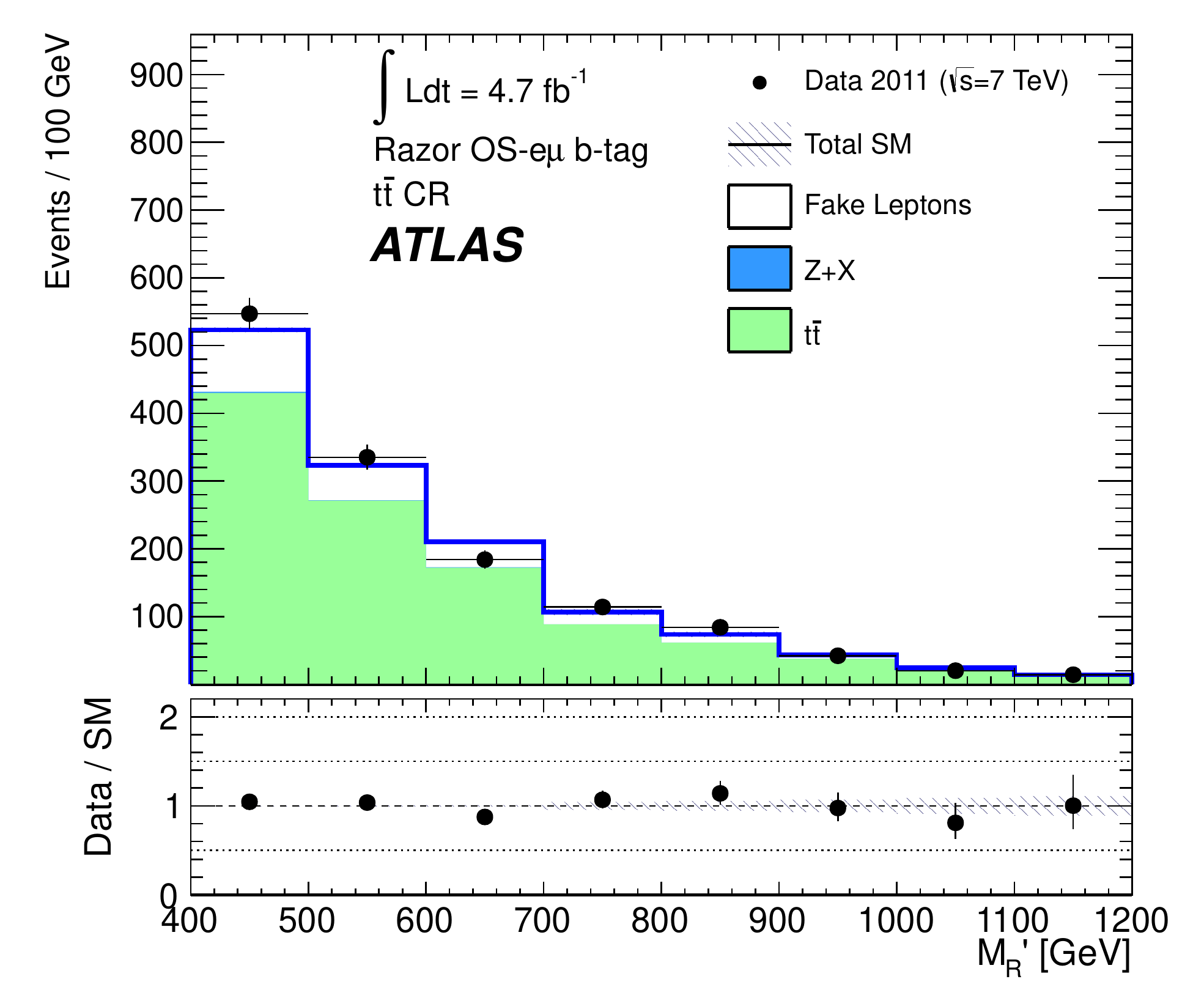}
\includegraphics[width=0.47\textwidth]{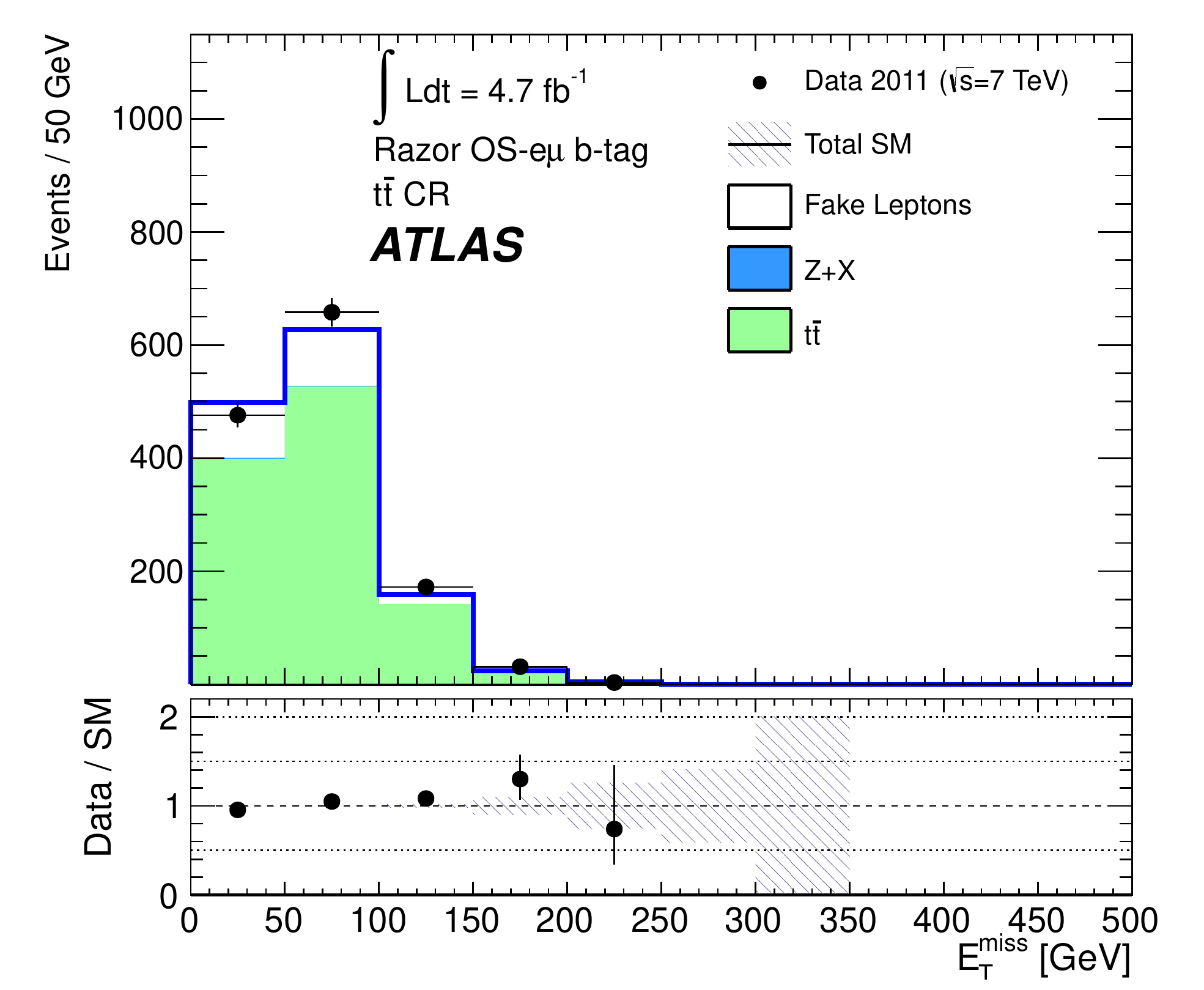}
\caption{
The distribution of $M_R'$ (left) and missing transverse momentum (right) in the two-lepton \ttbar\ control regions (dots with error bars), the expectation after the control region fit for various backgrounds (filled), and the systematic uncertainty (hatched).
\label{fig:tcr_mr}
}
\end{figure*}

Figure~\ref{fig:tcr_r} and~\ref{fig:tcr_mr} show the one-lepton and
two-lepton \ttbar\ control regions, respectively. The fit reduces the
normalisation of the \ttbar\ background in the one-lepton control
region by approximately 15--20\,\% with respect to the unmodified
expectation from MC simulation.  This shift predominantly affects
the semi-leptonic \ttbar\ background.  In the two-lepton analysis,
there is a significant contribution to the background expectation from
$Z$-boson events with heavy flavour, particularly at low \MET.  
The lowest $M_R'$ bin shows the most significant disagreement,
which demonstrates the importance of shape profiling by binning the
control regions.  Although in that lowest bin, particularly in the
two-muon channel, the MC simulation underestimates the amount of data,
a single-binned normalisation of the \ttbar\ background would result
in an overestimation of the background at high $M_R'$.
The distributions of missing transverse momentum are also shown 
for the two-lepton control regions.

\begin{figure*}[tbp]
\centering
\includegraphics[width=0.47\textwidth]{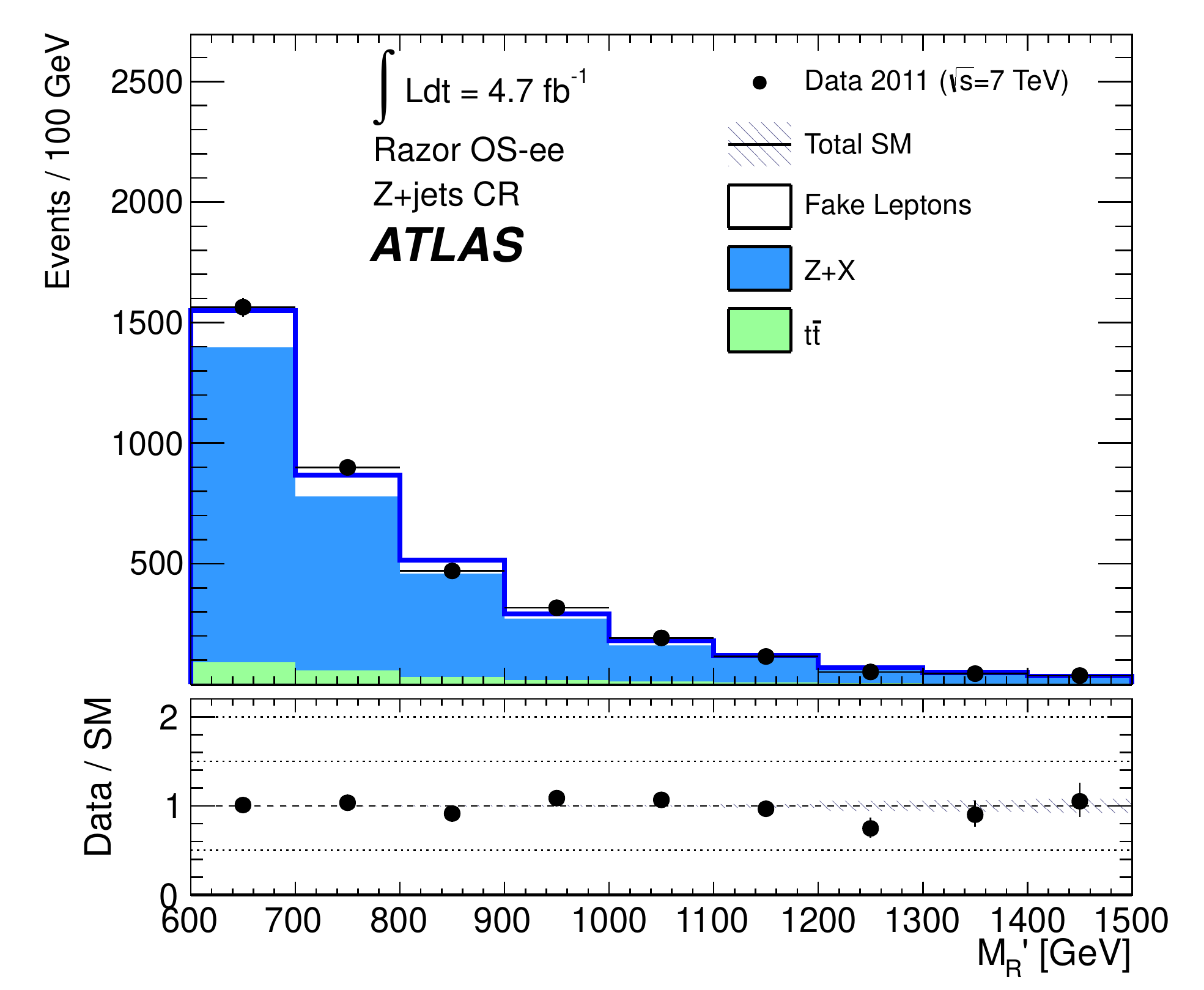}
\includegraphics[width=0.47\textwidth]{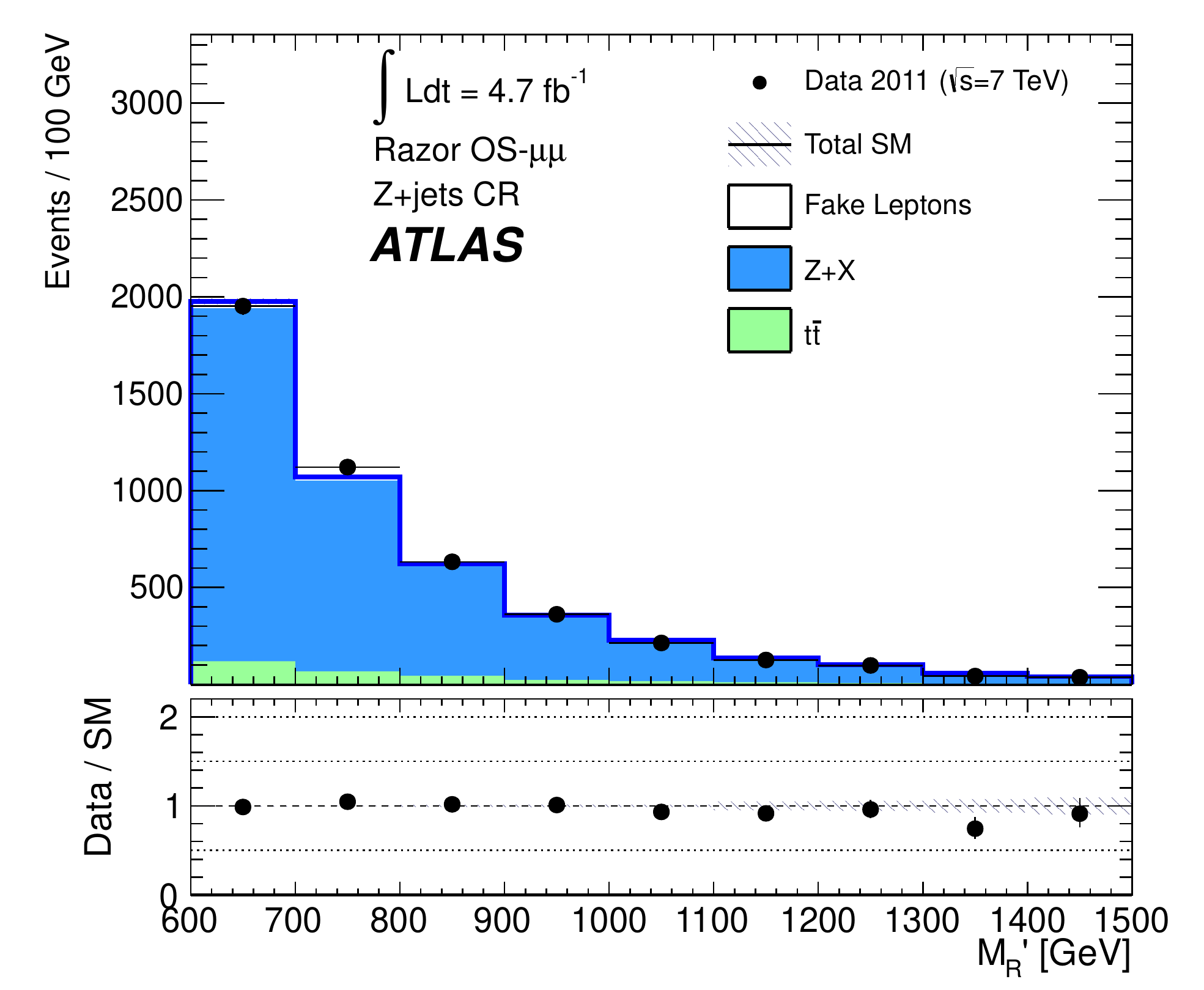}
\includegraphics[width=0.47\textwidth]{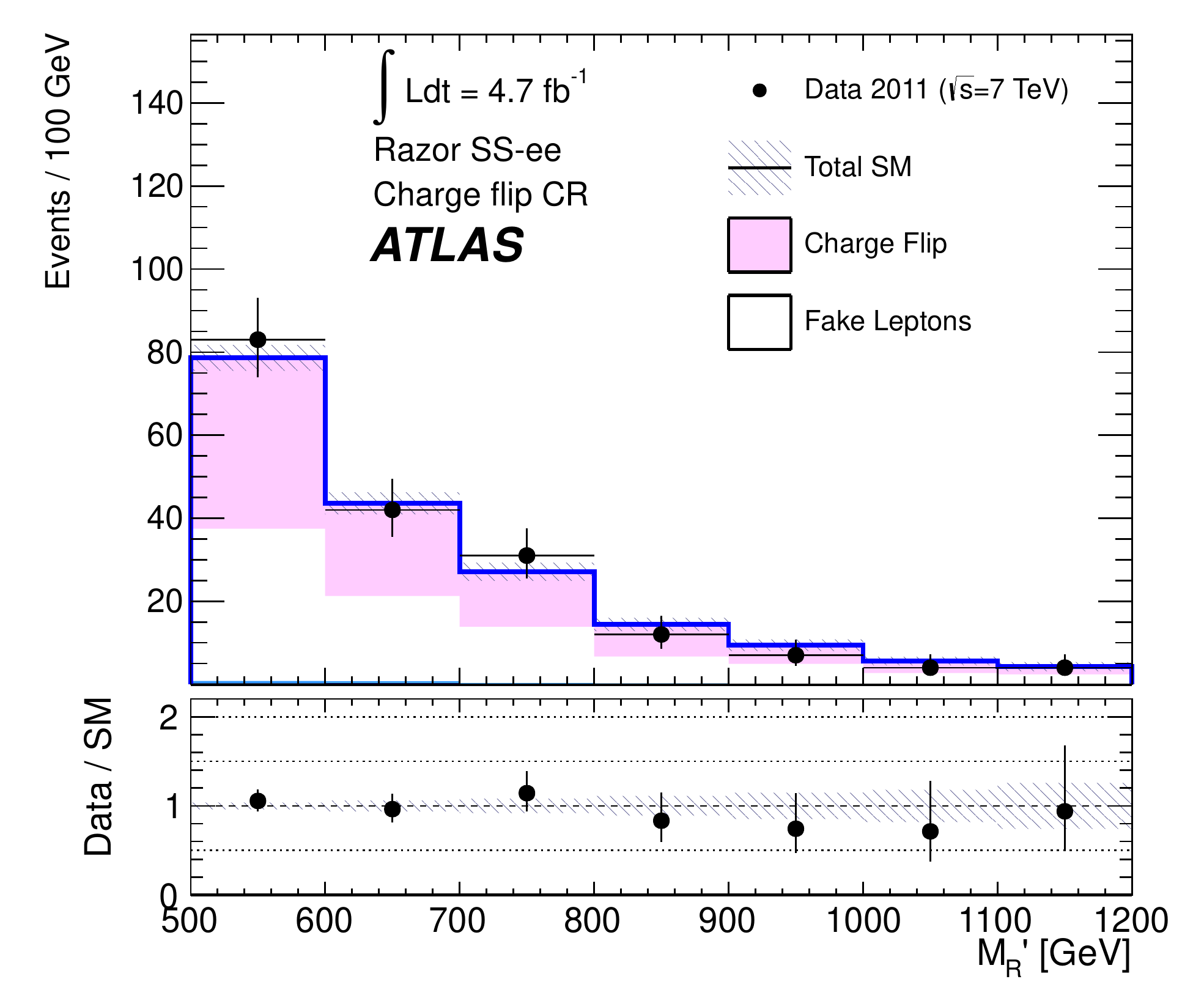}
\includegraphics[width=0.47\textwidth]{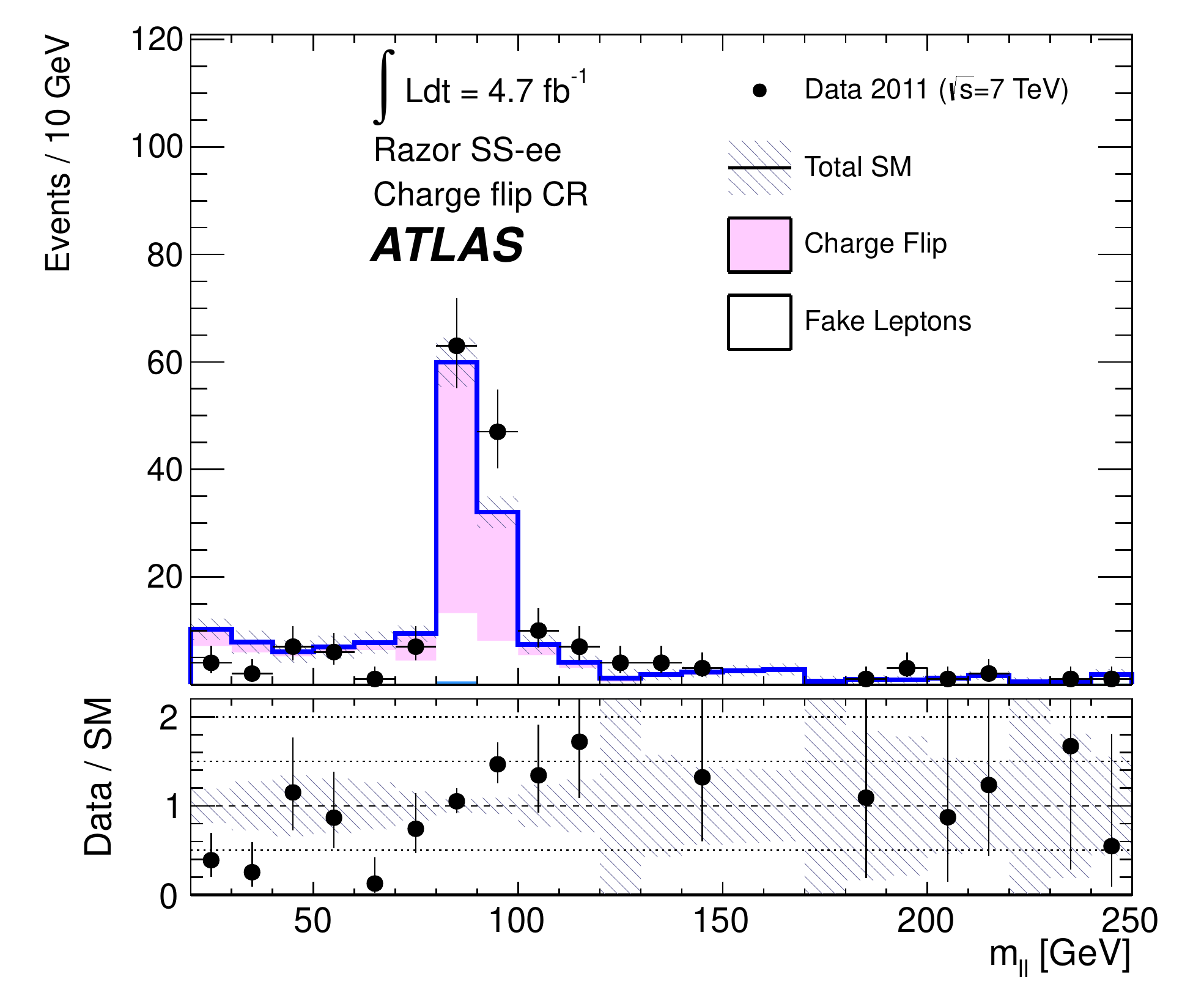}
\caption{
Top, the distribution of $M_R'$ in the $Z$+jets control regions.  Bottom, the distribution of $M_R'$ (left) and dilepton mass (right) in the charge flip control region (dots with error bars), the expectation after the control region fit for various backgrounds (filled), and the systematic uncertainty (hatched).
\label{fig:zcr_mr}}
\end{figure*}

The distributions of $M_R'$ for the two $Z$+jets 
control regions are shown in Fig.~\ref{fig:zcr_mr}.  In both the electron and
muon channel, good agreement after the control region fit is visible.  

The charge flip background is significant in the same-sign two-electron
channel.  Figure~\ref{fig:zcr_mr} also shows the distribution of
$M_R'$ in a charge flip enriched control region.  
There remain significant uncertainties on the
background even after the control region fit, since it is dominated by charge flip
and fake leptons, both of which have large systematic uncertainties
associated with them.  The distribution of dilepton mass is also shown.

The contributions to each of the control regions before and after the fit to the control regions are shown in Tables~\ref{table:CRfit01L} and~\ref{table:CRfit2L}.

\begin{table*}
\caption{Results of the background-only fit to the control regions in the zero- and one-lepton control regions, for an integrated luminosity of 4.7~fb$^{-1}$.
The errors shown are the statistical plus systematic uncertainties.
\label{table:CRfit01L} }
\begin{center}
\scalebox{0.88}{
\begin{tabular}{@{\extracolsep{\fill}}lrrrrrrrrrrrr}
\noalign{\smallskip}\hline\noalign{\smallskip}
{\bf  Control region}          & Had. $b$-veto Multijet & Had. $b$-tag Multijet & $e$ $W$+jets & $\mu$ $W$+jets & e \ttbar     & $\mu$ \ttbar  \\[-0.05cm]
\noalign{\smallskip}\hline\noalign{\smallskip}
Observed events         & $1032$            & $2153$            & $1833$            & $1413$            & $3783$            & $3479$            \\ 
\noalign{\smallskip}\hline\noalign{\smallskip}
Fitted background events    & $1030 \pm 30$ & $2150 \pm 50$ & $1840 \pm 40$ & $1410 \pm 30$ & $3820 \pm 60$ & $3470 \pm 50$ \\
\noalign{\smallskip}\hline\noalign{\smallskip}
Fitted top events           & $21 \pm 7$   & $170 \pm 19$   & $280 \pm 30$  & $290 \pm 30$  & $2800 \pm 60$ & $2800 \pm 60$ \\
Fitted $W$/$Z$ events       & $90 \pm 10$  & $26 \pm 4$     & $670 \pm 40$  & $690 \pm 50$  & $210 \pm 20$  & $240 \pm 30$  \\
Fitted $WW$ diboson events & $0.54 \pm 0.18$    & $0.14 \pm 0.05$      & $4.2 \pm 1.8$     & $4.5 \pm 1.9$     & $1.2 \pm 0.5$     & $1.0 \pm 0.4$     \\
Fitted multijet events     & $920 \pm 30$      & $1960 \pm 50$      & $0 \pm 0$      & $0 \pm 0$      & $0 \pm 0$      & $0 \pm 0$    \\
Fitted charge flip events   & $0 \pm 0$      & $0 \pm 0$        & $0 \pm 0$       & $0 \pm 0$       & $0 \pm 0$       & $0 \pm 0$       \\
Fitted fake lepton events   & $0 \pm 0$    & $0 \pm 0$      & $890 \pm 50$  & $430 \pm 40$  & $810 \pm 70$  & $440 \pm 60$  \\
\noalign{\smallskip}\hline\noalign{\smallskip}
\noalign{\smallskip}\hline\noalign{\smallskip}
Expected events                 & $990$    & $2670$   & $2110$     & $1560$     & $4300$     & $3790$   \\ 
\noalign{\smallskip}\hline\noalign{\smallskip}
MC exp. top events              & $52$     & $245$    & $450$      & $470$      & $3300$     & $3250$   \\ 
MC exp. $W$/$Z$ events          & $110$    & $28$     & $740$      & $760$      & $200$      & $220$    \\ 
MC exp. $WW$ diboson events    & $0.61$      & $0.18$      & $4.6$        & $4.6$        & $1.4$        & $1.1$      \\ 
MC exp. multijet events        & $830$      & $2400$      & $0$      & $0$      & $0$      & $0$   \\
Charge flip events (estimated from data) & $0$      & $0$      & $0$        & $0$        & $0$        & $0$      \\ 
Fake lepton events (estimated from data) & $0$      & $0$      & $910$      & $330$      & $800$      & $310$    \\ 
\noalign{\smallskip}\hline\noalign{\smallskip}
\end{tabular}
}
\end{center}
\end{table*}

\begin{table*}
\caption{Results of the background-only fit to the control regions in the two-lepton control regions, for an integrated luminosity of 4.7~fb$^{-1}$.
The errors shown are the statistical plus systematic uncertainties.
\label{table:CRfit2L} }
\begin{center}
\scalebox{0.98}{
\begin{tabular}{@{\extracolsep{\fill}}lrrrrrrrrrrrr}
\noalign{\smallskip}\hline\noalign{\smallskip}
{\bf  Control region}          & ee \ttbar      & $\mu\mu$ \ttbar     & $e\mu$ \ttbar       & $ee$ $Z$            & $\mu\mu$ $Z$            & ee Charge flip  \\[-0.05cm]
\noalign{\smallskip}\hline\noalign{\smallskip}
Observed events         & $272$            & $347$           & $1340$            & $3688$            & $4579$            & $183$ \\
\noalign{\smallskip}\hline\noalign{\smallskip}
Fitted background events     & $277 \pm 14$ & $310 \pm 10.$ & $1320 \pm 30$ & $3670 \pm 60$ & $4590 \pm 70$ & $183 \pm 13$   \\
\noalign{\smallskip}\hline\noalign{\smallskip}
Fitted top events            & $198 \pm 7$  & $237 \pm 8$  & $1090 \pm 30$ & $220 \pm 9$   & $281 \pm 11$  & $0.104 \pm 0.011$     \\   
Fitted $W$/$Z$ events        & $45 \pm 4$   & $51 \pm 5$   & $3.5 \pm 0.3$     & $3090 \pm 90$ & $4220 \pm 80$ & $1.06 \pm 0.11$     \\   
Fitted $WW$ diboson events  & $0.22 \pm 0.08$    & $0.10 \pm 0.15$    & $1.3 \pm 0.5$     & $6 \pm 3$     & $8 \pm 5$     & $1.2 \pm 0.6$     \\   
Fitted multijet events      & $0 \pm 0$      & $0 \pm 0$      & $0 \pm 0$      & $0 \pm 0$      & $0 \pm 0$      & $0 \pm 0$     \\
Fitted charge flip events    & $0 \pm 0$      & $0 \pm 0$      & $0 \pm 0$       & $0 \pm 0$       & $0 \pm 0$       & $94 \pm 14$     \\   
Fitted fake lepton events    & $34 \pm 15$  & $22 \pm 8$   & $220 \pm 40$  & $360 \pm 100$ & $80 \pm 50$   & $87 \pm 19$     \\   
\noalign{\smallskip}\hline\noalign{\smallskip}
\noalign{\smallskip}\hline\noalign{\smallskip}
Expected events                 & $305$     & $336$     & $1340$   & $3920$    & $5050$    & $148$       \\
\noalign{\smallskip}\hline\noalign{\smallskip}
MC exp. top events              & $225$     & $276$     & $1220$   & $278$     & $357$     & $0.094$         \\
MC exp. $W$/$Z$ events          & $41$      & $47$      & $3.1$      & $3360$    & $4600$    & $1.14$         \\
MC exp. $WW$ diboson events    & $0.21$       & $0.09$       & $1.2$      & $6$       & $8$       & $1.2$         \\
MC exp. multijet events        & $0$      & $0$      & $0$      & $0$      & $0$      & $0$          \\
Charge flip events (estimated from data) & $0$       & $0$       & $0$      & $0$       & $0$       & $94$        \\
Fake lepton events (estimated from data) & $39$      & $13$      & $120$    & $270$     & $80$      & $51$        \\
\noalign{\smallskip}\hline\noalign{\smallskip}
\end{tabular}
}
\end{center}
\end{table*}

\begin{table*}
\begin{center}
\caption{
Results of the background-only fit to the control regions in the zero- and one-lepton signal regions, for an integrated luminosity of 4.7~fb$^{-1}$.
The errors shown are the statistical plus systematic uncertainties.
The $p_0$-values and significances are given for single-bin signal regions with somewhat tighter $M_R'$ cuts, along with the 95\,\% Confidence Level upper limit on the number of and cross-section for non-Standard-Model production within each signal region.
\label{table:SRbreakdown01L}}
\scalebox{0.89}{
\begin{tabular}{@{\extracolsep{\fill}}lrrrrrrrrrrrrrrrrrrrrrrrr}
\noalign{\smallskip}\hline\noalign{\smallskip}
Signal region               & Had. $b$-veto          & Had. $b$-tag           & $e$ $b$-veto             & $e$ $b$-tag            & $\mu$ $b$-veto       & $\mu$ $b$-tag      \\
\noalign{\smallskip}\hline\noalign{\smallskip}
Observed events             & $4$                  & $30$                 & $6$                  & $13$               & $9$                & $4$              \\
\noalign{\smallskip}\hline\noalign{\smallskip}
Fitted background events    & $5.5 \pm 1.5$        & $39 \pm 7$           & $10 \pm 2$           & $6.6 \pm 1.7$      & $5.5 \pm 1.7$      & $4.4 \pm 1.3$    \\
\noalign{\smallskip}\hline\noalign{\smallskip}     
Fitted top events           & $0.40 \pm 0.14$      & $21 \pm 3$           & $2.7 \pm 0.9$        & $5.0 \pm 1.3$      & $1.7 \pm 0.6$      & $3.7 \pm 1.1$    \\
Fitted $W$/$Z$ events       & $4.9 \pm 1.3$        & $3.8 \pm 0.7$        & $7.2 \pm 1.7$        & $1.2 \pm 0.5$      & $3.8 \pm 1.3$      & $0.6 \pm 0.5$    \\
Fitted $WW$ diboson events & $0.03 \pm 0.02$      & $0.029 \pm 0.010$      & $0.01 \pm 0.02$      & $0.000 \pm 0.009$    & $0.001 \pm 0.008$    & $0.010 \pm 0.005$    \\
Fitted multijet events         & $0.25 \pm 0.10$      & $14 \pm 5$      & $0 \pm 0$      & $0 \pm 0$      & $0 \pm 0$      & $0 \pm 0$  \\
Fitted charge flip events   & $0 \pm 0$      & $0 \pm 0$      & $0 \pm 0$      & $0 \pm 0$      & $0 \pm 0$      & $0 \pm 0$    \\
Fitted fake lepton events   & $0 \pm 0$      & $0 \pm 0$      & $0.3 \pm 0.3$  & $0.5 \pm 0.6$  & $0 \pm 0$      & $0 \pm 0$    \\
\noalign{\smallskip}\hline\noalign{\smallskip}
\noalign{\smallskip}\hline\noalign{\smallskip}
Expected events                & $6.7$       & $55$        & $14$        & $8.5$       & $9.5$       & $5.1$   \\
\noalign{\smallskip}\hline\noalign{\smallskip} 
MC exp. top events             & $0.88$      & $30$        & $5.7$       & $6.3$       & $3.4$       & $4.6$   \\
MC exp. $W$/$Z$ events         & $5.6$       & $4.0$       & $8.5$       & $1.8$       & $6.1$       & $0.5$   \\
MC exp. $WW$ diboson events   & $0.04$      & $0.046$     & $0.01$      & $0.000$     & $0.012$     & $0.010$   \\
MC exp. multijet events       & $0.20$      & $21$      & $0$      & $0$      & $0$      & $0$ \\ 
Charge flip events (estimated from data) & $0$         & $0$         & $0$         & $0$         & $0$         & $0$   \\
Fake lepton events (estimated from data) & $0$         & $0$         & $0.3$       & $0.5$       & $0$         & $0$   \\
\noalign{\smallskip}\hline\noalign{\smallskip} 
\noalign{\smallskip}\hline\noalign{\smallskip}
Tight $M_R'$ cut (\GeV) & 600                                   & 1100                                             & 600                                             & 1100                                             & 600                                             & 1100                                               \\
Observed events      & 4             & 5          & 5             & 6             & 2             & 4             \\
Background events    & $6.2 \pm 1.8$ & $13 \pm 3$ & $5.3 \pm 1.6$ & $2.4 \pm 1.0$ & $2.4 \pm 1.0$ & $1.9 \pm 0.8$ \\
$p_0$-value (Gauss. $\sigma$)  & 0.72 ($-0.57$)                                     & 0.91 ($-1.35$)                                     & 0.53 ($-0.07$)                                     & 0.07 (1.50)                                     & 0.54 ($-0.10$)                                     & 0.16 (0.98)                                   \\
Upper limit on $N_\text{BSM}$   & 5.2 ($6.3^{\uparrow  9.4}_{\downarrow 4.3}$)  & 6.5 ($9.3^{\uparrow  12.9}_{\downarrow 6.9}$)  & 6.3 ($6.4^{\uparrow  9.5}_{\downarrow 4.4}$)  & 9.0 ($5.5^{\uparrow  8.4}_{\downarrow 3.7}$)  & 4.4 ($4.5^{\uparrow  7.1}_{\downarrow 3.0}$)  & 6.8 ($4.8^{\uparrow  7.5}_{\downarrow 3.2}$)  \\
Upper limit on $\sigma$ (fb)    & 1.1 ($1.3^{\uparrow  2.0}_{\downarrow 0.9}$)  & 1.4 ($2.0^{\uparrow  2.7}_{\downarrow 1.5}$)  & 1.3 ($1.4^{\uparrow  2.0}_{\downarrow 0.9}$)  & 1.9 ($1.2^{\uparrow  1.8}_{\downarrow 0.8}$)  & 0.9 ($1.0^{\uparrow  1.5}_{\downarrow 0.6}$)  & 1.4 ($1.0^{\uparrow  1.6}_{\downarrow 0.7}$)  \\
\noalign{\smallskip}\hline\noalign{\smallskip}
\end{tabular}
}
\end{center}
\end{table*}

\begin{table*}
\begin{center}
\caption{
Results of the background-only fit to the control regions in the two-lepton signal regions, for an integrated luminosity of 4.7~fb$^{-1}$.
The errors shown are the statistical plus systematic uncertainties.
The $p_0$-values and significances are given for single-bin signal regions, along with the 95\,\% Confidence Level upper limit on the number of and cross-section for non-Standard-Model production within each signal region.
\label{table:SRbreakdown2L}}
\scalebox{0.82}{
\begin{tabular}{@{\extracolsep{\fill}}lrrrrrrrrrrrrrrrrrrrrrrrr}
\noalign{\smallskip}\hline\noalign{\smallskip}
Signal region           & OS-$ee$            & OS-$\mu\mu$      & SS-$ee$           & SS-$\mu\mu$     & OS-$e\mu$        & SS-$e\mu$    \\[-0.05cm]
\noalign{\smallskip}\hline\noalign{\smallskip}
Observed events         & $10$             & $15$             & $11$            & $8$             & $18$             & $18$             \\
\noalign{\smallskip}\hline\noalign{\smallskip}
Fitted background events     & $12 \pm 2$        & $13 \pm 2$           & $6 \pm 4$            & $4 \pm 3$          & $20. \pm 3$      & $14 \pm 8$          \\
\noalign{\smallskip}\hline\noalign{\smallskip}
Fitted top events            & $10.2 \pm 1.5$    & $10.7 \pm 1.6$       & $0.12 \pm 0.04$      & $0.39 \pm 0.17$    & $19 \pm 2$       & $0.7 \pm 0.2$          \\
Fitted $W$/$Z$ events        & $0.54 \pm 0.10$   & $0.6 \pm 0.2$        & $0.16 \pm 0.04$      & $0.10 \pm 0.04$    & $0.26 \pm 0.04$  & $0.33 \pm 0.07$          \\
Fitted $WW$ diboson events  & $0.4 \pm 0.4$     & $0.4 \pm 0.5$        & $0.6 \pm 0.3$        & $0.6 \pm 0.5$      & $0.5 \pm 1.0$    & $1.2 \pm 0.7$          \\
Fitted multijet events      & $0 \pm 0$         & $0 \pm 0$            & $0 \pm 0$        & $0 \pm 0$          & $0 \pm 0$        & $0 \pm 0$          \\
Fitted charge flip events    & $0 \pm 0$         & $0 \pm 0$            & $1.6 \pm 0.4$        & $0 \pm 0$          & $0 \pm 0$        & $1.1 \pm 0.2$          \\
Fitted fake lepton events    & $1.2 \pm 1.3$     & $1.3 \pm 1.1$        & $3 \pm 4$            & $3 \pm 3$          & $0.6 \pm 0.6$    & $10. \pm 8$          \\
\noalign{\smallskip}\hline\noalign{\smallskip}      
\noalign{\smallskip}\hline\noalign{\smallskip}
Expected events                 & $15$      & $16$        & $6$         & $5$        & $24$       & $14$          \\
\noalign{\smallskip}\hline\noalign{\smallskip}
MC exp. top events              & $13.1$     & $14.7$     & $0.13$      & $0.49$     & $23$       & $0.6$         \\
MC exp. $W$/$Z$ events          & $0.67$     & $0.4$      & $0.19$      & $0.21$     & $0.27$     & $0.36$        \\
MC exp. $WW$ diboson events     & $0.4$      & $0.4$      & $0.7$       & $0.7$      & $0.5$      & $1.2$         \\
MC exp. multijet events         & $0$        & $0$        & $0$         & $0$        & $0$        & $0$           \\
Charge flip events (estimated from data) & $0$        & $0$        & $1.6$       & $0$        & $0$        & $1.0$         \\
Fake lepton events (estimated from data) & $1.2$      & $1.3$      & $3$         & $3$        & $0.6$      & $11$          \\
\noalign{\smallskip}\hline\noalign{\smallskip}
\noalign{\smallskip}\hline\noalign{\smallskip}
$p_0$-value (Gauss. $\sigma$)  & 0.71 ($-0.56$)                                      & 0.32 (0.46)                                      & 0.15 (1.05)                                       & 0.18 (0.93)                                      & 0.68 ($-0.48$)                                       & 0.29 (0.54)            \\ 
Upper limit on $N_\text{BSM}$   & 7.3 ($8.8^{\uparrow  12.9}_{\downarrow 6.2}$)  & 11.1 ($9.4^{\uparrow  13.7}_{\downarrow 6.6}$)  & 14.0 ($10.2^{\uparrow  14.4}_{\downarrow 7.4}$)  & 11.4 ($8.0^{\uparrow  11.4}_{\downarrow 5.7}$)  & 9.4 ($11.1^{\uparrow  16.0}_{\downarrow 7.8}$)  & 17.7 ($14.9^{\uparrow  20.8}_{\downarrow 10.8}$)  \\
Upper limit on $\sigma$ (fb)    & 1.6 ($1.9^{\uparrow  2.7}_{\downarrow 1.3}$)  & 2.4 ($2.0^{\uparrow  2.9}_{\downarrow 1.4}$)  & 3.0 ($2.2^{\uparrow  3.1}_{\downarrow 1.6}$)  & 2.4 ($1.7^{\uparrow  2.4}_{\downarrow 1.2}$)  & 2.0 ($2.4^{\uparrow  3.4}_{\downarrow 1.7}$)  & 3.8 ($3.2^{\uparrow  4.4}_{\downarrow 2.3}$)  \\
\noalign{\smallskip}\hline\noalign{\smallskip}
\end{tabular}
}
\end{center}
\end{table*}

Various tests of the fit are carried out in order to ensure
its stability.  As a test of the multijet background constraint and the
validity of fitting the $M_R'$ distributions in those control
regions, the control region fit is instead performed in the number of jets with
$\pt>30$~\GeV.  The $\pt$ cut is raised from the baseline selection to
make the fit less sensitive to pile-up effects.  The expectation for
the multijet background in the signal regions is consistent with the
main result.

\begin{figure*}[tbp]
\centering
\includegraphics[width=0.47\textwidth]{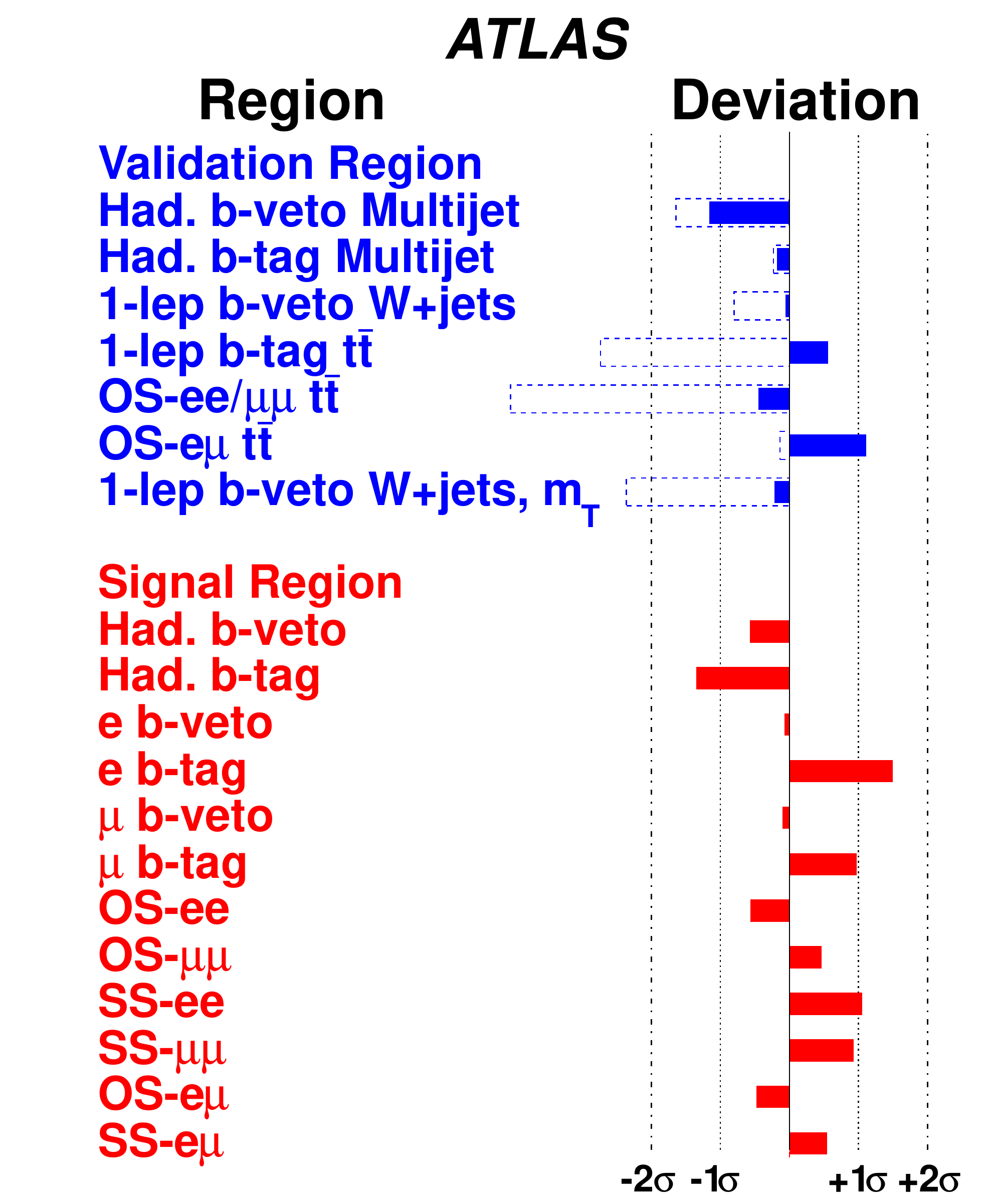}
\caption{
Pull distributions of the numbers of events in the validation regions (VR) and signal regions (SR).
The filled (dashed) bars show the agreement after (before) the background-only fit to the control regions has been performed.
\label{fig:pull_plot}
}
\end{figure*}

The yields and shapes of the validation regions show good agreement with 
the Standard Model expectation.  The significance of the deviation of the 
observation from the expectation in each of the signal and validation
regions are shown in Figure~\ref{fig:pull_plot}.  
There is some tension between the 
same-flavour and opposite-flavour dilepton \ttbar\ validation regions, but 
there is no indication of a systematic mis-modelling of any of the major backgrounds.
The yields of all validation regions are within $1.2\sigma$ of the SM expectations.

\begin{figure*}[tp]
\centering
\includegraphics[width=0.47\textwidth]{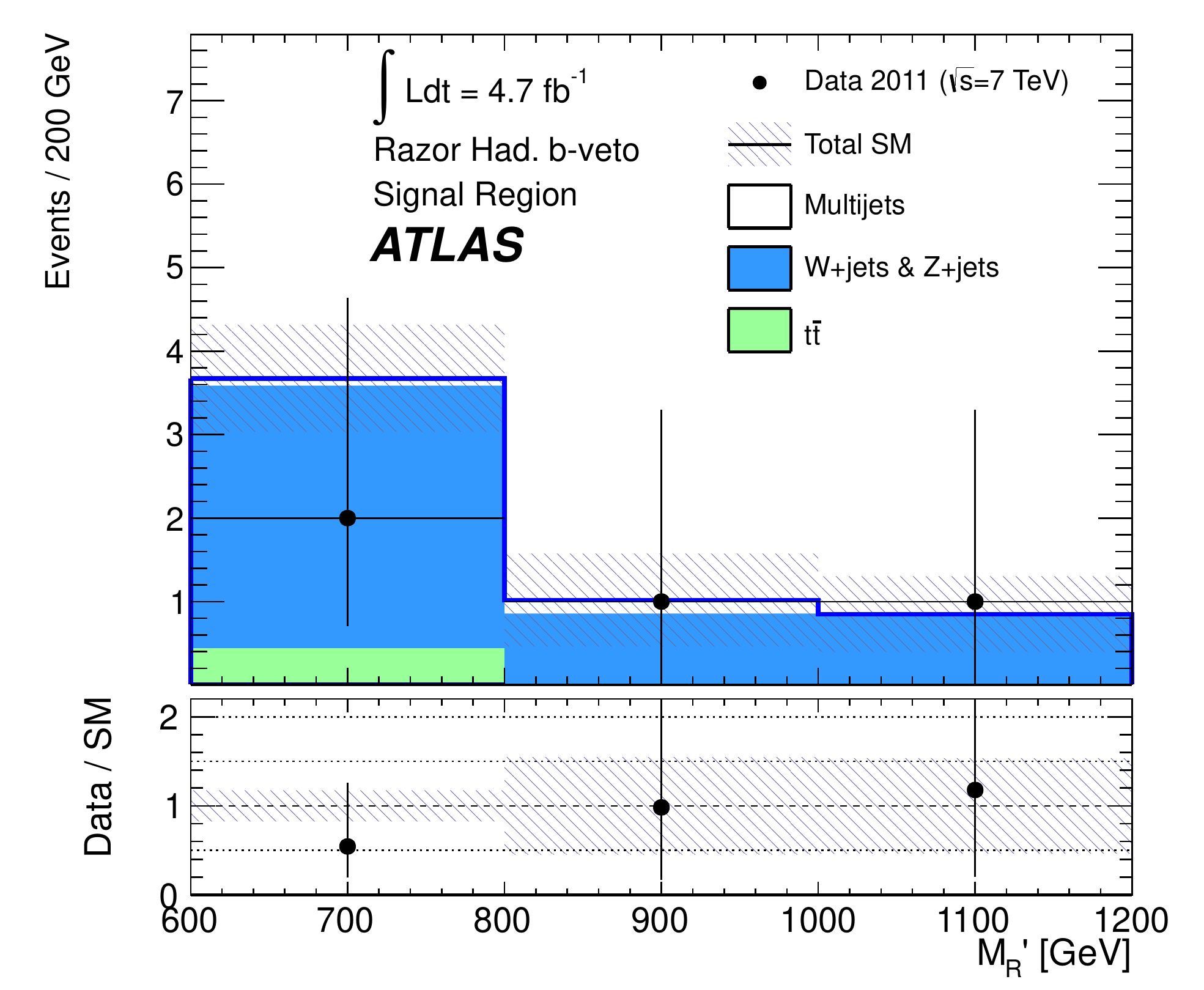}
\includegraphics[width=0.47\textwidth]{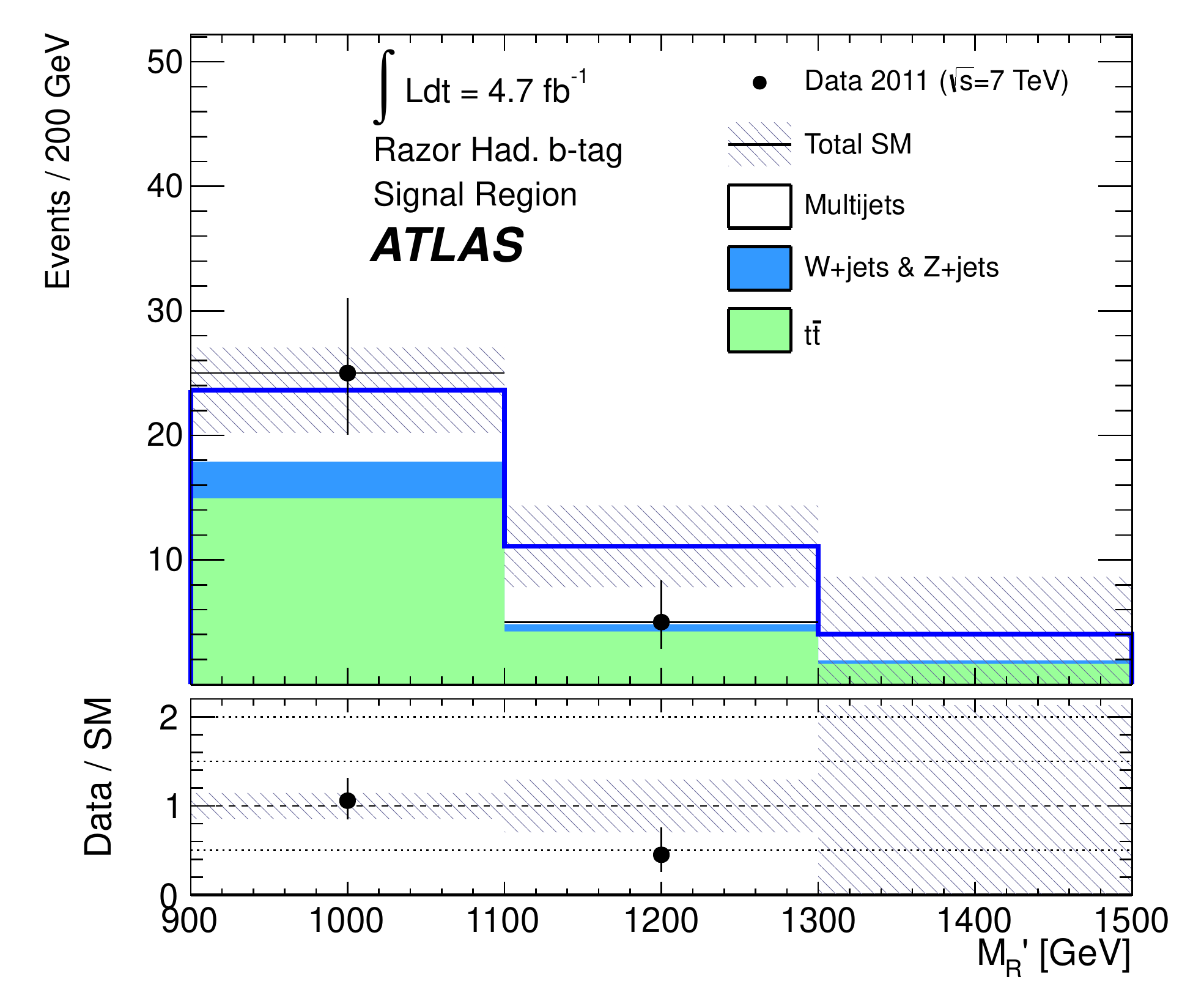}
\includegraphics[width=0.47\textwidth]{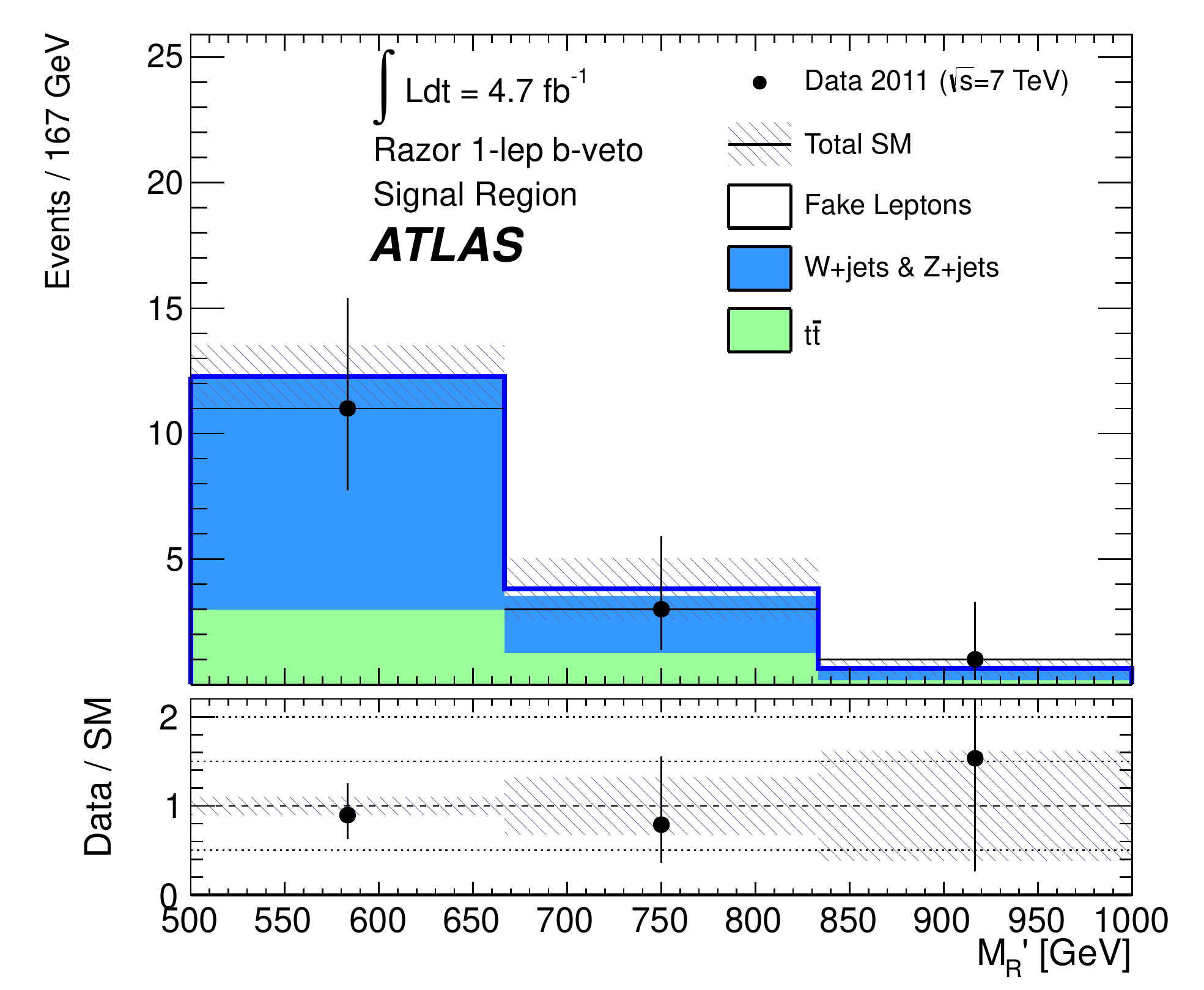}
\includegraphics[width=0.47\textwidth]{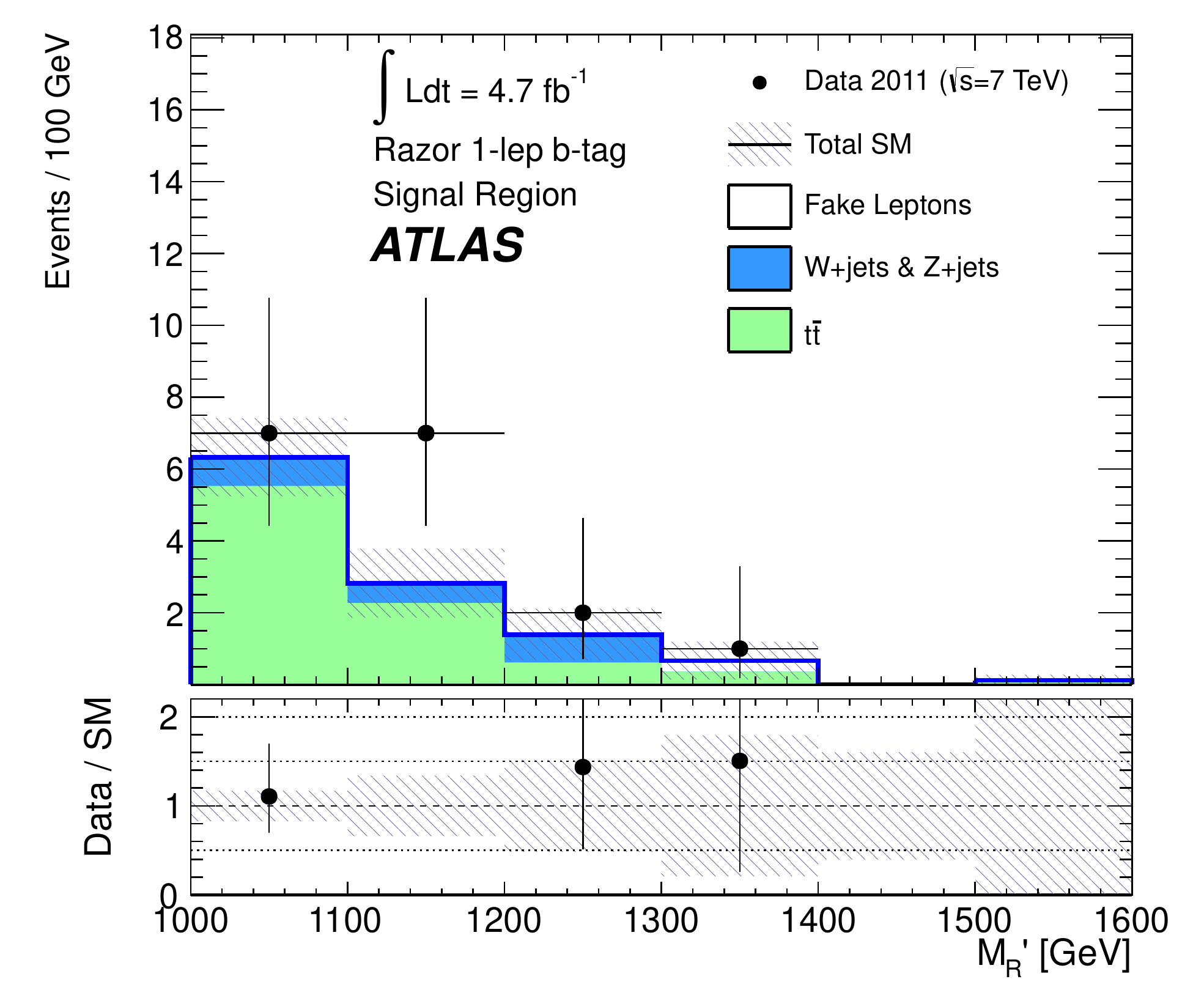}
\caption{
\label{fig:01LepSR}  
The all-hadronic (top) and one-lepton (bottom) signal regions with a $b$-tagged jet veto (left) and requirement (right), after the fit to the control regions has been performed.}
\end{figure*}

\begin{figure*}[tp]
\centering
\includegraphics[width=0.47\textwidth]{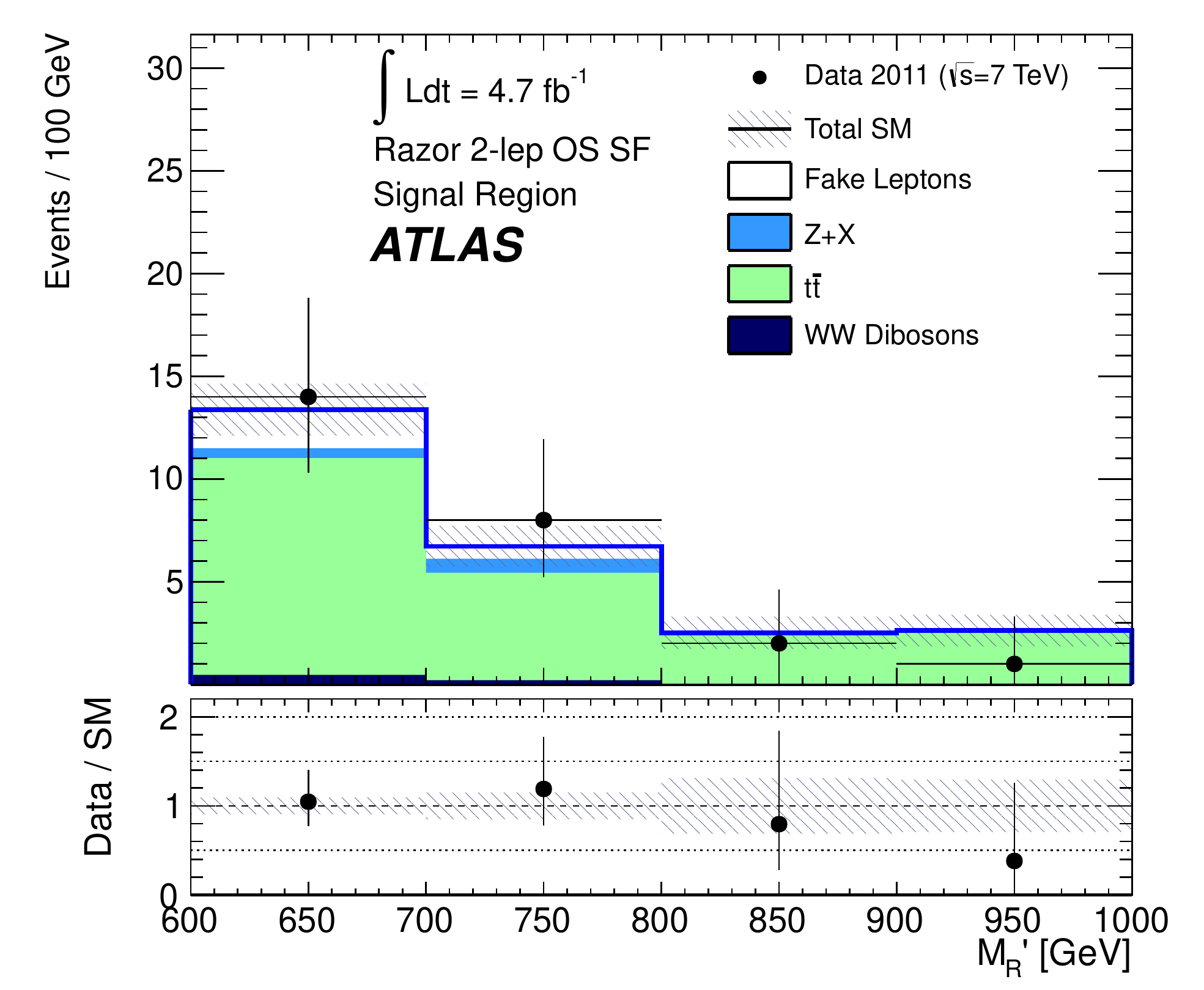}
\includegraphics[width=0.47\textwidth]{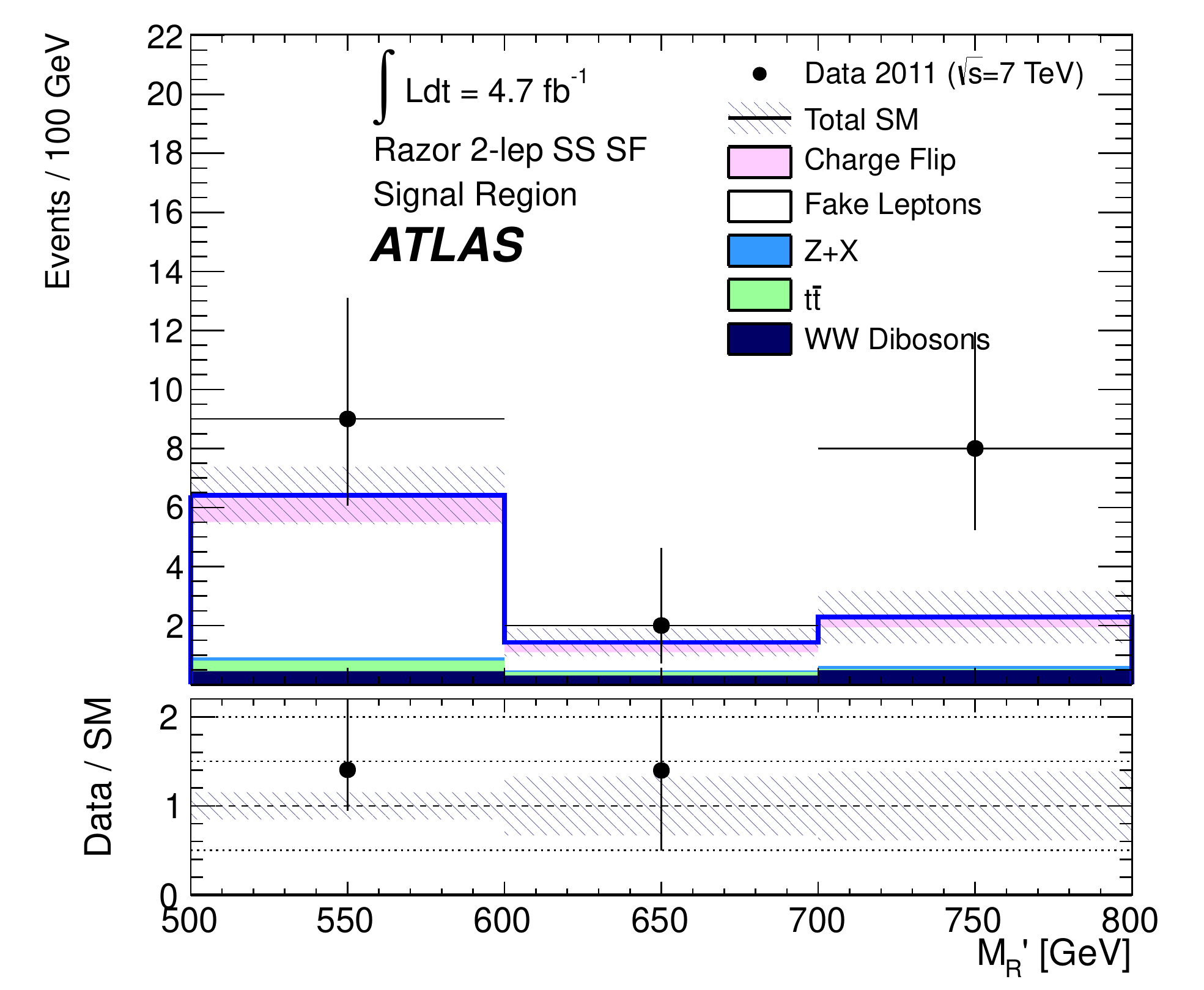}
\includegraphics[width=0.47\textwidth]{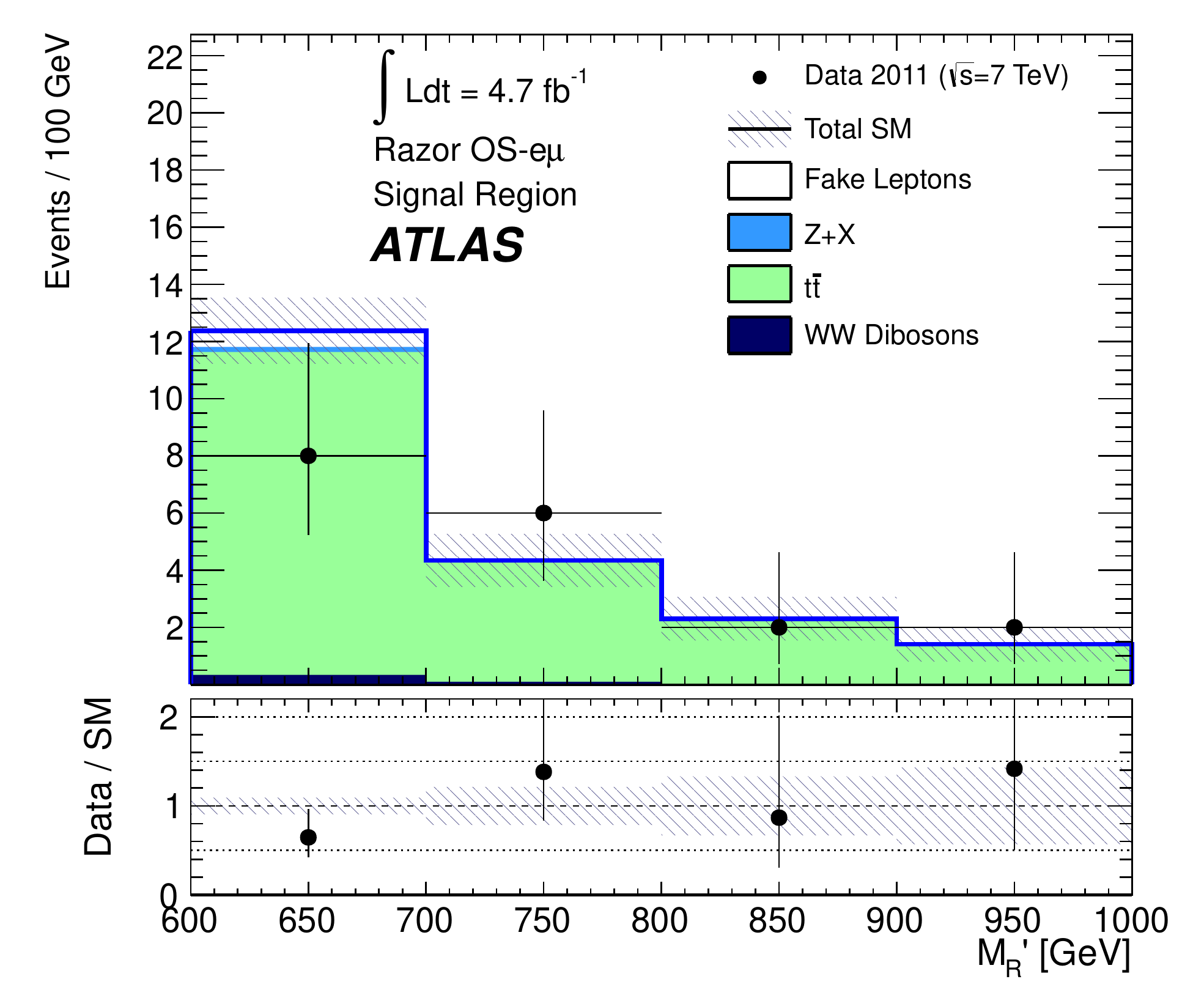}
\includegraphics[width=0.47\textwidth]{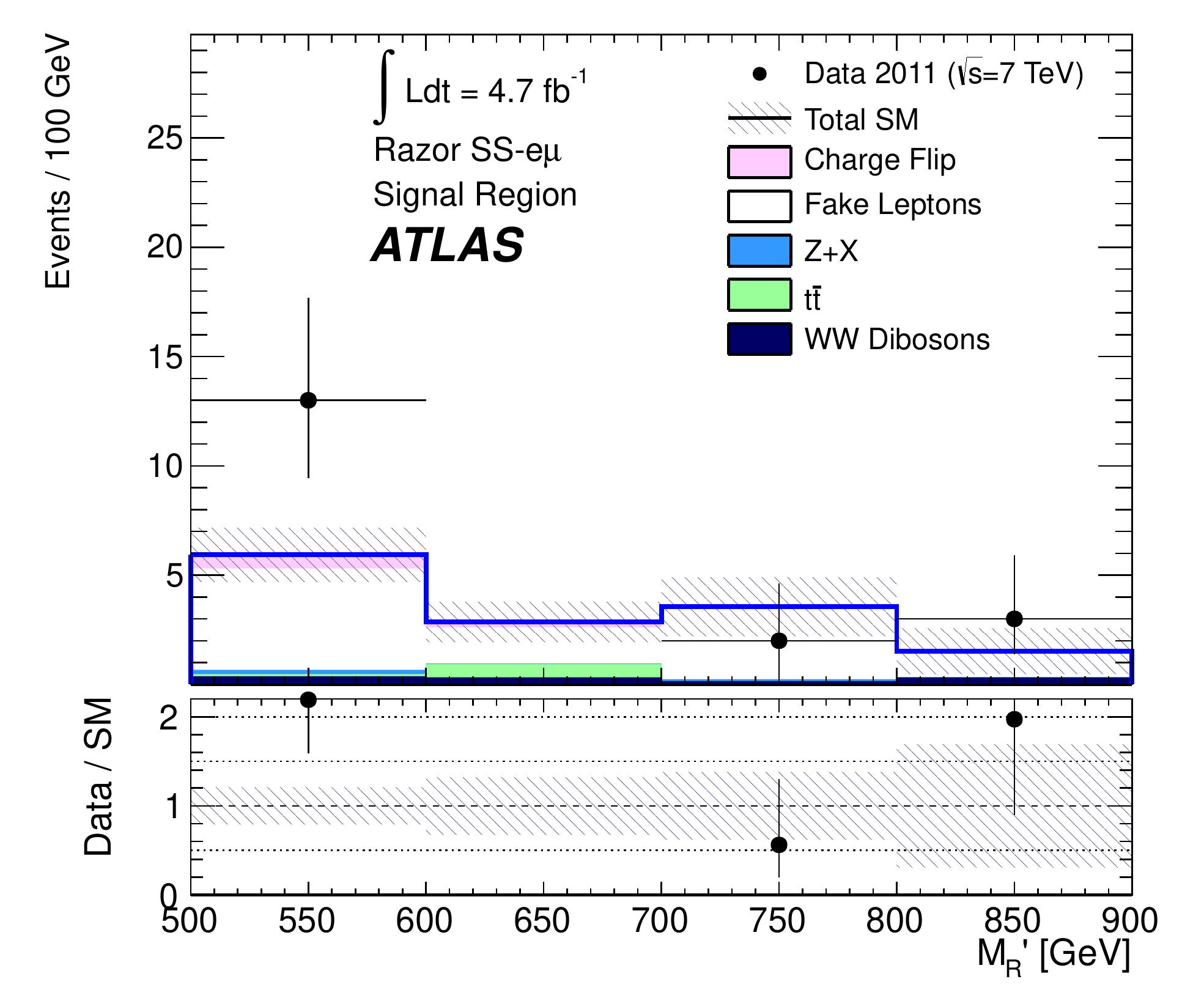}
\caption{
\label{fig:2LepSR}     
The two-lepton signal regions for same-flavour (top) and opposite-flavour (bottom) leptons of the same sign (left) and opposite sign (right), after the fit to the control regions has been performed.}
\end{figure*}

The numbers of expected events in each signal region before and after the fit to the control regions are shown in Tables~\ref{table:SRbreakdown01L} and~\ref{table:SRbreakdown2L}.
Additionally, the probability ($p_0$-value) that a background-only pseudo-experiment is more signal-like than observed is given for each individual signal region.
To obtain these $p_0$-values, the fit in the signal region proceeds in the same way as the control-region-only fit, except that the number of events observed in the signal region is included as an input to the fit.  
Then, an additional parameter for the non-Standard-Model signal strength, constrained to be non-negative, is fitted.
The shape of the distributions in the signal region is neglected in this fit.
Therefore, in order to provide tighter constraints on non-Standard-Model production, in some of the high-count signal regions the $M_R'$ requirements are tightened.
The significance of the excess is given, along with the model-independent upper limit on the number of events and cross-section times acceptance times efficiency from non-Standard-Model production.
The distributions in all signal regions as a function of $M_R'$ of background expectations after the fit to the control region has been performed are shown in Figs.~\ref{fig:01LepSR} and~\ref{fig:2LepSR}.
No significant deviations from the expected background are found.
The most significant excess is 1.50 standard deviations from the expectation, in the one electron, $b$-tagged jet signal region.

\section{Exclusion results}
\label{sec:exclusion}

Using these signal regions, the $CL_S$~\cite{CLS} prescription is applied to find 95\,\% 
Confidence Level (CL) one-sided limits on the production of SUSY events in various models.
A likelihood is constructed, taking into account signal shape information
provided by the binning of the signal regions.  All fitted nuisance parameters, 
with their correlations, are included  in the likelihood.
Because the typical $M_R'$ of the signal may vary across a signal grid, the 
use of shape information results in an observed exclusion that is not consistently
above or below the expected.
The observed limits for the separate zero-, one-, and two-lepton signal
regions are also constructed additionally, with all control
regions included as constraints.

\begin{figure*}[!t]
\centering
\includegraphics[width=0.87\textwidth]{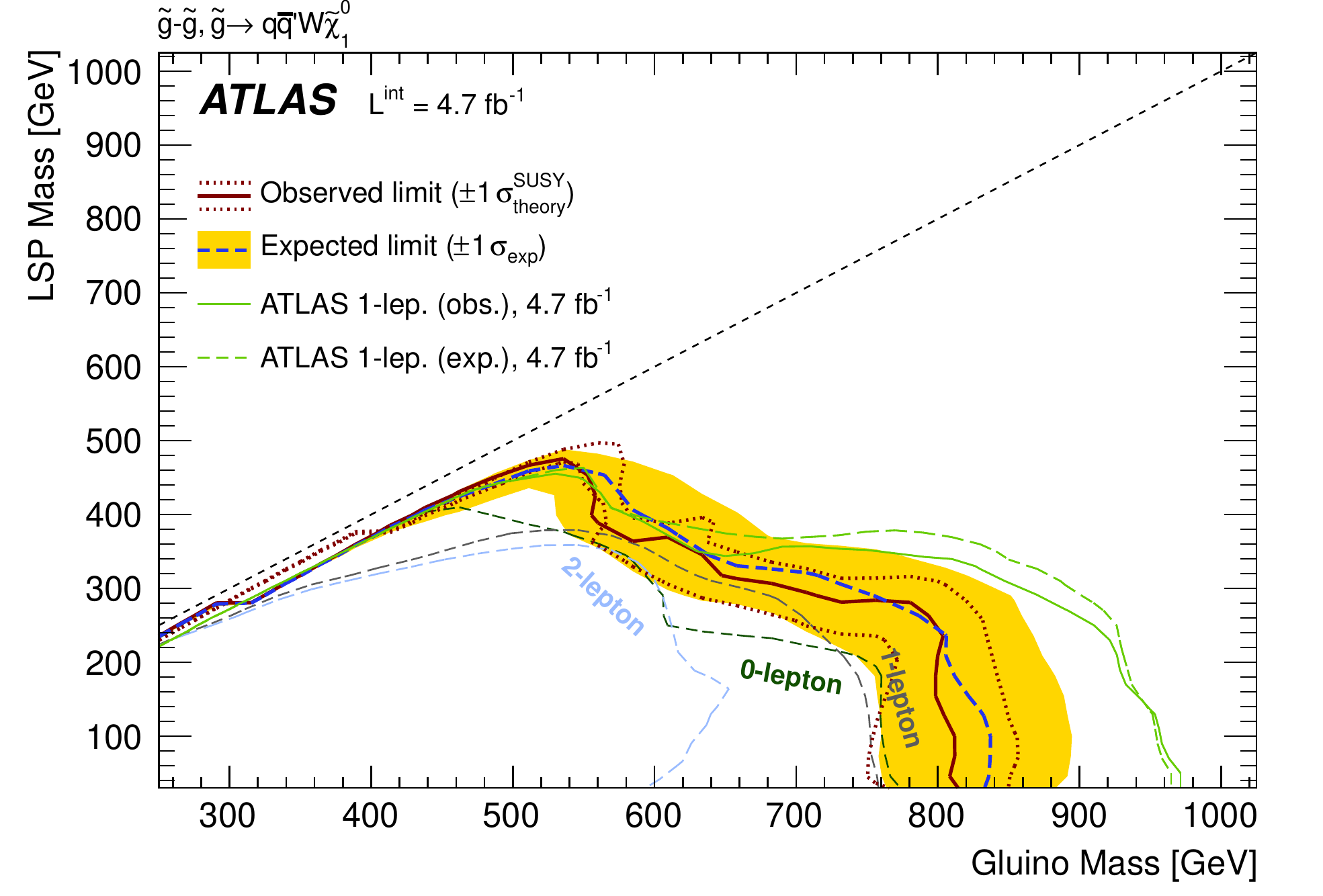}
\includegraphics[width=0.87\textwidth]{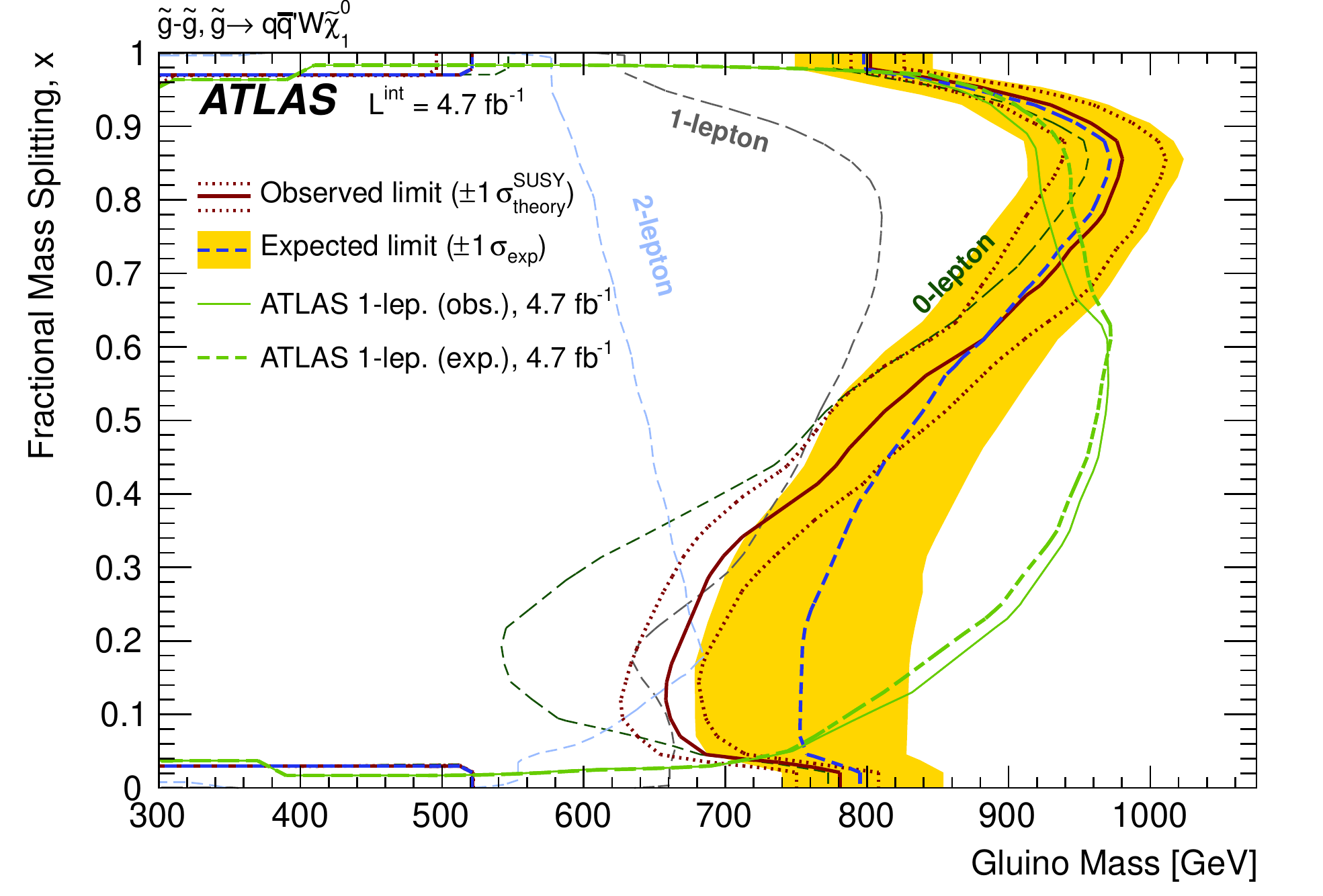}
\caption{
Exclusion in a simplified model with gluino pair-production, where the gluinos decay to a chargino via the emission of two quarks and the chargino decays to the LSP and a $W$ boson.
Top, for the chargino mass exactly between the gluino and LSP mass, in the gluino mass--LSP mass plane, and bottom, for the LSP mass fixed to 60~\GeV, in the gluino mass--$x$ plane, with $x=(m_\text{chargino}-m_\text{LSP})/(m_\text{gluino}-m_\text{LSP})$.
The exclusion is shown for the combination, as well as for each individual channel.
Overlaid is the exclusion result published in Ref.~\cite{multiLepMultiJetATLAS}.
\label{fig:OS_exclusion}
}
\end{figure*}

Figure~\ref{fig:OS_exclusion} shows exclusion contours for a
simplified model with gluino pair-production, where the gluinos decay
to a chargino and two quarks and the chargino subsequently decays to a
$W$ boson and the LSP.  Two planes are shown for this simplified
model.  The first fixes the chargino mass to be exactly between the
LSP and gluino mass and shows the exclusion in the gluino-mass--LSP-mass plane.  
The production cross-section falls smoothly and
exponentially with $m_\text{heavy}$, while $M_R'$ and
therefore the acceptance times
efficiency for a signal region typically rises with the mass
splitting, $m_\text{heavy}-m_\text{LSP}$.  In the second plane, the
LSP mass is fixed to 60~\GeV\ and the exclusion is shown in the gluino
mass--$x$ plane, where
$x=(m_\text{chargino}-m_\text{LSP})/(m_\text{gluino}-m_\text{LSP})$.
The zero- and one-lepton signal regions with a $b$-tagged jet
requirement do not contribute to the exclusion because these
simplified models have only light quarks in the matrix element final
state.

At high $x$, although the leptons have high-$\pt$, the larger branching
fraction of the $W$ to quarks allows the zero-lepton channel to dominate.
At moderate $x$, the leptons allow better discrimination between signal and background 
in the one- and two-lepton channels.  At low $x$, the leptons have
too low $\pt$, and the zero-lepton channel again dominates.
At high $x$, the limit set by this analysis exceeds somewhat that of the dedicated 
0-lepton and 1-lepton ATLAS searches~\cite{0leptonATLAS,multiLepMultiJetATLAS}, while at low $x$ the limit is
weaker.  This asymmetry is produced by the difference in kinematics
in these two regions of the plane.  At high $x$, the charginos are almost at
rest in the lab frame, and the event topology is dominated by a two-body decay, 
$\tilde{\chi}_1^\pm\rightarrow W^\pm\tilde{\chi}_1^0$.
At low $x$, on the other hand, the chargino is highly boosted, and the topology is
dominated by a three-body decay, $\tilde{g}\rightarrow q\bar{q}' \tilde{\chi}_1^\pm$.
Thus, the high-$x$ events have a higher typical $R$ than the low-$x$ events,
and the two have approximately the same $M_R'$ distribution.

\begin{figure*}[!t]
\centering
\includegraphics[width=0.87\textwidth]{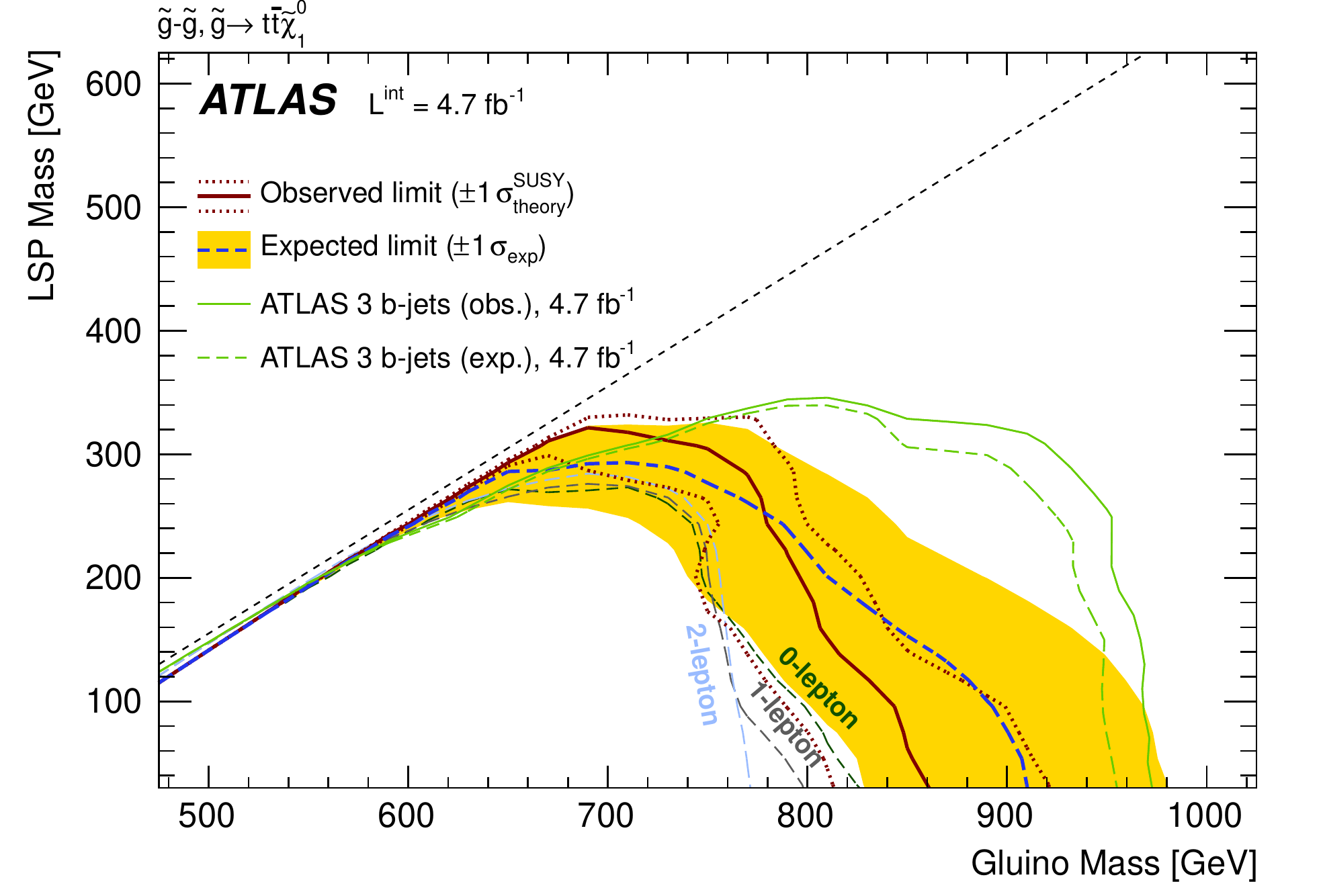}
\caption{
Exclusion in a simplified model with gluino pair-production, where the gluinos decay to the LSP via the emission of a \ttbar\ pair.
The exclusion is shown for the combination, as well as for each individual channel.
Overlaid is the exclusion result published in Ref.~\cite{0leptonATLASMBJ}.
\label{fig:Gtt_exclusion}
}
\end{figure*}

Figure~\ref{fig:Gtt_exclusion} shows exclusion contours in 
simplified models with gluino pair-production, where the gluinos decay
to the LSP via the emission of a \ttbar\ pair.  The exclusion is
presented in the gluino mass--LSP mass plane, and, since all top quarks
are required to be on-shell, only points with
$m_\text{gluino}>m_\text{LSP}+2\times m_\text{top}$ are considered.
The zero- and one-lepton signal regions with a $b$-tagged jet veto do
not contribute to the exclusion, because these models include four top
quarks per event.  At small mass splitting, the limits here are somewhat
stronger than the ATLAS multijet analysis~\cite{0leptonATLASMJ}, though
both are exceeded by those of the dedicated multi-$b$-jet analysis~\cite{0leptonATLASMBJ}. 
The three $b$-tagged jet requirement suppresses the background substantially 
while preserving the signal acceptance because of the four tops in the event.
The combined limit on LSP mass falls more
quickly than that of the multi-$b$-jet analysis because $M_R'$ 
is proportional to the mass splitting in the event,
here $m_\text{gluino}-m_\text{LSP}$.  The zero-lepton razor analysis is
limited in this case by the use of the $\HT$ trigger, which was chosen to 
avoid a bias in the $M_R'$ distribution.

\begin{figure*}[!t]
\centering
\includegraphics[width=0.87\textwidth]{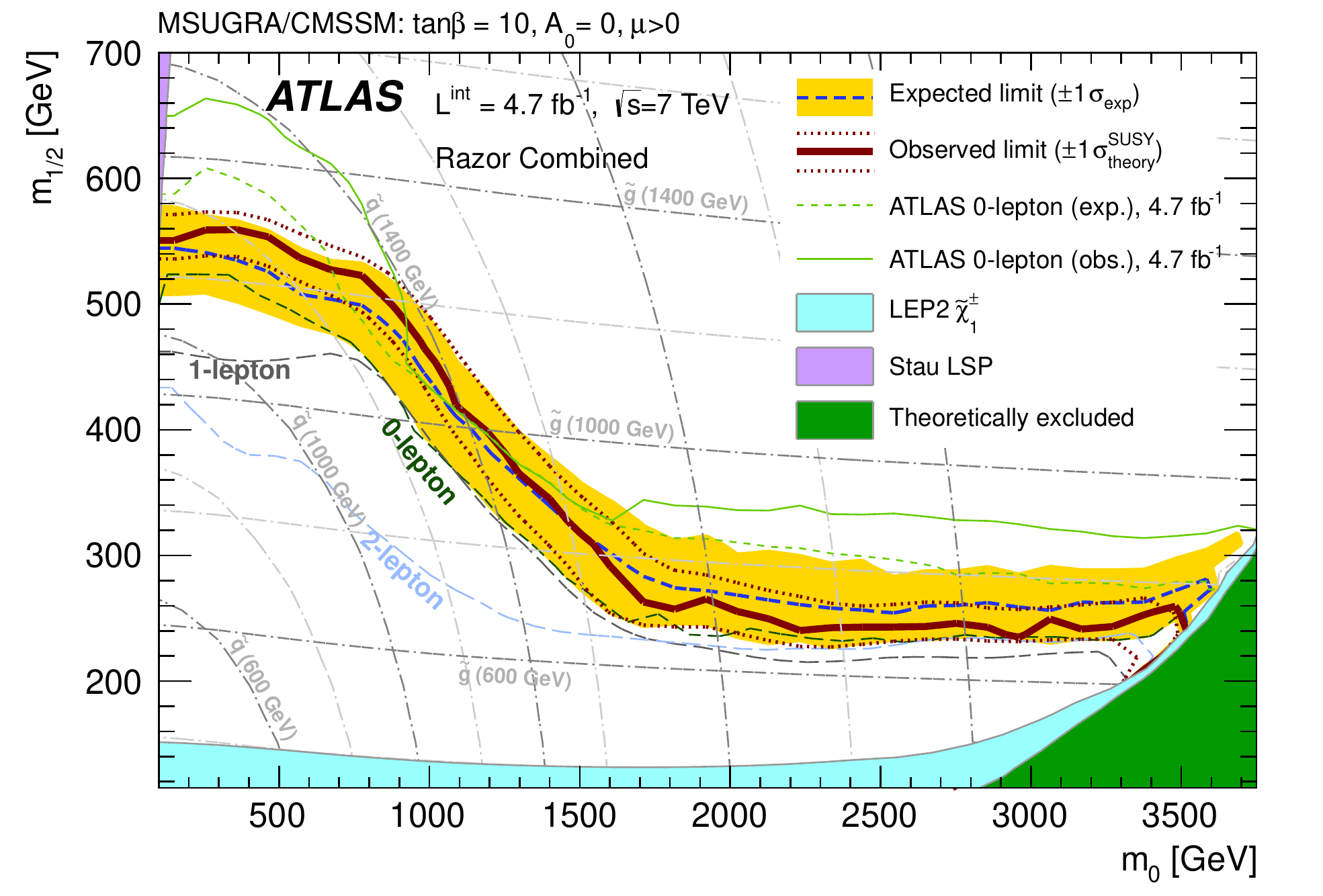}
\caption{
Exclusion in a plane of the constrained minimal supersymmetric model.
The exclusion is shown for the combination, as well as for each individual channel.
Overlaid is the exclusion result published in Ref.~\cite{0leptonATLAS}.
\label{fig:SU_exclusion}
}
\end{figure*}

Finally, Figure~\ref{fig:SU_exclusion} shows exclusion contours in a plane of MSUGRA with $\tan(\beta)=10$, $A_0=0$, and $\mu>0$.
At low $m_0$, where squark pair-production is dominant, the zero-lepton channel dominates the exclusion, although it is affected somewhat by the jet multiplicity requirement that is not applied in the dedicated signal region of Ref.~\cite{0leptonATLAS}, which therefore has more stringent limits.
The leptonic channels enter at high $m_0$, particularly where longer decay chains are common.
The robustness of the individual limits have also been cross checked by removing some of the control regions.
For example, removing the zero- and one-lepton control regions from the calculation of the two-lepton limit, the MSUGRA limit changes by less than 20~\GeV\ in $m_{1/2}$.
These limits are consistent with those of earlier ATLAS analyses~\cite{0leptonATLAS,0leptonATLASMJ,multiLepMultiJetATLAS} in the $m_\text{gluino}\approx m_\text{squark}$ region.
In this region, the single mass-splitting scale of the main strong production modes should produce a somewhat sharper peak in $M_R'$,
allowing an improved limit in the shape fit.
At large $m_0$, the high $M_R'$ requirement of the all-hadronic signal regions, resulting from the $\HT$ trigger use, produce a somewhat weaker limit than the ATLAS multijet analysis~\cite{0leptonATLASMJ}.

The complementarity of a search using razor variables can be quantified 
by studying the overlap of the signal regions with the dedicated
searches.  Various signal models have been studied to understand
this overlap, including both simplified models and full SUSY production models.
The overlap between the signal regions presented here and other searches in
ATLAS~\cite{0leptonATLAS,multiLepMultiJetATLAS,0leptonATLASMJ,multiLepATLAS} is
typically 10--50\,\%, with similar overlaps in the data. 
The signal regions of this
search access kinematic regions that are different from those of
the standard searches. In simplified models in particular,
the overlap between the dominant signal regions in the standard ATLAS analyses
and the signal regions presented here is below 10--15\,\%.  Thus, the regions
of SUSY parameter space and kinematic phase space excluded by this search 
complement those excluded by earlier ATLAS searches using the same data sample.

\section{Summary}
\label{sec:summary}

A search for supersymmetry including final states with zero, one, and two leptons, with and without $b$-tagged jets, in 4.7~fb$^{-1}$ of $\sqrt{s}=7$~\TeV\ $pp$ collisions has been presented.
Mutually exclusive signal regions exploiting these final states are combined with the use of variables that include both transverse and longitudinal event information.
No significant excess of events beyond the Standard Model background expectation was observed in any signal region.
Fiducial cross section upper limits on the production of new physics beyond the Standard Model are shown.
Exclusion contours at 95\,\% CL are provided in SUSY-inspired simplified models and in the constrained minimal supersymmetric extension of the Standard Model.

\begin{acknowledgements}


We thank CERN for the very successful operation of the LHC, as well as the
support staff from our institutions without whom ATLAS could not be
operated efficiently.

We acknowledge the support of ANPCyT, Argentina; YerPhI, Armenia; ARC,
Australia; BMWF and FWF, Austria; ANAS, Azerbaijan; SSTC, Belarus; CNPq and FAPESP,
Brazil; NSERC, NRC and CFI, Canada; CERN; CONICYT, Chile; CAS, MOST and NSFC,
China; COLCIENCIAS, Colombia; MSMT CR, MPO CR and VSC CR, Czech Republic;
DNRF, DNSRC and Lundbeck Foundation, Denmark; EPLANET, ERC and NSRF, European Union;
IN2P3-CNRS, CEA-DSM/IRFU, France; GNSF, Georgia; BMBF, DFG, HGF, MPG and AvH
Foundation, Germany; GSRT and NSRF, Greece; ISF, MINERVA, GIF, DIP and Benoziyo Center,
Israel; INFN, Italy; MEXT and JSPS, Japan; CNRST, Morocco; FOM and NWO,
Netherlands; BRF and RCN, Norway; MNiSW, Poland; GRICES and FCT, Portugal; MERYS
(MECTS), Romania; MES of Russia and ROSATOM, Russian Federation; JINR; MSTD,
Serbia; MSSR, Slovakia; ARRS and MVZT, Slovenia; DST/NRF, South Africa;
MICINN, Spain; SRC and Wallenberg Foundation, Sweden; SER, SNSF and Cantons of
Bern and Geneva, Switzerland; NSC, Taiwan; TAEK, Turkey; STFC, the Royal
Society and Leverhulme Trust, United Kingdom; DOE and NSF, United States of
America.

The crucial computing support from all WLCG partners is acknowledged
gratefully, in particular from CERN and the ATLAS Tier-1 facilities at
TRIUMF (Canada), NDGF (Denmark, Norway, Sweden), CC-IN2P3 (France),
KIT/GridKA (Germany), INFN-CNAF (Italy), NL-T1 (Netherlands), PIC (Spain),
ASGC (Taiwan), RAL (UK) and BNL (USA) and in the Tier-2 facilities
worldwide.

\end{acknowledgements}


\bibliographystyle{spphys}       
\bibliography{razor-paper}

\onecolumn
\clearpage 
\begin{flushleft}
{\Large The ATLAS Collaboration}

\bigskip

G.~Aad$^{\rm 48}$,
T.~Abajyan$^{\rm 21}$,
B.~Abbott$^{\rm 111}$,
J.~Abdallah$^{\rm 12}$,
S.~Abdel~Khalek$^{\rm 115}$,
A.A.~Abdelalim$^{\rm 49}$,
O.~Abdinov$^{\rm 11}$,
R.~Aben$^{\rm 105}$,
B.~Abi$^{\rm 112}$,
M.~Abolins$^{\rm 88}$
% ...
\bigskip
\\
% ...
$^{11}$ Institute of Physics, Azerbaijan Academy of Sciences, Baku, Azerbaijan\\
$^{12}$ Institut de F{\'\i}sica d'Altes Energies and Departament de F{\'\i}sica de la Universitat Aut{\`o}noma de Barcelona and ICREA, Barcelona, Spain\\
% ...
$^{21}$ Physikalisches Institut, University of Bonn, Bonn, Germany\\
% ...
$^{48}$ Fakult{\"a}t f{\"u}r Mathematik und Physik, Albert-Ludwigs-Universit{\"a}t, Freiburg, Germany\\
$^{49}$ Section de Physique, Universit{\'e} de Gen{\`e}ve, Geneva, Switzerland\\
% ...
$^{88}$ Department of Physics and Astronomy, Michigan State University, East Lansing MI, United States of America\\
% ...
$^{105}$ Nikhef National Institute for Subatomic Physics and University of Amsterdam, Amsterdam, Netherlands\\
% ...
$^{111}$ Homer L. Dodge Department of Physics and Astronomy, University of Oklahoma, Norman OK, United States of America\\
$^{112}$ Department of Physics, Oklahoma State University, Stillwater OK, United States of America\\
% ...
$^{115}$ LAL, Universit{\'e} Paris-Sud and CNRS/IN2P3, Orsay, France\\
% ...
\end{flushleft}


\end{document}